\documentclass[aps,prc,twocolumn,nofootinbib,floatfix,superscriptaddress]{revtex4-2}

\usepackage{graphicx}
\usepackage{dcolumn}
\usepackage{amsmath}
\usepackage{amssymb}
\usepackage{psfrag}
\usepackage{epsfig}
\usepackage{epsf}
\usepackage{float}
\usepackage{slashed}
\usepackage{wasysym}

\usepackage{color}

\usepackage{multirow}

\definecolor{linkcolor}{rgb}{0,0,0.40} 
\usepackage[pdfpagelabels, pdfencoding=auto, psdextra]{hyperref}
\hypersetup{%
    pdfsubject=Paper,
    unicode = true,
    breaklinks = true,
    colorlinks = true,
    linkcolor = linkcolor,
    citecolor = linkcolor,
    menucolor = linkcolor,
    urlcolor = linkcolor
}

\graphicspath{{./figures/}}

\newcommand{\nmax}{\ensuremath{N_{\rm max}}}
\newcommand{\hw}{\ensuremath{\hbar\omega}}
\newcommand{\beq}{\begin{equation}}
\newcommand{\eeq}{\end{equation}}
\newcommand{\beqa}{\begin{eqnarray}}
\newcommand{\cbar}[0]{\bar c}
\newcommand{\eeqa}{\end{eqnarray}}
\newcommand{\nn}{\nonumber \\ }

\newcommand{\chead}[1]{\multicolumn{1}{c}{#1}}

\begin{document}


\title{Nuclear properties with semilocal momentum-space regularized chiral interactions beyond N$^2$LO}


%
\author{P.~Maris}
\email[]{pmaris@iastate.edu}
\affiliation{Department of Physics and Astronomy, Iowa State
  University, Ames, Iowa 50011, USA}

\author{R.~Roth}
\affiliation{Institut f\"ur Kernphysik, Technische Universit\"at 
Darmstadt, 64289 Darmstadt, Germany}
\affiliation{Helmholtz Forschungsakademie Hessen f\"ur FAIR, GSI Helmholtzzentrum, 64289 Darmstadt, Germany}
  
\author{E.~Epelbaum}
\affiliation{Ruhr-Universit\"at
  Bochum, Fakult\"at f\"ur Physik und Astronomie, Institut f\"ur Theoretische Physik II, D-44780 Bochum, Germany}

\author{R.J.~Furnstahl}
\affiliation{Department of Physics, The Ohio State University, 
Columbus, Ohio 43210, USA}

\author{J.~Golak}
\affiliation{M.~Smoluchowski Institute of Physics, Jagiellonian
University,  PL-30348 Krak\'ow, Poland}

\author{K.~Hebeler}
\affiliation{Institut f\"ur Kernphysik, Technische Universit\"at 
Darmstadt, 64289 Darmstadt, Germany}
\affiliation{ExtreMe Matter Institute EMMI, GSI Helmholtzzentrum für Schwerionenforschung GmbH, 
64291 Darmstadt, Germany}
\affiliation{Max-Planck-Institut f\"ur Kernphysik, Saupfercheckweg 1, 69117 Heidelberg, Germany}

\author{T.~H\"uther}
\affiliation{Institut f\"ur Kernphysik, Technische Universit\"at 
Darmstadt, 64289 Darmstadt, Germany}

\author{H.~Kamada}
\affiliation{Department of Physics, Faculty of Engineering,
Kyushu Institute of Technology, Kitakyushu 804-8550, Japan}

\author{H.~Krebs}
\affiliation{Ruhr-Universit\"at
  Bochum, Fakult\"at f\"ur Physik und Astronomie, Institut f\"ur Theoretische Physik II, D-44780 Bochum, Germany}

\author{H.~Le}
\affiliation{Institut f\"ur Kernphysik, Institute for Advanced Simulation 
and J\"ulich Center for Hadron Physics, Forschungszentrum J\"ulich, 
D-52425 J\"ulich, Germany}

\author{Ulf-G.~Mei{\ss}ner}
\affiliation{Helmholtz-Institut~f\"{u}r~Strahlen-~und~Kernphysik~and~Bethe~Center~for~Theoretical~Physics,
~Universit\"{a}t~Bonn,~D-53115~Bonn,~Germany}
\affiliation{Institut f\"ur Kernphysik, Institute for Advanced Simulation 
and J\"ulich Center for Hadron Physics, Forschungszentrum J\"ulich, 
D-52425 J\"ulich, Germany}
\affiliation{Tbilisi State University, 0186 Tbilisi, Georgia}
\affiliation{CASA, Forschungszentrum J\"ulich, D-52425 J\"ulich, Germany}

\author{J.A.~Melendez}
\affiliation{Department of Physics, The Ohio State University, 
Columbus, Ohio 43210, USA}

\author{A.~Nogga}
\affiliation{Institut f\"ur Kernphysik, Institute for Advanced Simulation 
and J\"ulich Center for Hadron Physics, Forschungszentrum J\"ulich, 
D-52425 J\"ulich, Germany}
\affiliation{CASA, Forschungszentrum J\"ulich, D-52425 J\"ulich, Germany}

\author{P.~Reinert}
\affiliation{Ruhr-Universit\"at
  Bochum, Fakult\"at f\"ur Physik und Astronomie, Institut f\"ur Theoretische Physik II, D-44780 Bochum, Germany}

\author{R.~Skibi\'nski}
\affiliation{M.~Smoluchowski Institute of Physics, Jagiellonian
University,  PL-30348 Krak\'ow, Poland}

\author{J.P.~Vary}
\affiliation{Department of Physics and Astronomy, Iowa State
  University, Ames, Iowa 50011, USA}

\author{H.~Wita{\l}a}
\affiliation{M.~Smoluchowski Institute of Physics, Jagiellonian
University,  PL-30348 Krak\'ow, Poland}

\author{T.~Wolfgruber}
\affiliation{Institut f\"ur Kernphysik, Technische Universit\"at 
Darmstadt, 64289 Darmstadt, Germany}

\collaboration{LENPIC Collaboration}


\date{\today}

\begin{abstract}
We present a comprehensive investigation of few-nucleon systems as
well as light and medium-mass nuclei up to $A=48$ using the current
Low Energy Nuclear Physics International Collaboration two-nucleon
interactions in combination with the third-order (N$^2$LO)
three-nucleon forces. To address the systematic overbinding
of nuclei starting from $A \sim 10$ found in our earlier study
utilizing the N$^2$LO two- and three-nucleon forces, we take into
account higher-order corrections to the two-nucleon potentials up through
fifth order in chiral effective field theory. The resulting Hamiltonian can be completely 
determined using the $A=3$ binding energies and selected 
nucleon-deuteron cross sections as input. It is then shown to predict other
nucleon-deuteron scattering observables and spectra
of light $p$-shell nuclei, for which a detailed correlated truncation error
analysis is performed, in agreement with experimental data.
Moreover, the predicted ground state energies of nuclei in the oxygen isotopic
chain from $^{14}$O to $^{26}$O as well as $^{40}$Ca and $^{48}$Ca show
a remarkably good agreement with experimental values, given that the Hamiltonian is fixed completely from the $A \leq 3$ data,
once the fourth-order (N$^3$LO) corrections to the two-nucleon
interactions are taken into account. On the other hand, the charge
radii are found to be underpredicted by $\sim
10\%$ for the oxygen isotopes and by almost $20\%$ for $^{40}$Ca and
$^{48}$Ca.  
\end{abstract}



\maketitle

\section{Introduction} 
\label{sec:intro}

Chiral effective field theory (EFT) and ab initio few- and many-body
methods play a key role in the quest for precision nuclear
theory \cite{Epelbaum:2008ga,Machleidt:2011zz,Epelbaum:2019kcf,Hammer:2019poc,Tichai:2020dna,Marcucci:2019hml,Tews:2020hgp,Piarulli:2020mop,Lazauskas:2020jlw,Lee:2020meg,Gandolfi:2020pbj,Soma:2020xhv,Hergert:2020bxy,Lahde:2019npb}. For the simplest nuclear system involving just two nucleons,
chiral EFT has already reached a high level of maturity in terms of
accuracy and precision. In particular, the latest-generation 
semilocal momentum-space regularized (SMS) nucleon-nucleon (NN)
potentials of Ref.~\cite{Reinert:2020mcu} at the highest available order N$^4$LO$^+$ provide, for
the regulator values $\Lambda = 450$ and $500$~MeV, a
nearly perfect description of mutually compatible neutron-proton and
proton-proton scattering data below $E_{\rm lab} = 300$~MeV with $\chi^2_{\rm datum} = 1.01$ \cite{Epelbaum:2022cyo}. 
This qualifies the N$^4$LO$^+$ NN potentials to be regarded as partial wave analysis from the point of view of NN data description, and also puts them among the most accurate and precise NN interactions to date. 
The determination of the pion-nucleon coupling constants
$f_0^2$, $f_p^2$ and $f_c^2$ from NN scattering data in Ref.~\cite{Reinert:2020mcu} and the
calculation of the deuteron structure radius $
r_{\rm str} = 1.9729^{+0.0015}_{-0.0012}$~fm in Ref.~\cite{Filin:2020tcs} provide
additional examples of recent chiral EFT calculations at a sub-percent accuracy level. 

Maintaining a comparable level of accuracy and precision beyond the NN
sector is currently not feasible because of both computational limitations
and unavailability of consistently regularized many-body forces and
exchange current operators beyond third order (N$^2$LO) of the chiral
EFT expansion \cite{Epelbaum:2019kcf}, which represents the main limiting factor by restricting the calculations to N$^2$LO. An
overview of ongoing efforts towards developing consistent many-body
forces and nuclear current operators can be found in Ref.~\cite{Epelbaum:2022cyo}.  In
the meantime, a series of detailed investigations of low-energy three-nucleon scattering
observables and selected properties of light and medium-mass nuclei at
low orders in chiral EFT has been performed by the
Low Energy Nuclear Physics International Collaboration (LENPIC)
using different variants of chiral EFT NN interactions \cite{Epelbaum:2014efa,Epelbaum:2014sza,Reinert:2017usi}
with and without the three-nucleon force (3NF) at N$^2$LO
\cite{vanKolck:1994yi,Epelbaum:2002vt}, see Refs.~\cite{LENPIC:2015qsz,LENPIC:2018lzt,LENPIC:2018ewt,Epelbaum:2019zqc,Maris:2020qne}. Some of the most interesting conclusions from these studies
include the explicit and implicit (i.e., based on the discrepancies between
calculated observables and experimental data) verification of the 3NF
effects being compatible with the expected size of N$^2$LO
corrections, in line with the Weinberg power counting \cite{Weinberg:1990rz,Weinberg:1991um}, as well
as insights into the convergence pattern of chiral EFT for nuclear
systems and implications for uncertainty quantification. 

In our recent paper \cite{Maris:2020qne}, we have calculated selected
nucleon-deuteron (Nd) elastic scattering and breakup observables,
properties of the $A=3$ and $A=4$ nuclei as well as spectra of
$p$-shell nuclei up to $A=16$ using the SMS potentials at leading
(LO), next-to-leading (NLO) and N$^2$LO from Ref.~\cite{Reinert:2017usi} in
combination with the 3NF at N$^2$LO regularized in the same way as the
SMS NN potentials. While the obtained predictions at N$^2$LO were
generally found to be consistent with experimental data within errors,
a systematic overbinding of nuclei was found starting from $A \sim
10$ and increasing with $A$. Furthermore, a slight underprediction was
observed for the $^4$He structure radius, which however still came out
consistently with the experimental value at the $95\%$ confidence
level, while the radii of heavier nuclei were not considered. 

The main purpose of this paper is to shed light on the origin of the
significant (even at the  $95\%$ confidence 
level) overbinding of heavier $p$-shell nuclei at N$^2$LO found in our
previous study \cite{Maris:2020qne}. To clarify whether this discrepancy is related 
to deficiencies of the NN force at N$^2$LO or rather has to be resolved by
higher-order corrections to the 3NF, we perform a series of
calculations based on the higher-order SMS NN potentials (N$^3$LO,
N$^4$LO and N$^4$LO$^+$) in combination with the 3NF at
N$^2$LO. While the obtained predictions are still accurate only at
the N$^2$LO level due to the missing contributions to the many-body
forces at N$^3$LO and beyond, we demonstrate that the overbinding
issue is resolved by including higher-order contributions to the NN
force. Moreover, we extend the results of Ref.~\cite{Maris:2020qne} to heavier
nuclei by performing calculations for the oxygen and calcium isotope
chains and study the convergence pattern of chiral EFT for the
corresponding charge radii. Last but not least, the large generated
set of calculated energy levels allows us to perform a more
detailed error analysis of the correlated excitation energies of the
considered nuclei. 

Our paper is organized as follows. In Sec.~\ref{sec:3Nbreakup} we
focus on the 3N systems and 
discuss the determination of the low-energy constants (LECs) entering
the N$^2$LO 3NF, a selected range of Nd elastic and
breakup scattering observables along with properties of $A=3$
nuclei. The main focus of Sec.~\ref{sec:light} is on light $p$-shell
nuclei. For $^4$He, we benchmark the calculations using the
No-Core Configuration Interaction (NCCI) method, applied to the Hamiltonians
softened by means of a Similarity Renormalization Group (SRG)
transformation, with the results obtained by solving the Yakubovsky
equations with bare interactions. Heavier $p$-shell nuclei considered
in this section are
calculated using the NCCI approach. We also present the analysis of the correlated truncation errors for the calculated spectra. 
Finally, our results for heavier nuclei up to $A=48$ obtained using
the in-medium no-core shell model (IM-NCSM)  approach are presented in
Sec.~\ref{sec:mediummass}, while the main conclusions are summarized in
Sec.~\ref{conclusion}.

\section{Three-nucleon systems}
\label{sec:3Nbreakup}

\subsection{Neutron-deuteron elastic scattering: Total cross section}
\label{sec:3N_1}

\begin{figure*}
\includegraphics[width=2.\columnwidth]{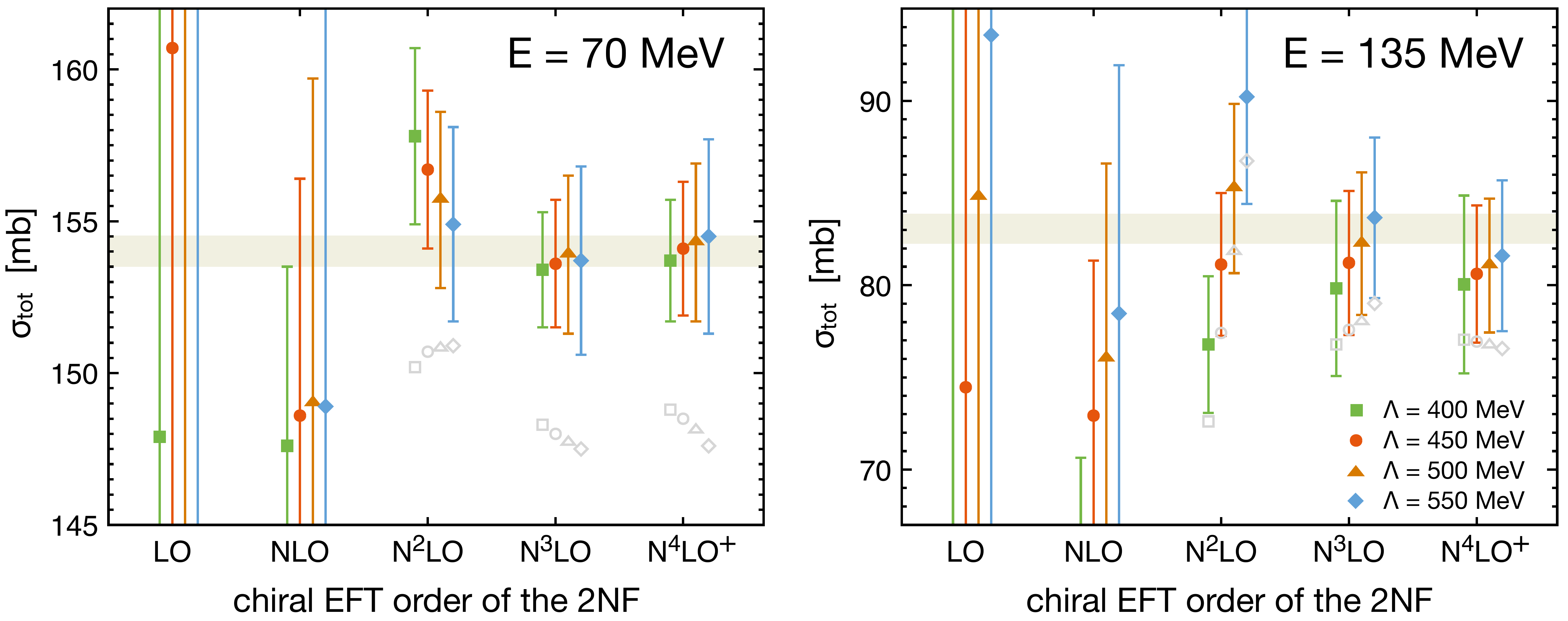}
  \caption{Predictions for the neutron-deuteron total cross-section at 70 MeV (left panel) and 135 MeV (right panel)
based on the semilocal momentum-space regularized chiral interactions at different orders
(shown by solid symbols with error bars). Three-nucleon force is included at N$^2$LO only.
Error bars show the EFT truncation uncertainty calculated using the Bayesian model $\bar C_{0.5-10}^{650}$
from Ref.~\cite{Epelbaum:2019zqc} (68\% DoB intervals). For the incomplete calculations at N$^3$LO and N$^4$LO, the quoted errors
correspond to  the N$^2$LO truncation uncertainties. Gray open symbols without error bars show the results
based on the two-nucleon forces only. Horizontal bands are experimental data from Ref.~\cite{Abfalterer:1998zz}.}
\label{TotalCS}
\end{figure*}

In this section, we present our results for 3N observables. We start with the 3N continuum and will discuss the bound states below. The continuum results  were obtained within the Faddeev approach, which is a well established method
of studying 3N processes. In brief, starting from a given two-nucleon potential, first the solution of the Lippmann-Schwinger equation
for the transition matrix (t-matrix) was obtained. Next, this t-matrix together with a 3N interaction entered the Faddeev equation 
for the auxiliary state $T$.
After solving this equation the transition amplitudes, both for the elastic and inelastic Nd scattering were computed.
Finally, these amplitudes were used to obtain observables: the differential cross sections and various 
polarization observables. All computations were performed using the partial wave decomposition in momentum space. 
That approach has been described in detail in Refs.~\cite{Gloeckle:1995jg,glockle1983quantum}. 

In Fig.~\ref{TotalCS}, we show the results for the total neutron-deuteron cross-section at the laboratory energies of $E = 70$~MeV and  $E = 135$~MeV for various chiral orders and all
available cutoff values. At LO and NLO, the results are based on the two-nucleon (NN) force only. Starting
from N$^2$LO, we also include the three-nucleon force (3NF) at N$^2$LO. The regularized form of the
employed 3NF is specified in Eq.~(1) of Ref.~\cite{Maris:2020qne}. Following this paper, the
two low-energy constants (LECs) $c_D$ and $c_E$ entering the 3NF are fixed from the triton binding
energy and the differential cross-section minimum at $E = 70$~MeV.\footnote{Notice that the determination
of $c_D$ and $c_E$ needs to be carried out for each value of the cutoff $\Lambda$ and for every order of
the NN force starting from N$^2$LO.} The results for the total cross-section shown in  Fig.~\ref{TotalCS},
therefore, come out as predictions. We do not show the results for the total cross-section
at low energies since it is governed by the S-wave contribution and known to be correlated with the
triton binding energy. Also, the effects of the 3NF we are interested in here start becoming significant
at intermediate energies above $E \sim 50$~MeV. 

The obtained results show a number of interesting features. First, as already pointed out in Ref.~\cite{LENPIC:2015qsz}
based on a different version of the chiral potentials, the NLO predictions appear to underestimate the total cross-section,
and the size of the discrepancy with the experimental data is roughly consistent with the NLO truncation errors.
Naturally, softer NLO NN interactions with smaller values of the cutoff $\Lambda$ show 
larger deviations from the data at high energies, which is most pronounced for $\Lambda = 400$~MeV.
Including higher-order corrections to the NN force up through N$^4$LO$^+$, the results for $\sigma_{\rm tot}$
tend to converge to values that underestimate the cross-section data by $\sim 4$\% ($\sim 7$\%) at $E = 70$~MeV
($E = 135$~MeV). These observations are in line with the systematics found using high-precision
phenomenological NN potentials \cite{Abfalterer:1998zz}, the feature that should not come as a surprise given the nearly perfect
description of NN data at N$^4$LO$^+$ \cite{Reinert:2017usi}.

The discrepancy between the predicted Nd scattering observables based on NN interactions only
and experimental data are expected to be resolved by the 3NF. In line with the chiral power counting,
the leading 3NF at N$^2$LO indeed brings the calculated total cross-section in agreement with the
data within N$^2$LO truncation errors. Also the magnitude of the 3NF effects appears to be consistent
with the expectations based on the power counting, see also Ref.~\cite{LENPIC:2015qsz} for a related discussion.
These findings are consistent with the results shown in Fig.~2 of Ref.~\cite{Maris:2020qne}.

The cutoff dependence of the obtained predictions also reveals interesting insights into the
convergence pattern of the chiral expansion. In particular, one observes that the
rather significant $\Lambda$-dependence of the N$^2$LO results at the larger energy
of $E = 135$~MeV is mostly absorbed into the ``running'' of the N$^3$LO NN contact
interactions. The remaining cutoff dependence of the predictions at N$^3$LO and N$^4$LO$^+$, both
with and without the 3NF, is significantly smaller than the N$^2$LO truncation error. This might be explained by the expectation for the residual
cutoff dependence to be taken care of by short-range 3NF operators that appear
at N$^4$LO.\footnote{Notice, however, that the strength of some of the short-range terms
is enhanced by a factor of $m/\Lambda_b$ \cite{Girlanda:2020pqn}, where $m$ and $\Lambda_b$ refer to the nucleon mass
and the breakdown scale of chiral EFT in the few-nucleon sector, respectively. This is because for the SMS NN interactions of Ref.~\cite{Reinert:2017usi}, a specific choice was made to remove the redundant (off-shell) N$^3$LO
contact interactions. However, the largely universal results for the Nd total cross-section based
on a broad class of different high-precision NN potentials seem to indicate that
this observable is almost insensitive to off-shell ambiguities of the NN force.}

\subsection{Nucleon-deuteron elastic and breakup scattering}
\label{sec:3N_2}

\begin{figure*}
\includegraphics[width=2.\columnwidth]{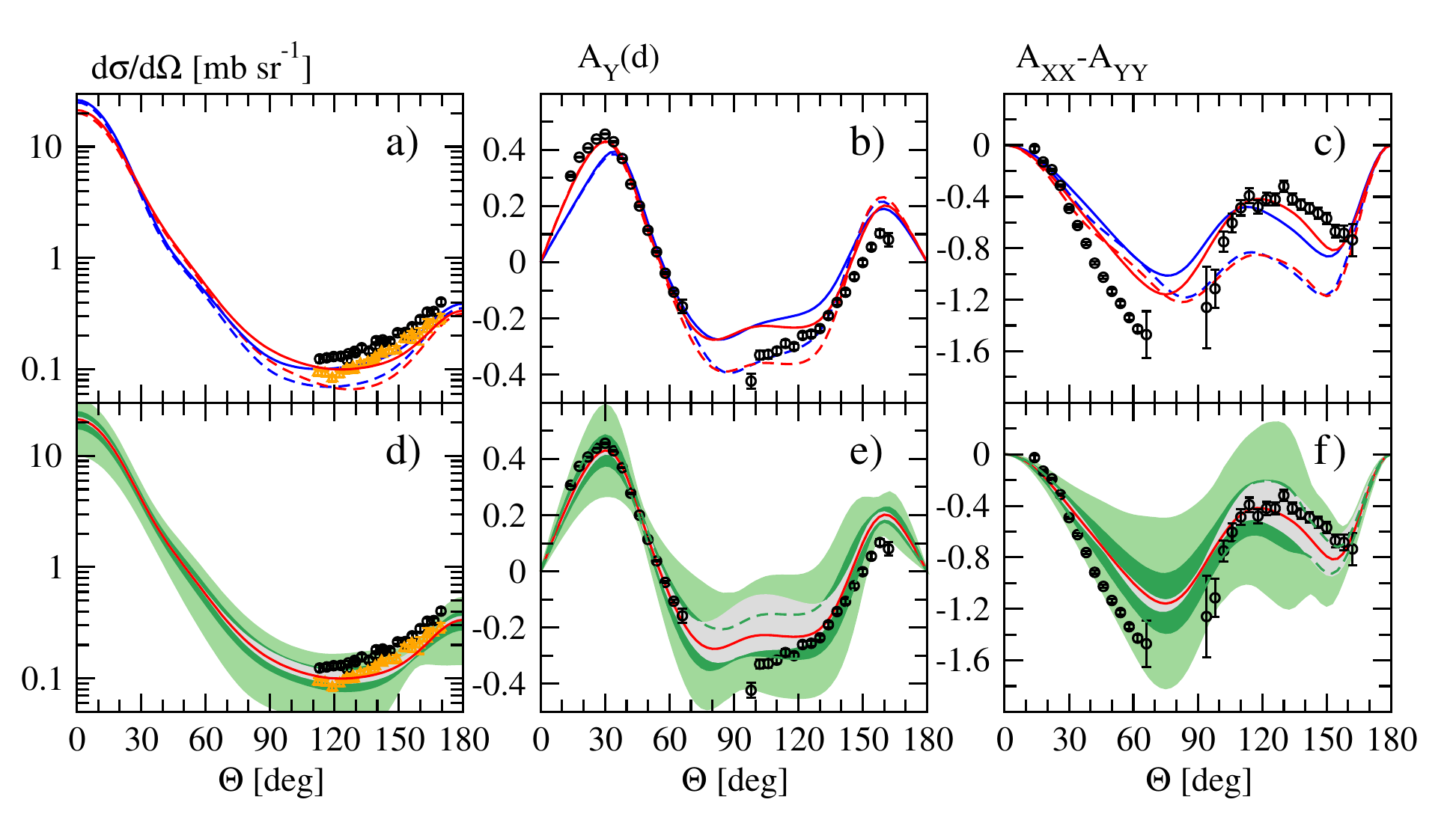}
  \caption{The center-of-mass differential cross section $\frac{{\rm d}\sigma}{{\rm d}\Omega}$, the deuteron vector analyzing power A$_Y{\rm(d)}$ and the the tensor analyzing power A$_{XX}-$A$_{YY}$ for the neutron-deuteron elastic scattering at incoming neutron lab. energy $E=200$~MeV. In the top panels the dashed blue (red) curve represents predictions based on the two-nucleon N$^2$LO (N$^4$LO$^+$) forces. The solid blue curve represents complete results at N$^2$LO and the solid red curve stands for predictions of N$^4$LO$^+$ NN interaction supplemented by N$^2$LO 3NF. In all cases, the cutoff $\Lambda=450$~MeV is used. In the bottom panels, the light (dark) green band shows the size of the truncation error at 95\% (68\%) DoB. The grey band shows a spread of the N$^4$LO$^+$ NN + N$^2$LO 3N force based predictions due to the value of $\Lambda$ regulator, in the range of $\Lambda \in [400-550]$~MeV. The red curve is the same as in the upper panels. In e) and f), the dashed dark-green curve shows the borders of the dark green band. Data in a) and d) are from ~\cite{IGO197233}: black circles for $E=181$~MeV and orange triangles for $E=216.5$~MeV. Data in b), c), e) and f) are from~\cite{Przewoski-135and200data}.}
  \label{fig:Ndelastic200}
\end{figure*}

Let us now turn to other observables in the elastic Nd scattering process. In this case 
predictions obtained at N$^2$LO with 2N and 3N interactions
for lower incoming nucleon kinetic energies ($E=65$~MeV and $E=135$~MeV) 
were shown in Ref.~\cite{Maris:2020qne}. 
Having now at our disposal the N$^4$LO$^+$ NN interaction 
we decided to investigate a higher energy case, which we choose to be $E=200$~MeV.
In the top panel of Fig.~\ref{fig:Ndelastic200}, we compare these new predictions, obtained with the N$^4$LO$^+$ NN interaction supplemented by the N$^2$LO 3NF, 
with the strict  N$^2$LO results. In addition, we show predictions solely based  on the N$^2$LO and N$^4$LO$^+$ NN interactions. 
For the differential cross section, taking into account higher terms in the NN interaction slightly modifies predictions 
at the center-of-mass scattering angles $\theta >80^{\circ}$. The effects of the 3NF 
only indirectly (through the values of the LECs $c_D$ and $c_E$)
depend on the order of the chiral NN force used, and the whole 
difference between N$^2$LO and N$^4$LO$^+$ NN predictions transfers to those for NN+3N forces. The data remain underpredicted 
in both cases, which is similar to the observations made for phenomenological forces~\cite{Witala_2001:PhysRevC.63.024007}. 
Among all possible polarization observables, which are more sensitive to details of the nuclear interactions, various situations 
can be found. In Figs.~\ref{fig:Ndelastic200}b and~\ref{fig:Ndelastic200}c, we show two examples: for the vector analyzing power A$_{\rm Y}$(d), the 3NF acts in a similar way
if combined with the N$^2$LO or N$^4$LO$^+$ NN potential, but the 3N force effects for the tensor analyzing power 
A$_{\rm XX}$-A$_{\rm YY}$ depend on the order of the NN force. Combining 
the N$^4$LO$^+$ NN interaction with the N$^2$LO 3NF delivers a slightly better data description, but definitely leaves room for improvement. 
The lower panels of Fig.~\ref{fig:Ndelastic200} shows theoretical uncertainties for the N$^4$LO$^+$ NN 
+ N$^2$LO 3N force predictions. At this rather high energy both 95\% and 68\% degree of belief (DoB) intervals 
for truncation errors remain wide but the data are at least in the first of these two intervals. 
The cut-off dependence, represented by the grey band comprising predictions with the regulator $\Lambda \in [400-550]$~MeV, is also 
significant at some scattering angles and comparable to the 68\% DoB truncation errors.

\begin{figure*}[t]
  \includegraphics[width=2.\columnwidth]{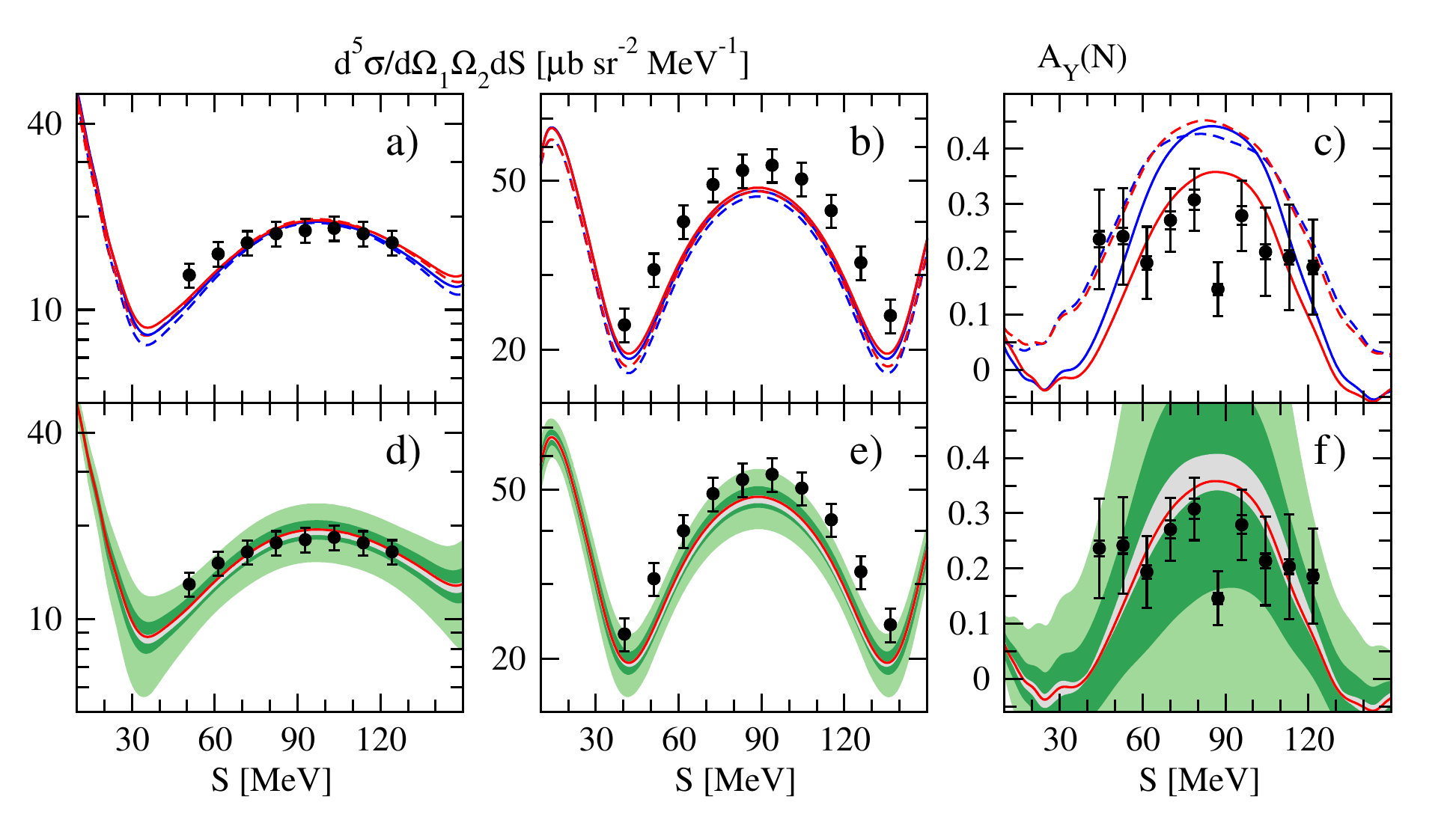}
  \caption{The differential cross section and the nucleon analyzing power A$_{\rm Y}$(N) at incoming neutron lab. energy $E=135$~MeV. The directions of momenta of outgoing neutrons are: in a) and d): $\theta_1=20^{\circ}, \theta_2=16^{\circ},$ and $\phi_{12}=180^{\circ}$, in b) and e): $\theta_1=28^{\circ}, \theta_2=28^{\circ},$ and $\phi_{12}=180^{\circ}$, and in c) and f): $\theta_1=20^{\circ}, \theta_2=16^{\circ},$ and $\phi_{12}=20^{\circ}$. In the top panels the dashed blue (red) curve represents predictions based on the two-nucleon N$^2$LO (N$^4$LO$^+$) forces. The solid blue curve represents complete results at N$^2$LO and the solid red curve stands for predictions of N$^4$LO$^+$ NN interaction supplemented by N$^2$LO 3NF. In all cases, the cutoff $\Lambda=450$~MeV is used. In the bottom panels, light (dark) green band shows size of truncation error at 95\% (68\%) DoB at $\Lambda=450$~MeV. The grey band shows a spread of the N$^4$LO$^+$ NN + N$^2$LO 3NF based predictions due to the variation of $\Lambda$  in the range $\Lambda \in [400-550]$~MeV. The red curve is the same as in the upper panels.
  Data are from Ref.~\cite{Tavakoli_data135breakup}. 
  }
  \label{fig:Ndbreakup135}
  \end{figure*}

\begin{figure*}[t]
  \includegraphics[width=2.\columnwidth]{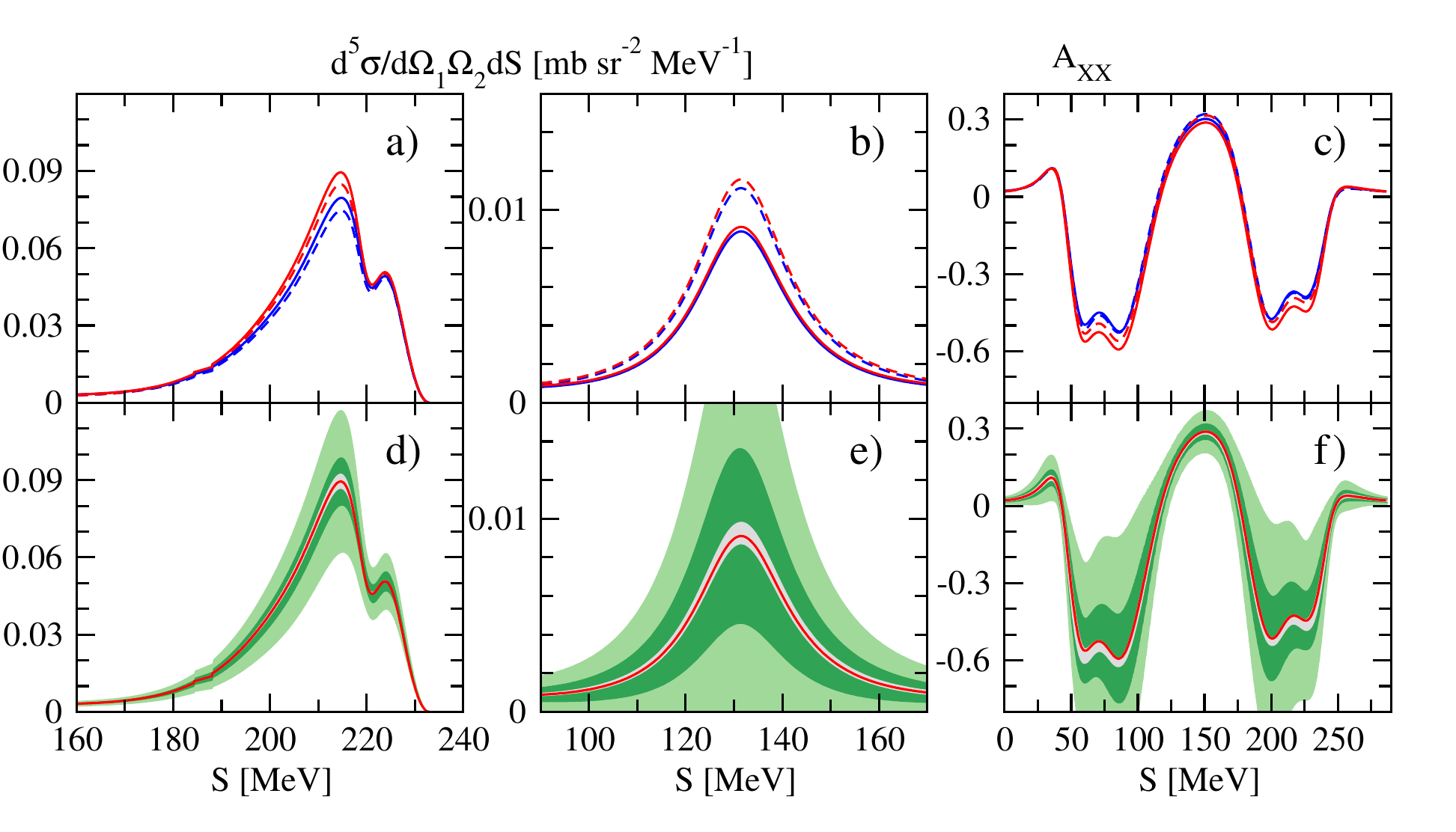}
  \caption{The differential cross section and the deuteron analyzing power A$_{\rm{XX}}$ at incoming neutron lab. energy $E=200$~MeV. The directions of momenta of outgoing neutrons are: in a) and d): $\theta_1=100^{\circ}, \theta_2=25^{\circ},$ and $\phi_{12}=180^{\circ}$, in b) and e): $\theta_1=15^{\circ}, \theta_2=10^{\circ},$ and $\phi_{12}=0^{\circ}$, and in c) and f): $\theta_1=50^{\circ}, \theta_2=50^{\circ},$ and $\phi_{12}=160^{\circ}$. The curves and bands are the same as in Fig.~\ref{fig:Ndbreakup200}.
  }
  \label{fig:Ndbreakup200}
  \end{figure*}

Next, in Figs.~\ref{fig:Ndbreakup135} and~\ref{fig:Ndbreakup200}, we show a few results for the differential cross section and the analyzing 
powers for selected kinematical configurations defined by the directions of two final 
proton momenta and the position on the S-curve~\cite{Gloeckle:1995jg}. 
In the case of the cross section at both energies ($E=135$~MeV in Fig.~\ref{fig:Ndbreakup135}, and $E=200$~MeV in Fig.~\ref{fig:Ndbreakup200}), the situation is
similar to elastic scattering: there are small differences in the predictions when replacing the NN forces.
These differences remain when the 3NF is included. At both energies, the cut-off dependence remains 
visibly smaller than the truncation errors. Depending on the kinematical configuration, the data description is satisfactory, 
or small discrepancies persist. The nucleon vector analyzing power A$_{\rm Y}$(N) shown in Fig.~\ref{fig:Ndbreakup135}c is characterized
by a strong effect of the 3NF when combined with the N$^4$LO$^+$ NN interaction, while the strictly N$^2$LO predictions
are insensitive to the 3NF for $70^{\circ} < \theta < 115^{\circ}$. Clearly, the 3NF combined with the N$^4$LO$^+$ NN force moves predictions towards 
the data, however large experimental errors do not allow us to
go beyond qualitative conclusions. At $E=200$~MeV, we show the tensor analyzing power A$_{\rm XX}$, for 
which, at both minimum points (around $\theta =75^{\circ}$ and $\theta =220^{\circ}$) the 3N force effects 
depend on the order of the NN interaction. As for the differential cross section, the truncation errors 
for the analyzing powers shown here are much bigger than the uncertainty related to the value of the cut-off.  

Summarizing, we find that the N$^4$LO$^+$ NN interaction, supplemented by the N$^2$LO 3N force yields a 
satisfactory description of the Nd continuum data, leaving however room for corrections from
higher orders of the three-nucleon interaction, see Ref.~\cite{Witala:contactN4LO} for recent work in this direction.

\subsection{Binding energies}
\label{sec:3N_3}

\begin{table*}[t]
  \begin{ruledtabular}
    \begin{tabular*}{\textwidth}{@{\extracolsep{\fill}}llcrrrrrrrrrrrrr}
         &                                   & \chead{$\Lambda$} &  \chead{$E$} &    \chead{$\langle H \rangle$} &   \chead{$\langle T \rangle$}&  \chead{$\langle V_{NN} \rangle$} &  \chead{$\langle V_{3NF} \rangle$} &  \chead{$\langle T_{CSB} \rangle$} &  \chead{$\langle \Psi | \Psi \rangle$} & \chead{P(S)} & \chead{P(P)} &  \chead{P(D)} & \chead{\quad$r_p$} & \chead{$r_n$} \\
\hline
 \multirow{6}{*}{$^3$H}   &   LO    &  \multirow{6}{*}{450}   &   $-$12.22\phantom{1} &      $-$12.24\phantom{1} &  52.38 &   $-$64.61 &   ---      & $-$10.51 &  1.0000 &   96.25 &   0.019 &    3.73 &    1.250 &    1.319 \\
   &   NLO                &                                   &   $-$8.515 &      $-$8.521 &  34.31 &   $-$42.82 &   ---      &  $-$5.80 &  0.9999 &   94.79 &   0.028 &    5.19 &    1.556 &    1.702 \\
    &   N$^2$LO     &                                  &   $-$8.483 &      $-$8.489 &  36.13 &   $-$44.16 &    $-$0.459 &  $-$5.84 &  0.9995 &   92.54 &   0.077 &    7.38 &    1.576 &    1.725 \\
    &   N$^3$LO  &                                       &  $-$8.483 &       $-$8.489 &  35.60 &   $-$43.56 &   $-$0.520 &     $-$5.72 &  0.9996 &    92.53 &   0.078 &     7.39 &    1.579 &    1.729 \\    
    &   N$^4$LO &                                       &  $-$8.483  &      $-$8.489 &   35.35 &  $-$43.40 &   $-$0.430 &     $-$5.75 &  0.9996 &    92.77 &   0.078 &     7.16 &    1.579 &    1.728  \\    
    &   N$^4$LO$^+$  &                                      &  $-$8.483 &      $-$8.489 &   35.46 &   $-$43.49 &   $-$0.460 &     $-$5.75 &  0.9996 &    92.64 &   0.079 &     7.28 &    1.580 &    1.729 \\    
\hline
\multirow{6}{*}{$^3$H}   &   LO  &  \multirow{6}{*}{500}      &   $-$12.52\phantom{1} &      $-$12.53\phantom{1} &  57.84 &   $-$70.36 &   ---      & $-$11.53 &  0.9999 &   94.96 &   0.036 &    5.01 &    1.224 &    1.286 \\
   &   NLO                &                                   &   $-$8.325 &      $-$8.332 &  35.87 &   $-$44.19 &   ---      &  $-$6.15 &  0.9998 &   94.29 &   0.032 &    5.68 &    1.575 &    1.725 \\
    &   N$^2$LO       &                                  &   $-$8.482 &      $-$8.488 &  40.27 &   $-$48.09 &   $-$0.660  &   $-$6.24 &  0.9992 &   91.39 &   0.109 &    8.50 &    1.581 &    1.731 \\
    &   N$^3$LO  &                                       &  $-$8.483 &       $-$8.489 &  37.83 &   $-$45.59 &   $-$0.724 &     $-$5.93 &  0.9994 &    91.80 &   0.103 &     8.10 &    1.580 &    1.731 \\  
    &   N$^4$LO  &                                       &  $-$8.483 &       $-$8.489 &   37.86 &   $-$45.72 &  $-$0.628 &     $-$6.07 &  0.9994 &    92.02 &   0.106 &     7.87 &    1.580 &    1.730 \\    
    &   N$^4$LO$^+$  &                                      &  $-$8.484 &      $-$8.490 &   38.08 &    $-$45.89 &   $-$0.672 &     $-$6.07 &  0.9994 &    91.84 &   0.108 &     8.05 &    1.582 &    1.731 \\    
\hline
Expt. $^3$H & & & $-$8.482 &  $-$8.482 &  --- &  --- &  --- &  --- &  --- &  --- &  --- &  --- &  1.604(96) &  ---  \\
\hline
 \multirow{6}{*}{$^3$He}   &      LO  &  \multirow{6}{*}{450} &   $-$11.34\phantom{1} &      $-$11.33\phantom{1} &  51.45 &   $-$62.79 &   ---      &   9.85 &  1.0000 &   96.24 &   0.019 &    3.75 &    1.342 &    1.264 \\
   &   NLO               &                                    &   $-$7.751 &      $-$7.745 &  33.55 &   $-$41.30 &   ---      &    5.22 &  0.9998 &   94.79 &   0.027 &    5.18 &    1.744 &    1.579 \\
   &   N$^2$LO      &                                    &   $-$7.734 &      $-$7.729 &  35.37 &   $-$42.65 &   $-$0.452  &    5.26 &  0.9995 &   92.57 &   0.076 &    7.35 &    1.766 &    1.598 \\
    &   N$^3$LO  &                                       &  $-$7.737   &     $-$7.732 &  34.85 &   $-$42.08 &   $-$0.509 &      5.15 &  0.9995 &    92.55 &   0.076 &     7.37 &    1.770 &    1.601  \\    
    &   N$^4$LO  &                                       &  $-$7.739   &     $-$7.734 &   34.61 &  $-$41.93 &   $-$0.423 &      5.18 &  0.9995 &    92.78 &   0.076 &     7.14 &    1.769 &    1.601  \\    
    &   N$^4$LO$^+$  &                                      &  $-$7.740 &       $-$7.734 &   34.72 &  $-$42.01 &   $-$0.452 &      5.18 &  0.9995 &    92.66 &   0.078 &     7.26 &    1.770 &    1.602\\    
\hline
 \multirow{6}{*}{$^3$He}   &   LO  &  \multirow{6}{*}{500}    &   $-$11.63\phantom{1} &      $-$11.62\phantom{1} &  56.88 &   $-$68.51 &   ---      &  10.87 &  0.9999 &   94.94 &   0.036 &    5.02 &    1.308 &    1.237 \\
   &   NLO               &                                    &   $-$7.574 &      $-$7.568 &  35.07 &   $-$42.65 &   ---      &     5.56 &  0.9997 &   94.30 &   0.031 &    5.67 &    1.768 &    1.598 \\
   &   N$^2$LO     &                                    &   $-$7.739 &      $-$7.733 &  39.44 &   $-$46.54 &   $-$0.641  &    5.65 &  0.9991 &   91.43 &   0.107 &    8.47 &    1.772 &    1.602 \\
    &   N$^3$LO &                                      &  $-$7.738 &        $-$7.733 &   37.04 &  $-$44.07 &   $-$0.705 &      5.34 &  0.9993 &    91.83 &   0.101 &     8.07 &    1.772 &    1.602  \\    
    &   N$^4$LO  &                                      &  $-$7.743 &        $-$7.737 &   37.08 &  $-$44.21 &   $-$0.615 &      5.49 &  0.9993 &    92.05 &   0.104 &     7.85 &    1.771 &    1.602 \\    
    &   N$^4$LO$^+$ &                                     &  $-$7.744 &        $-$7.738 &   37.29 &  $-$44.38 &   $-$0.658 &      5.49 &  0.9993 &    91.87 &   0.106 &     8.02 &    1.772 &    1.603 \\    
\hline 
Expt. $^3$He & & & $-$7.718 &  $-$7.718 &  --- &  --- &  --- &  --- &  --- &  --- &  --- &  --- &  1.792(17) &  --- \\
    \end{tabular*}
\caption{\label{tab:3H3He} Summary of energies and wave function properties for $^3$H/$^3$He for NN  interactions up to N$^4$LO$^+$ (including N$^2$LO 3NFs starting from N$^2$LO). Energies and cutoffs are given in MeV except for $\langle T_{CSB} \rangle$ which is given in keV. Radii are given in fm and the S,P and D-state probabilities are given in \%.
For the experimental values of the point-proton radii, we use the structure radii.}
\end{ruledtabular}   
\end{table*}

We now turn to the predictions for binding energies for $^3$He and $^3$H. The energies have been obtained 
by solving Faddeev equations in momentum space using a partial wave decomposition as described in \cite{Maris:2020qne}. For these calculations, the NN subsystem angular momenta are restricted 
to $j_{12} \le 5$. In order to take the full charge dependence into account, the 3N states include 
the dominant isospin $T=1/2$ and a small $T=3/2$ component. This is sufficient to obtain energies 
with a numerical uncertainty of 1~keV. For $^3$He also the point proton-proton Coulomb interaction 
is taken into account. The results are summarized in Table~\ref{tab:3H3He}. 
For the calculation of the energies and also for the fitting of the LECs $c_D$ and $c_E$ 
an averaged proton-neutron mass 
was employed. Afterwards the change $\left\langle T_{CSB} \right\rangle$ 
of the kinetic energy due to using physical proton and neutron masses is perturbatively 
estimated. In most cases, it is approximately of the order for 5-7~keV. This contribution is included in the 
results for the expectation value $\left\langle H \right\rangle$ but not in the energy $E$ obtained 
by solving the Faddeev equations. When excluding the contribution from the nucleon mass difference, 
the expectation value and the energy agree within 1~keV which is a non-trivial confirmation of the 
numerical accuracy. 

The results for LO, NLO and N$^2$LO have already been presented in  \cite{Maris:2020qne}.  
LO and NLO  overpredict the binding energy. It was also found in \cite{Maris:2020qne} that, 
at N$^2$LO, the prediction without 3NFs is slightly underbinding $^3$H and $^3$He. 
This also holds true for the higher order NN interactions. Therefore, the properly 
adjusted 3NF contributes attractively to the 3N systems. The contribution is comparable to the 
expectation values $\left\langle V_{3NF} \right\rangle$  of the 3NF. The deviations of the
N$^2$LO to N$^4$LO$^+$ results for the $^3$H energy from experiment are mostly due to the 
contribution of the proton-neutron mass difference that was not taken into account when 
fitting the 3NF parameters. The good agreement with experiment is of course no prediction but  
by construction. When $^3$H is used to fit the 3NF, $^3$He is slightly overbound compared 
to experiment. The difference is of the order of 20~keV, which is comparable to the 
contribution expected for charge-symmetry breaking NN forces 
and other electromagnetic contributions \cite{Nogga:2002qp}, which are not included here.

For symmetric operators, we exploit the faster convergence with respect to partial waves of Faddeev components 
compared to wave functions for the evaluation of expectation values and also for the normalization. 
Therefore, wave functions are not normalized to 1. The deviation of $\langle \Psi | \Psi \rangle$ from one 
is a measure  of higher partial wave contributions to the wave function and is below 0.1\%. We also give 
the probabilities P(S), P(P)  and P(D) for the 3N system being in a total orbital momentum $L=0,1$ and 2 
state. As expected, the S-state dominates the $A=3$ nuclei. Generally, the P-state contribution is tiny.
The D-state is more significant. A direct comparison to calculations without 3NF (not shown) reveals that 3NFs 
increase the D-state contribution. It also is enhanced for larger cutoffs $\Lambda$ as can be seen 
in the table. This is in line with results for phenomenological interactions \cite{Nogga:2002qp}. Finally, 
we also give the proton and neutron radii assuming point-like protons and neutrons (referred to as point-proton/neutron radii in the following). These quantities are correlated with the binding 
energies. It is reassuring that the results are quite independent of the cutoff and order once 
the 3NFs have been added. For a comparison to experiment, we include the 
structure radius defined by 
\begin{equation}
\label{eq:strucrad}
    r^2_{str} = r^2_c - \left( R_p^2+\frac{3}{4m_p^2} + \frac{N}{Z}R^2_n \right) 
\end{equation}
where $r_c$ is the charge radius of the nucleus, $R_p$ the proton charge radius and $R^2_n$ the
neutron charge radius squared. $N$ and $Z$ are the neutron and proton numbers. The values given here
have been obtained in \cite{Maris:2020qne} using the values of CODATA-2018 and the current PDG values 
for $R_p$ and $R_n$, respectively. The structure radius differs from the non-observable point-proton radius 
due to relativistic corrections and exchange charge density and similar contributions. For $^3$H,
the structure radius is in agreement with the point-proton radius within the experimental error bar. For $^3$He, 
we observe a slight underprediction of experiment.

\section{Light Nuclei}
\label{sec:light}

\subsection{Helium-4}

\begin{table*}[t]
    \begin{ruledtabular}
    \begin{tabular*}{\textwidth}{@{\extracolsep{\fill}}llcrrrrrrrrrrrrr}
      & \chead{$\Lambda$} &  \chead{$E$} &    \chead{$\langle H \rangle$} &   \chead{$\langle T \rangle$}&  \chead{$\langle V_{NN} \rangle_1$} &  \chead{$\langle V_{NN} \rangle_2$} &  \chead{$\langle V_{3NF} \rangle$} &  \chead{$\langle \Psi | \Psi \rangle_1$} & \chead{$\langle \Psi | \Psi \rangle_2$} & \chead{P(S)} & \chead{P(P)} &  \chead{P(D)} & \chead{\quad$r_p$} & \chead{$r_n$} \\
\hline
LO &    \multirow{6}{*}{450} & $-49.98$ & $-49.98$ & $124.44$ & $-174.42$ & $-174.43$ &  --- & 0.99994 & 0.99998 & $95.72$ & 0.070 &  $4.21$ & 0.991 & 0.988 \\
NLO &     & $-29.35$ & $-29.35$ &  $71.48$ & $-100.83$ & $-100.83$ &  --- & 0.99975 & 0.99962 & $92.03$ & 0.129 &  $7.84$ & 1.378 & 1.372 \\
N$^2$LO &     & $-28.61$ & $-28.61$ & $75.74$ & $-101.97$ & $-101.96$ & $-2.38$ & 0.99954 & 0.99927 & $86.72$ & 0.462 & $12.82$ & 1.426 & 1.420 \\
 N$^3$LO &     & $-28.35$ & $-28.35$ & $73.60$ &  $-99.43$ &  $-99.42$ & $-2.52$ & 0.99958 & 0.99935 & $87.10$ & 0.429 & $12.47$ & 1.433 & 1.427 \\
  N$^4$LO  &     & $-28.29$ & $-28.28$ & $73.04$ &  $-99.25$ &  $-99.25$ & $-2.07$ & 0.99958 & 0.99936 & $87.41$ & 0.429 & $12.16$ & 1.435 & 1.429 \\
N$^4$LO$^+$ &     & $-28.31$ & $-28.31$ & $73.45$ &  $-99.48$ &  $-99.47$ & $-2.27$ & 0.99958 & 0.99935 & $87.07$ & 0.446 & $12.48$ & 1.435 & 1.430 \\
\hline
  LO &    \multirow{6}{*}{500} & $-51.46$ & $-51.46$ & $139.21$ & $-190.67$ & $-190.67$ &  --- & 0.99991 & 0.99994 & $93.74$ & 0.147 &  $6.11$ & 0.957 & 0.954 \\
  NLO  &     & $-28.14$ & $-28.13$ &  $74.56$ & $-102.69$ & $-102.68$ &  --- & 0.99941 & 0.99895 & $90.96$ & 0.153 &  $8.89$ & 1.411 & 1.405 \\
 N$^2$LO &       & $-28.71$ & $-28.68$ & $86.73$ & $-111.93$ & $-111.88$ & $-3.48$ & 0.99895 & 0.99816 & $85.06$ & 0.598 & $14.35$ & 1.427 & 1.421 \\
 N$^3$LO &     & $-28.56$ & $-28.55$ & $80.30$ & $-105.00$ & $-104.97$ & $-3.85$ & 0.99929 & 0.99883 & $85.51$ & 0.568 & $13.92$ & 1.430 & 1.424 \\
 N$^4$LO  &     & $-28.48$ & $-28.47$ & $80.56$ & $-105.60$ & $-105.56$ & $-3.42$ & 0.99924 & 0.99872 & $85.79$ & 0.581 & $13.63$ & 1.433 & 1.428 \\
 N$^4$LO$^+$ &     & $-28.52$ & $-28.50$ & $81.20$ & $-105.97$ & $-105.94$ & $-3.73$ & 0.99922 & 0.99868 & $85.30$ & 0.606 & $14.10$ & 1.434 & 1.429 \\
 \hline 
 Expt.    &      &     $-$28.28   & $-$28.28 & ---        & ---        & ---        & ---        & ---        & ---        & ---        & ---        & ---        & 1.462(6)       & ---       
     \end{tabular*}                                                                                                                                                                                        
\caption{\label{tab:4He} Summary of energies and wave function
  properties for $^4$He for NN interactions up to N$^4$LO$^+$ (including N$^2$LO 3NFs starting from N$^2$LO). See text for explanations. Energies and cutoffs
  are given in MeV. The point-proton and neutron radii $r_p$ and $r_n$ are given in fm and the S, P and D-state
  probabilities are given in \%. For the experimental value of the point-proton radius, we use the structure 
  radius. The tiny numerical difference from the previous work \cite{Maris:2020qne} is due to the calculation including not only the isospin $T = 0$ component but also $T = 1$ and $T = 2$ ones.}
\end{ruledtabular}   
\end{table*}

We also performed Yakubovsky calculations in momentum space for $^4$He. The approach has been briefly described in \cite{Maris:2020qne}.
We present our results in Table~\ref{tab:4He}. For these calculations, we truncate the partial waves in several ways. First, the two-body 
subsystem total angular momentum is restricted to $j_{12} \le 5$, then the orbital angular momentum of the third 
and fourth nucleon or between two two-nucleon clusters is restricted to $l_i \le 6$. Finally, the sum of 
all orbital angular momenta is smaller than or equal to 10. For the calculations shown here, we also take the small admixtures of isospin $T=1$
and 2 to the dominant $T=0$ component into account. With these restrictions, our numerical accuracy is better than 10~keV for the 
binding energy and energy expectation values. Again, results up to N$^2$LO have already been shown in \cite{Maris:2020qne}. Note the small 
differences compared to the previous work that are due to the isospin $T=1$ and 2 components of the $^4$He state which were omitted 
in \cite{Maris:2020qne}. The calculations were performed using an averaged nucleon mass. For $^4$He, the contribution 
of the proton-neutron mass difference is tiny and is omitted. Due to the correlation of the 3N and 4N binding energy, we again find 
a considerable overbinding in LO and NLO. Starting from N$^2$LO, the 3N system is correctly bound. Nevertheless, 
there are still variations of the $^4$He binding energy 
of the order of 400~keV when the cutoff and/or order 
of the NN interaction is changed.   
The changes of energy at N$^4$LO are only of the order of 60~keV. The remaining deviations of the 
energies at the two cutoffs and the deviation from 
experiment can therefore be expected to be explained by the missing three- and four-nucleon forces at order N$^3$LO. Based on the contribution from 
NN interactions, we can expect to predict energies 
with an accuracy of 60 keV once a complete 
calculation up to N$^3$LO is performed. 

For the four-body system, using Yakubovsky equation 
has the advantage that the 4N states are simultaneously 
expanded in coordinates that single out a 3N subsystem 
(``3+1''coordinates) and coordinates that single out 
two two-body cluster (``2+2'' coordinates). 
The wave function can be represented in both kinds of coordinates. Similarly to the 3N system, 
we normalize the wave function using overlaps of 
Yakubovsky components and the wave function. 
Therefore, for our wave functions, the norm  
in ``3+1'' and ``2+2'' coordinates,  $\langle \Psi | \Psi \rangle_1$  and $\langle \Psi | \Psi \rangle_2$, respectively,
differ from 1. As can be seen in the table, the deviations are again less than 0.1\%. Generally, the norms indicate 
that the wave function is better represented in ``3+1'' coordinates. This also 
holds true for the evaluation of the expectation values of the NN interactions,  $\langle V_{NN} \rangle_1$ and  $\langle V_{NN} \rangle_2$. 
Therefore, we use the result in ``3+1'' coordinates $\langle V_{NN} \rangle_1$ for the evaluation of  $\langle H \rangle$. 
Note also that the expectation values of the 3NF $\langle V_{3NF} \rangle$ of the order of a few MeV 
are consistent with the power counting expectation. Compared to the 3N system, S-state probabilities are smaller while the P- and D-state probabilities 
are larger. Still, the S-state is dominating as can be expected for $^4$He. 

Finally,
we briefly comment on the non-observables point-proton and neutron radii. 
In the isospin $T=0$ approximation, both radii agree. The small 
difference is therefore entirely due to the $T=1$ and 2 components that are now included. 
We again compare the point-proton to the structure radius as defined in Eq.~\eqref{eq:strucrad}.
For $^4$He, we observe a small but visible underprediction of this value indicating 
that either the left out subleading 3NFs or the relativistic corrections or contributions of the exchange charge 
density are non-negligible for a high order prediction of the radii. Work in this direction is in progress. 


\subsection{Calculations for $p$-shell nuclei}

For selected $p$-shell, we use the No-Core Configuration Interaction 
(NCCI) approach~\cite{Barrett:2013nh} to calculate the ground states and 
low-lying narrow excited states.  In the NCCI approach we expand the
wave function $\Psi$ of a nucleus consisting of $A$ nucleons in an
$A$-body basis of Slater determinants $\Phi_k$ of single-particle wave
functions $\phi_{nljm}(\vec{r})$.  Here, $n$ is the radial quantum
number, $l$ the orbital motion, $j$ the total spin from orbital motion
coupled to the intrinsic nucleon spin, and $m$ the spin-projection.
The Hamiltonian ${\bf \hat H}$ is also expressed in this basis and
thus the many-body Schr\"odinger equation becomes a matrix eigenvalue
problem; for an NN potential plus 3NFs, this matrix is sparse for $A>4$.  
The eigenvalues of this sparse matrix are approximations to the energy
levels, to be compared to the experimental energy levels.

We use the conventional harmonic oscillator (HO) basis with energy
parameter $\hbar\omega$ for the single-particle wave functions, in
combination with a truncation on the total number of HO quanta in 
the system: the basis is limited to many-body basis states with 
$\sum_{A} N_i \le N_0 + \nmax$, with $N_0$ the minimal number of 
quanta for that nucleus and \nmax\ the truncation parameter.
In order to improve the convergence of the basis space expansion, 
we first apply a Similarity Renormalization Group (SRG)
transformation~\cite{Bogner:2007rx,Bogner:2009bt,Roth:2013fqa} to
soften these interactions.  All of the results for $p$-shell nuclei
presented here have been evolved to $\alpha=0.08$~fm$^4$ and all 
of them include (induced) 3NFs.  

Numerical convergence toward the exact results for a given 
Hamiltonian is obtained with increasing \nmax, and is marked by 
approximate \nmax\ and \hw\ independence.  
Furthermore, we apply the same procedure as in Refs.~\cite{Binder:2018pgl,Epelbaum:2018ogq,Maris:2020qne} to extrapolate
the approximate energy levels in finite bases to the complete (but infinite-dimensional) space~\cite{Maris:2008ax,Coon:2012ab,Furnstahl:2012qg,More:2013rma,Wendt:2015nba}.
Most of the actual numerical calculations to obtain the lowest eigenvalues 
of the increasingly large but sparse matrices were performed with 
the NCCI code MFDn~\cite{doi:10.1002/cpe.3129,SHAO20181} on the 
Cray XC40 Theta at the Argonne Leadership Computing Facility (ALCF), 
with additional calculations performed on the Cray XC40 Cori at the 
National Energy Scientific Computing Center (NERSC).

\subsection{Correlated truncation errors for spectra}
\label{sec:correlated_model}

For scattering observables, we have used a \emph{pointwise} Bayesian statistical model that estimates uncertainties learned from the order-by-order convergence pattern of the chiral expansion, but with each observable treated independently.
For the ground states and low-lying spectra in light nuclei presented in Tables~\ref{tab:EgsA4A6A8} through \ref{tab:Ex12C}, we take into account correlations between the convergence patterns of different observables.
This is particularly important for excitation energies, as they are a difference between excited- and ground-state energies that if treated independently would lead to individual errors added in quadrature.
As known from experience and the treatment in Ref.~\cite{Maris:2020qne},
these excitation energies are generally much better determined than energies of the individual levels.
Therefore, to avoid overestimating the truncation errors it is essential to apply a \emph{correlated} error model.

\begin{figure}[tbh]
  \includegraphics[width=0.85\columnwidth]{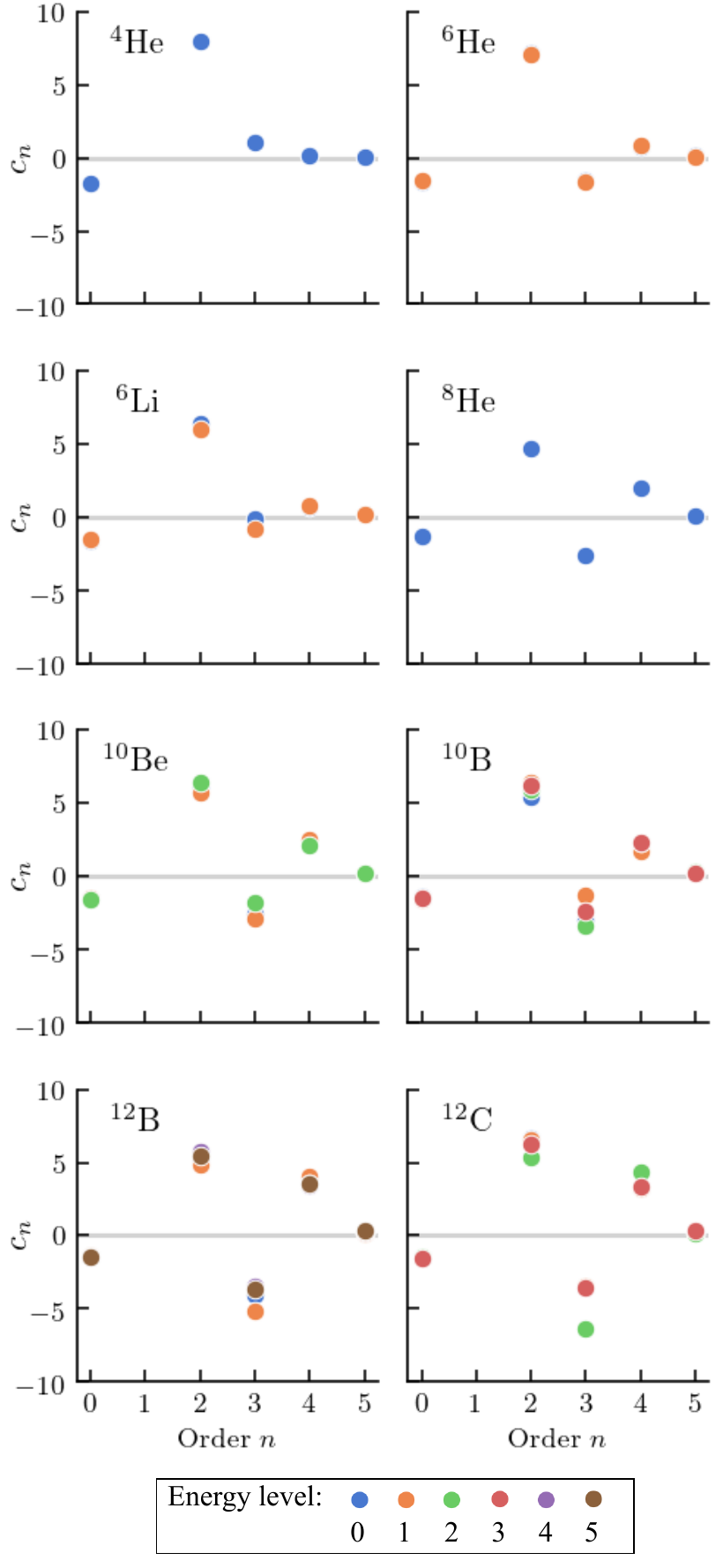}
  \caption{\label{fig:coefficients_450MeV} (Color online)
  Expansion coefficients defined as in Eq.~\eqref{eq:X_expansion} for the individual of the light nuclei listed with $\alpha = 0.08$~fm$^4$ and $\Lambda=450$~MeV. These are extracted with a fixed value of $Q \approx 0.31$ and $X_{\rm ref}$ taken from experiment~\cite{TILLEY20023,TILLEY2004155,KELLEY201771} (or the N$^2$LO result for the $0^+$ in $^8$Li).}
\end{figure}

\begin{figure}[tbh]
  \includegraphics[width=0.95\columnwidth]{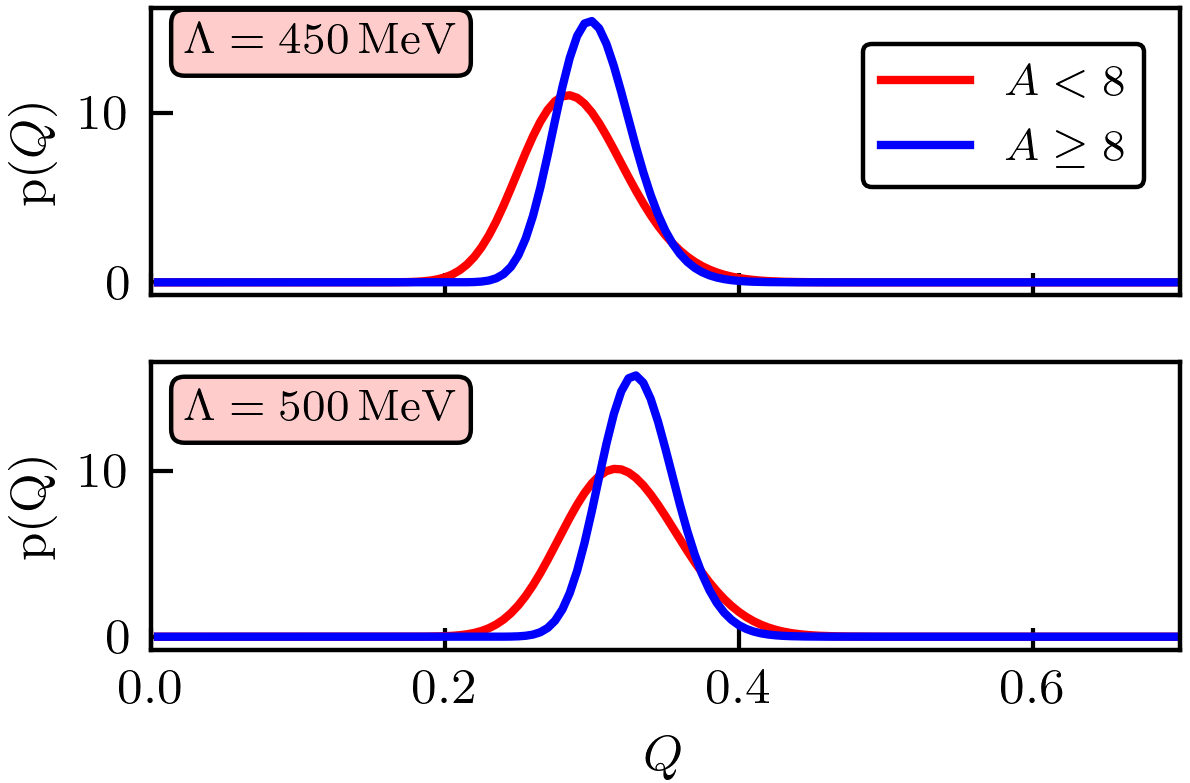}
  \caption{\label{fig:Q_posteriors} (Color online)
  Posteriors for the dimensionless expansion parameter $Q$ learned from the order-by-order coefficients for the ground and excited-state energies of the nuclei in Tables~\ref{tab:EgsA4A6A8} and \ref{tab:EgsA10A12}. 
  These coefficients encode the convergence pattern of the chiral expansion for these nuclei.
  The posteriors were extracted separately for nuclei with $A<8$ and $A\geq 8$ because of the differing degree of correlation between the coefficients in these two groups (see Fig.~\ref{fig:coefficients_450MeV}).
  The top panel is for $\Lambda=450\,\text{MeV}$ and the bottom
  panel for $\Lambda=500\,\text{MeV}$, both using $\alpha=0.08\,\text{fm}^4$.}
\end{figure}

A Bayesian model for correlated truncation errors based on Gaussian processes (GPs) was  developed in Ref.~\cite{Melendez:2019izc} and applied to infinite matter in Refs.~\cite{Drischler:2020hwi,Drischler:2020yad}. 
The adaptation of this model in Ref.~\cite{Maris:2020qne} to $p$-shell excitation energies was able to still use GPs for the discrete spectra, with every finite number of inputs having a joint Gaussian distribution.
The covariance structure between discrete energy levels and nuclei is learned from the observed pattern of order-by-order expansion coefficients $c_i$, which are defined for an observable $X$ by
\beqa \label{eq:X_expansion}
X &=& X^{(0)} + \Delta X^{(2)} + \Delta X^{(3)} + \dots \nn
  &=:& X_{\rm ref} \left( c_0 + c_2 Q^2 + c_3 Q^3 + \dots \right) \,.
\eeqa
Here $\Delta X^{(2)} := X^{(2)} - X^{(0)}$ and $\Delta X^{(3)} := X^{(3)} - X^{(2)}$, $Q$ is the expansion parameter,
the superscripts denote the chiral order $Q^n$, the ellipses refer to terms beyond N$^2$LO,
the quantity $X_{\rm ref} $ sets the overall scale and  
the $c_i$ are dimensionless.

There is a subtle but important complication to the Bayesian model for chiral EFT truncation errors in the present case, where the orders beyond N$^2$LO are incomplete because they include only the NN contributions.
It is clear that the expected error at N$^3$LO and higher orders should be counted the same as the expected error at N$^2$LO because in all cases there are omitted terms of N$^2$LO order.
But how should we \emph{extract} the $c_4$ and higher-order coefficients?
Naively one might argue that the differences of higher-order NN-only terms in \eqref{eq:X_expansion} should come with increasing powers of $Q$.
However, the LEC fitting of the NN interaction at each individual order to scattering data means only that they are two-body on-shell equivalent; there is an off-shell ambiguity for $A>2$.
With a complete N$^2$LO calculation, the off-shell ambiguity is resolved at the three-body level by the 3N contributions.
For N$^3$LO and higher, the residual N$^2$LO ambiguity persists, and in the absence of prior information it must be assumed that the difference in calculations of observables inherit an uncertainty of the same order as this ambiguity. Therefore we extract $c_4$ and higher using the same $Q$ counting as at N$^2$LO.

The correlations among the $c_i$ coefficients are manifested in Fig.~\ref{fig:coefficients_450MeV} through plots of the $c_i$'s  for ground states in Tables~\ref{tab:EgsA4A6A8} and \ref{tab:EgsA10A12} and the excited states in Tables~\ref{tab:ExA6A10} through \ref{tab:Ex12C}.
For this visualization, we extract the $c_i$s using a fixed $Q = M_\pi^{\rm  eff} / \Lambda_b = 200 / 650 \approx 0.31$ for all the states, with $X_{\rm ref}$ taken from experiment.
As already noted in Ref.~\cite{Maris:2020qne},
we see high correlation as expected between observable coefficients for the spectra of a given nucleus but also between nuclei, now continued beyond N$^2$LO.

To model these correlations, we introduce a covariance matrix and determine it empirically~\cite{scikit-learn}.
We emphasize that the correlations shown beyond $c_0$ are for the \emph{corrections} to the observables.
The truncation error model is contingent on the expansion parameter $Q$ and the characteristic variance $\cbar^2$ of the observable expansion coefficients $c_i$.
Unlike in Ref.~\cite{Maris:2020qne},
here we learn both $Q^2$ and $\cbar^2$ from the order-by-order calculations together with the prior expectations for each, which 
is possible because we now have enough higher-order coefficients for good statistics.
This is the case even though we omit low orders (namely $c_0$ and $c_2$) that obscure the order-by-order convergence pattern for the spectra of light nuclei because of the strong cancellation between kinetic and potential energies~\cite{Binder:2018pgl}.
(In Ref.~\cite{Maris:2020qne} only the $c_3$ coefficients were used to learn $\cbar^2$.)

The posteriors for the expansion parameter $Q$ that are learned from two sets of light nuclei are shown in Fig.~\ref{fig:Q_posteriors}. 
By ``learning'' we mean obtaining a statistical solution to the inverse problem of determining the distribution the coefficients come from (which is characterized by $\cbar^2$ and $Q$).
We use the hierarchical model from Appendix~A of Ref.~\cite{Melendez:2019izc}, which is computationally efficient and enables us to both parameterize our prior expectations and easily marginalize (i.e., integrate over) the hyperparameters to reduce sensitivity.
In Ref.~\cite{Maris:2020qne}, with only up to N$^2$LO available, we were sensitive to the choice of priors, but with higher orders included, this sensitivity is greatly reduced.
For the analysis here we use the scaled inverse-$\chi^2$ conjugate prior proposed in Ref.~\cite{Melendez:2019izc} with hyperparameters $\nu_0 = 1.5$ and $\tau_0 = 1.5$.

As seen in Fig.~\ref{fig:Q_posteriors}, the posteriors for $Q$ peak close to the value expected a priori ($Q\approx 0.3$), although the width of the posteriors is significant (and the $\Lambda$ = 500\,MeV results are slightly higher).
These fits were done separately for $A < 8$ and $A\geq 8$ nuclei because of different correlation patterns among the coefficients, as is evident in Fig.~\ref{fig:coefficients_450MeV}.
We expect in general that $Q$ should increase with the increasing average kinetic energy (the use of the non-observable kinetic energy in estimating $Q$ is discussed in Ref.~\cite{Binder:2018pgl}). 
This is consistent with the systematic trends of increasing $Q$ with increasing $A$ and $\Lambda$ in Fig.~\ref{fig:Q_posteriors}, but the broad widths preclude definitive conclusions.

The resulting Bayesian 95\% confidence intervals for the ground-state and excitation energies are given in Tables~\ref{tab:EgsA4A6A8}--\ref{tab:oxygen} and Figs.~\ref{fig:EgsA4A6A8}--\ref{fig:oxygen}.
The reduced error bars for excitation energies can be understood quantitatively through the formula for the variance of the difference of two Gaussian-distributed variables $A$ and $B$ with correlation coefficient $\rho$:
\beq
  \sigma^2_{A-B} = \sigma^2_A + \sigma^2_B - 2\rho \sigma_A \sigma_B 
  . \label{eq:Gaussian_difference}
\eeq
The learned values of $\rho$ were mostly between 0.85 and 0.9, which by \eqref{eq:Gaussian_difference} implies the correlated excitation-energy error bars are about 0.3--0.4 times the values from adding in quadrature.

\subsection{Ground state energies of $p$-shell nuclei}

\begin{table*}[t]
    \begin{ruledtabular}
    \begin{tabular*}{\textwidth}{lllll}
                    & \chead{$^{4}$He $(0^+)$} & \chead{$^{6}$He $(0^+)$} & \chead{$^{8}$He $(0^+)$} & \chead{$^{6}$Li $(1^+)$} \\
\hline \\[-9pt]
\multicolumn{5}{c}{$\Lambda = 450$ MeV} \\
\hline \\[-9pt]
 LO                 &  $-49.73(0.01)(*)$    &  $-46.7(0.3)(*)$      &  $-41.6(0.9)(*)$      &  $-50.4(0.3)(*)$  \\
 NLO                &  $-29.37(0.01)(3.6)$  &  $-27.86(0.14)(3.7)$  &  $-28.2(0.7)(6.0)$    &  $-31.93(0.09)(4.1)$  \\
 N$^2$LO            &  $-28.53(0.01)(1.0)$  &  $-29.04(0.07)(1.0)$  &  $-30.42(0.20)(1.8)$  &  $-32.04(0.06)(1.2)$  \\
 N$^3$LO            &  $-28.38(0.01)(1.0)$  &  $-28.39(0.08)(1.0)$  &  $-28.69(0.23)(1.8)$  &  $-31.41(0.06)(1.2)$  \\
 N$^4$LO            &  $-28.29(0.01)(1.0)$  &  $-28.28(0.08)(1.0)$  &  $-28.62(0.24)(1.8)$  &  $-31.28(0.06)(1.2)$  \\
 N$^4$LO$^+$           &  $-28.29(0.01)(1.0)$  &  $-28.33(0.07)(1.0)$  &  $-28.75(0.24)(1.8)$  &  $-31.32(0.06)(1.2)$  \\
 \hline \\[-9pt]
\multicolumn{5}{c}{$\Lambda = 500$ MeV} \\
\hline \\[-9pt]
 LO                 &  $-51.17(0.01)(*)$    &  $-47.6(0.4)(*)$      &  $-41.6(1.0)(*)$      &  $-51.1(0.3)(*)$  \\
 NLO                &  $-28.12(0.01)(3.6)$  &  $-27.39(0.10)(3.8)$  &  $-26.3(0.6)(6.9)$    &  $-31.45(0.06)(4.1)$  \\
 N$^2$LO            &  $-28.63(0.01)(1.2)$  &  $-29.21(0.06)(1.2)$  &  $-30.92(0.15)(2.3)$  &  $-32.29(0.04)(1.3)$  \\
 N$^3$LO            &  $-28.45(0.01)(1.2)$  &  $-28.54(0.08)(1.2)$  &  $-29.06(0.20)(2.3)$  &  $-31.61(0.05)(1.3)$  \\
 N$^4$LO            &  $-28.31(0.01)(1.2)$  &  $-28.37(0.07)(1.2)$  &  $-28.91(0.18)(2.3)$  &  $-31.41(0.05)(1.3)$  \\
 N$^4$LO$^+$           &  $-28.30(0.01)(1.2)$  &  $-28.41(0.07)(1.2)$  &  $-29.04(0.17)(2.3)$  &  $-31.43(0.05)(1.3)$  \\
\hline\\[-9pt]
Expt.               & $-28.296$    & $-29.27$       &  $-31.41$      &  $-31.99$
     \end{tabular*}
\caption{\label{tab:EgsA4A6A8} Ground-state energies of Helium isotopes and $^6$Li 
obtained with the NCCI approach for different chiral orders and cutoffs, 
SRG evolved to $\alpha = 0.08$~fm$^4$. Numbers in parenthesis indicate first the 
estimated extrapolation uncertainties and then the chiral truncation uncertainties at the 95\% confidence level, $(*)$ indicating no chiral truncation uncertainties at LO.}
\end{ruledtabular}   
\end{table*}
Our results for the stable Helium isotopes, as well as $^6$Li, are given in Table~\ref{tab:EgsA4A6A8} for different chiral orders and two values of the regulators, both SRG evolved to $\alpha = 0.08$~fm$^4$.  Induced 3NFs from the SRG evolution are included in all calculations, and starting from N$^2$LO, the explicit N$^2$LO 3NFs are also included; however, induced four-nucleon (and higher, for $A \ge 6$) are neglected in these calculations.  

For $^4$He we can compare the NCCI results using SRG evolved interactions with the Yakubovsky calculations of Table~\ref{tab:4He}, which are obtained without SRG evolution.  The numerical uncertainties in both the Yakubovsky results and the NCCI results are of the order of 10 keV or less.  Therefore, any differences in the ground state energies of $^4$He in Tables~\ref{tab:4He} and \ref{tab:EgsA4A6A8} beyond 10 keV are due to missing induced four-nucleon interactions.  For $\Lambda=450$~MeV, this difference is small, in most cases only of the order of 20~keV, but, for $\Lambda=500$~MeV, the SRG evolution leads to changes of the energy of about 200~keV.  This is still smaller than the expected contribution of N$^3$LO 3NFs, but will possibly become relevant once these subleading 3NFs are included in complete N$^3$LO calculations in the future.  

Based on the comparison with the Yakubovsky calculations for $^4$He, we anticipate an uncertainty of about 0.5\% to 1\% for $A \ge 6$ nuclei, due the SRG evolution to $\alpha = 0.08$~fm$^4$.   This SRG dependence is similar in magnitude to the estimated extrapolation uncertainties listed in Table~\ref{tab:EgsA4A6A8}.  Note that the effect of the omitted induced four-nucleon forces depends not only on the actual SRG parameter $\alpha$, but also on the interaction (that is, for these calculations, on the chiral order and on the regulator $\Lambda$), as well as on the nucleus. 

\begin{figure}[t]
  \includegraphics[width=0.53\columnwidth]{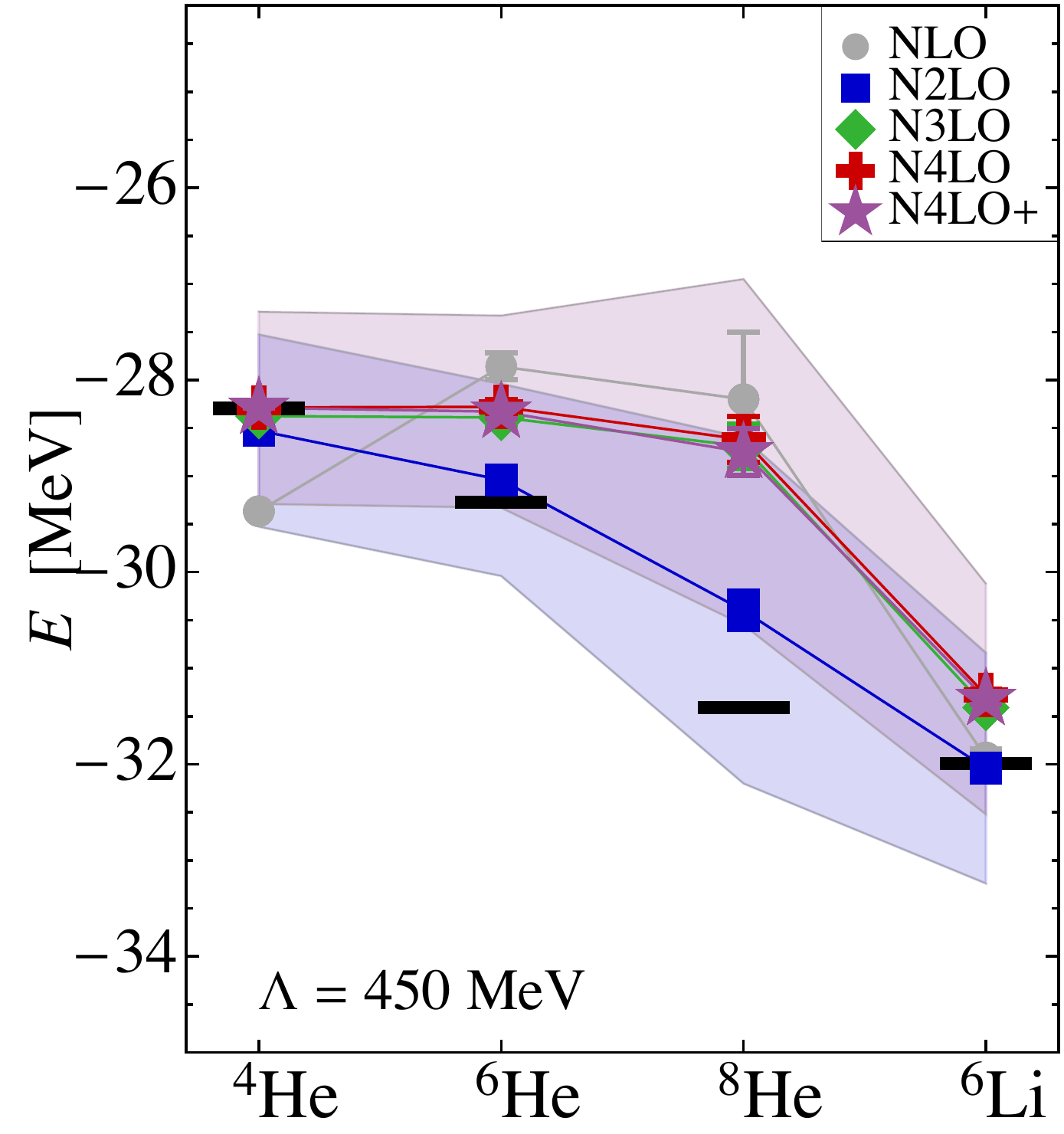}\hspace*{-2pt}
  \includegraphics[width=0.53\columnwidth]{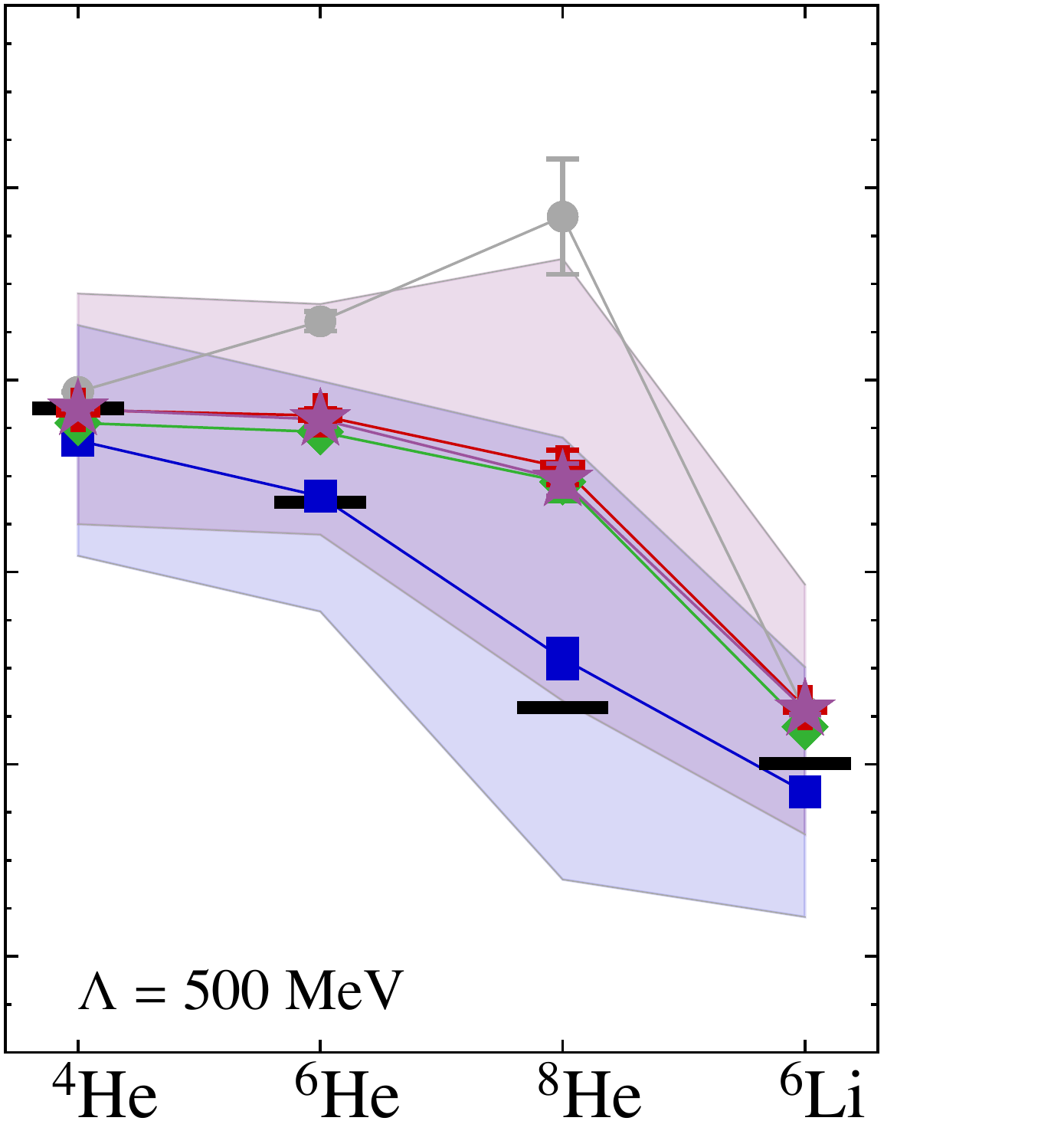}\\
  \caption{\label{fig:EgsA4A6A8} (Color online)
  Ground-state energies for $^4$He, $^6$He, $^8$He, and $^6$Li with SMS interactions 
  from NLO to N$^4$LO$^+$ with $\Lambda=450\,\text{MeV}$ (left-hand panel) and
  $\Lambda=500\,\text{MeV}$ (right-hand panel), both using $\alpha=0.08\,\text{fm}^4$.
  Error bars indicate the NCSM model-space uncertainties and shaded bands indicate 
  the chiral truncation uncertainties at the 95\% confidence level for the N$^2$LO (blue) 
  and N$^4$LO$^+$ (purple) NN potentials. 
  Horizontal bars show the experimental ground-state energies.}
\end{figure}

The second set of uncertainties in Table~\ref{tab:EgsA4A6A8} correspond 
to our Bayesian 95\% confidence intervals as discussed in the previous subsection;
and these results are shown in Fig.~\ref{fig:EgsA4A6A8}, with error bars for the 
numerical uncertainties, and the 95\% confidence intervals for the N$^2$LO NN 
potential (blue shaded band) and for the N$^4$LO$^+$ NN potential (purple shaded band).
This clearly shows that the chiral truncation uncertainties are both noticeably 
larger than the numerical uncertainties, and larger than the estimated 
SRG uncertainties of about 1\%.  With the exception of $^8$He, our predictions
for the ground state energies of these nuclei agree with the experimental
data within the estimated chiral uncertainty, for each of the N$^2$LO, 
N$^3$LO, N$^4$LO, and N$^4$LO$^+$ NN potentials in combination with the N$^2$LO 3NF.  
However, the central values with the N$^2$LO NN potential are noticeably closer 
to the experimental ground state energies of these light nuclei than those obtained 
with the N$^3$LO, N$^4$LO, or N$^4$LO$^+$ NN potential; while the difference between  
N$^3$LO, N$^4$LO, and N$^4$LO$^+$ results is of the same order as the estimated numerical 
and SRG uncertainties.  For $^8$He however, only the N$^2$LO NN potential agrees
with the experimental ground state energy within the 95\% confidence interval,
whereas the higher-order NN potentials lead to an underprediction of the 
$^8$He ground state energy.  Furthermore, note that the central values
for the He isotopes with the N$^3$LO, N$^4$LO, and N$^4$LO$^+$ NN potentials 
suggest increasing underbinding as one moves away from $N=Z$.  It remains 
to be seen how this changes when consistent N$^3$LO 3NFs are incorporated in future work.

\begin{figure}[t]
  \includegraphics[width=0.53\columnwidth]{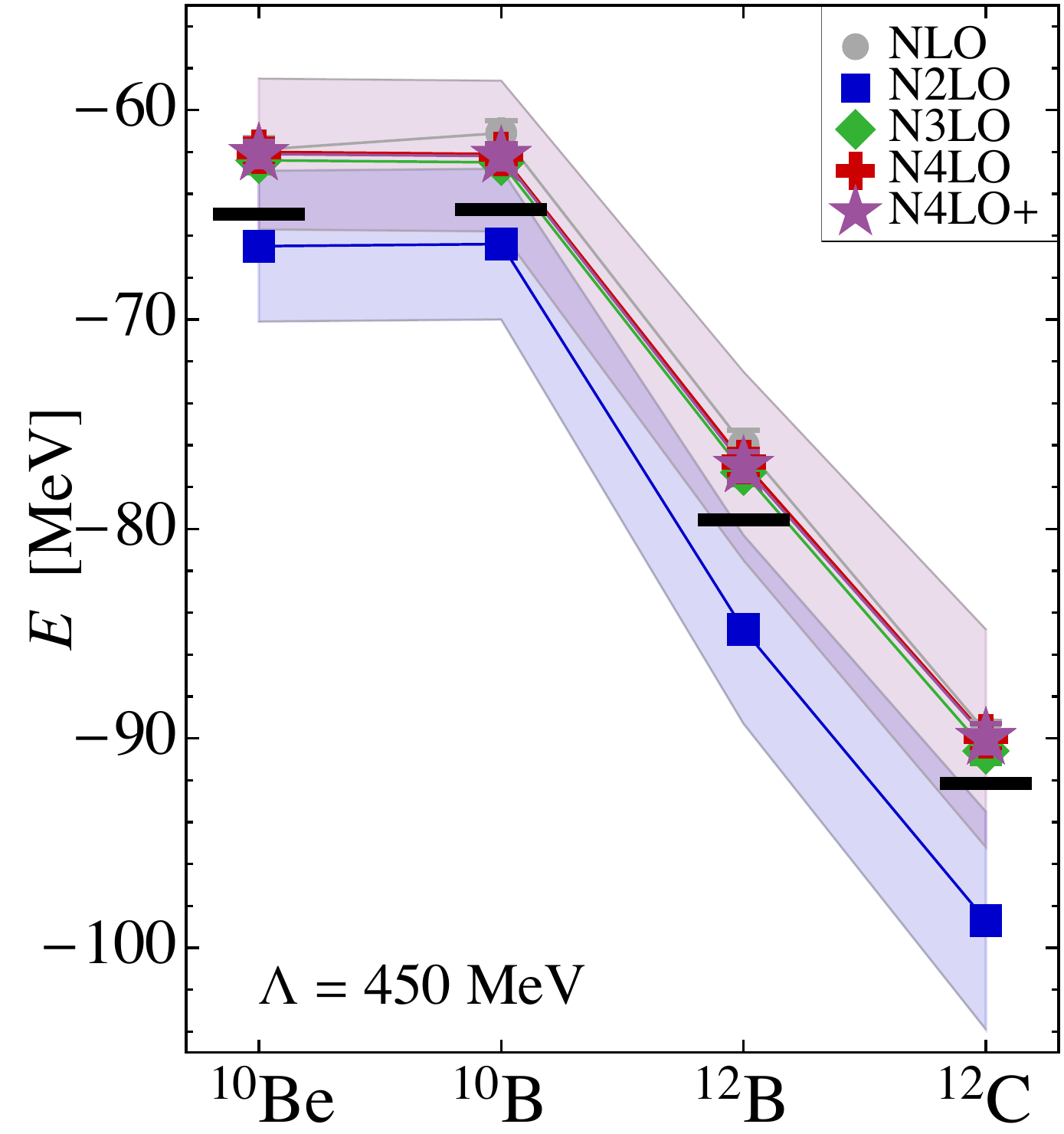}\hspace*{-2pt}
  \includegraphics[width=0.53\columnwidth]{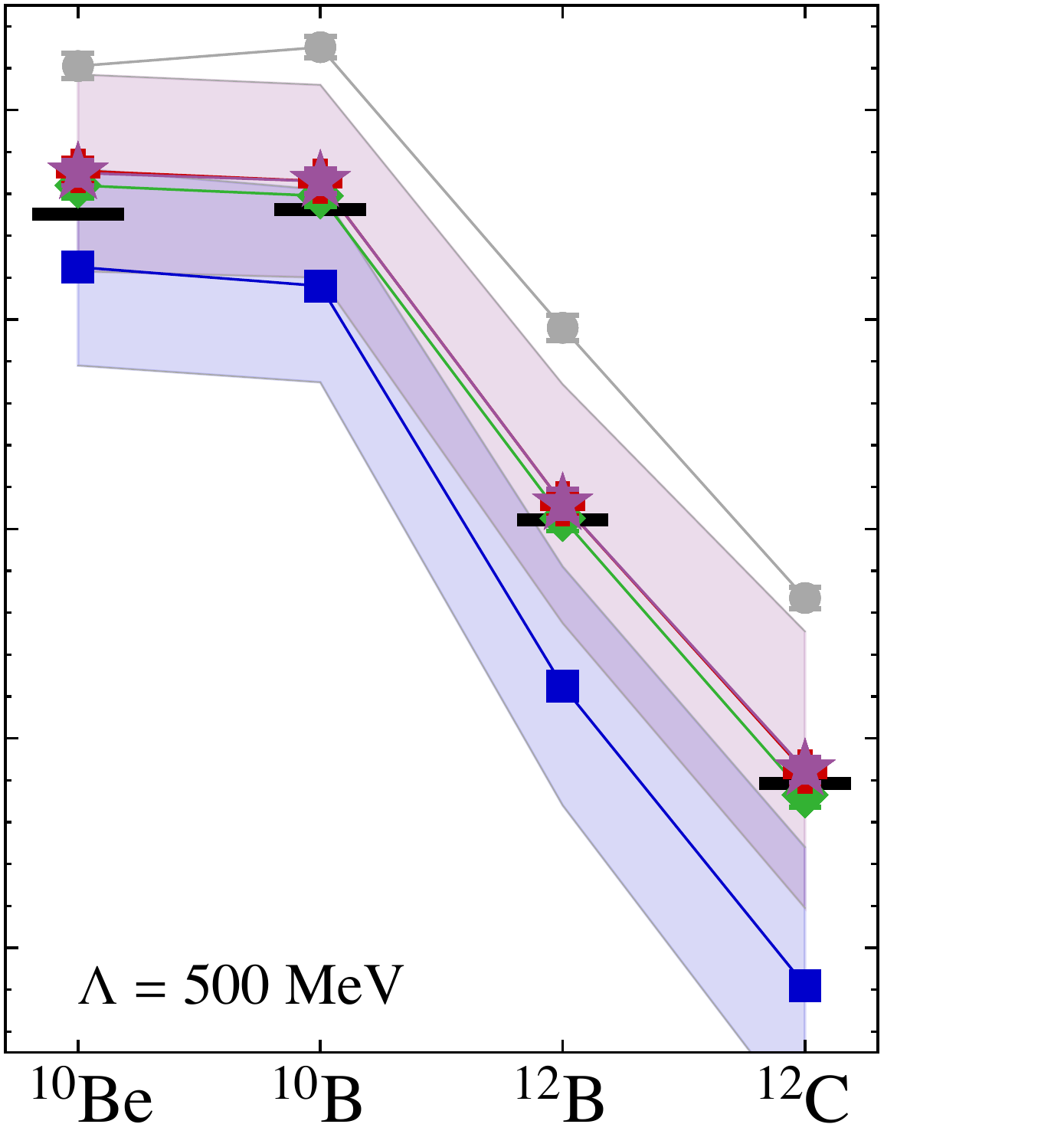}\\
  \caption{\label{fig:EgsA10A12} (Color online)
  Ground-state energies for $^{10}$Be, $^{10}$B, $^{12}$B, and $^{12}$C with SMS interactions 
  from NLO to N$^4$LO$^+$ with $\Lambda=450\,\text{MeV}$ (left-hand panel) and
  $\Lambda=500\,\text{MeV}$ (right-hand panel), both using $\alpha=0.08\,\text{fm}^4$.
  Error bars and bands are the same as in Fig.~\ref{fig:EgsA4A6A8}.
  Horizontal bars show the experimental ground-state energies.}
\end{figure}

\begin{table*}[t]
    \begin{ruledtabular}
    \begin{tabular*}{\textwidth}{lcccccc}
                    & $^{10}$Be $(0^+)$ & $^{10}$B $(3^+)$ & $^{12}$B $(1^+)$ & $^{12}$C $(0^+)$ & $^{14}$O $(0^+)$\\
\hline \\[-9pt]
\multicolumn{5}{c}{$\Lambda = 450$ MeV} \\
\hline \\[-9pt]
 LO                 &  $-97.7(1.5)(*)\phantom{1.}$    &  $-92.8(1.6)(*)\phantom{1.}$    &  $-113.7(1.3)(*)\phantom{11.}$     &  $-145.0(0.9)(*)\phantom{11.}$   & $-152.2(0.7)(*)\phantom{11.} $\\
 NLO                &  $-61.9(0.6)(12.)$  &  $-61.1(0.6)(12.)$  &  $ -76.0(0.7)(15.)$   &  $ -89.7(0.5)(17.)$ & $ -98.1(0.7)(32.) $ \\
 N$^2$LO            &  $-66.5(0.5)(3.6)$  &  $-66.4(0.4)(3.6)$  &  $ -84.8(0.4)(4.5)$   &  $ -98.7(0.4)(5.2)$ & $-113.1(0.4)(11.) $ \\
 N$^3$LO            &  $-62.4(0.6)(3.6)$  &  $-62.5(0.5)(3.6)$  &  $ -77.3(0.6)(4.5)$   &  $ -90.6(0.6)(5.2)$ & $ -99.8(0.8)(11.) $ \\
 N$^4$LO            &  $-62.0(0.6)(3.6)$  &  $-62.1(0.5)(3.6)$  &  $ -76.8(0.6)(4.5)$   &  $ -89.9(0.6)(5.2)$ & $ -99.0(0.8)(11.) $ \\
 N$^4$LO$^+$           &  $-62.1(0.6)(3.6)$  &  $-62.2(0.6)(3.6)$  &  $ -77.0(0.7)(4.5)$   &  $ -90.0(0.7)(5.2)$ & $ -99.2(0.8)(11.)$  \\
 \hline \\[-9pt]
\multicolumn{5}{c}{$\Lambda = 500$ MeV} \\
\hline \\[-9pt]
 LO                 &  $-98.1(1.7)(*)\phantom{1.}$    &  $-92.5(2.0)(*)\phantom{1.}$    &  $-111.7(1.6)(*)\phantom{11.}$     &  $-144.6(1.3)(*)\phantom{11.}$   & $-148.2(0.9)(*)\phantom{11.}$ \\
 NLO                &  $-57.9(0.6)(14.)$  &  $-57.0(0.5)(14.)$  &  $ -70.4(0.6)(17.)$   &  $ -83.3(0.5)(20.)$ &  $-89.6(0.6)(39.)$  \\
 N$^2$LO            &  $-67.5(0.4)(4.7)$  &  $-68.4(0.4)(4.6)$  &  $ -87.5(0.4)(5.7)$   &  $-101.8(0.4)(6.6)$ & $-116.9(0.4)(13.)$  \\
 N$^3$LO            &  $-63.6(0.6)(4.7)$  &  $-64.1(0.5)(4.6)$  &  $ -79.5(0.6)(5.7)$   &  $ -92.7(0.6)(6.6)$ & $-103.3(0.8)(13.)$  \\
 N$^4$LO            &  $-62.9(0.6)(4.7)$  &  $-63.4(0.6)(4.6)$  &  $ -78.8(0.6)(5.7)$   &  $ -91.6(0.6)(6.6)$ & $-102.0(0.9)(13.)$  \\
 N$^4$LO$^+$           &  $-63.0(0.6)(4.7)$  &  $-63.4(0.6)(4.6)$  &  $ -78.8(0.7)(5.7)$   &  $ -91.5(0.6)(6.6)$ & $-101.9(0.9)(13.)$  \\
\hline \\[-9pt]
Expt.               &   $-64.98$       &   $-64.75$     &   $-79.58$       &  $-92.16$ & $-98.7$
     \end{tabular*}
\caption{\label{tab:EgsA10A12} Ground-state energies of $A=10$, $12$, and $^{14}$O nuclei obtained in the NCSM for different chiral orders and cutoffs, SRG evolved to $\alpha = 0.08$~fm$^4$. Numbers in parenthesis indicate first the estimated extrapolation uncertainties and then the chiral truncation uncertainties at the 95\% level, $(*)$ indicating no chiral truncation uncertainties at LO.} 
\end{ruledtabular}   
\end{table*}

Our results for the ground state energies of $A=10$ and $12$ nuclei are shown in Fig.~\ref{fig:EgsA10A12}, 
and tabulated in Table~\ref{tab:EgsA10A12}; in the latter we also include our results for $^{14}$O.   
Again, the estimated numerical extrapolation uncertainties are of the order of 0.5\% to 1\%, that is, 
of the same order as the estimated dependence on the SRG parameter, 
whereas the chiral uncertainty estimate, at the 95\% level, is significantly larger.
Here in the middle of the $p$-shell, we see that for $A=10$ the N$^2$LO NN potential
overbinds, whereas using the NN potentials at N$^3$LO and higher lead to underbinding.
As we go up in the $p$-shell and increase $A$ further, we see that for $A=12$ 
the overbinding with the N$^2$LO NN potential increases, whereas the higher-order
NN potentials give excellent agreement with the data.  This appears to be a 
systematic trend, which continues for $^{14}$O and beyond, as will be discussed 
in more detail in the next section.

For all of these ground state energies, we see that the cutoff of $\Lambda = 500$~MeV leads to larger 
chiral truncation uncertainties than the smaller cutoff of $\Lambda = 450$~MeV, which is to be expected,
and also in agreement with the posteriors for the expansion parameter $Q$ shown in Fig.~\ref{fig:Q_posteriors}.

\subsection{Excitation spectra of $p$-shell nuclei}

In addition to the ground-state energies, we have also calculated the low-lying spectra of these $p$-shell nuclei, limiting ourselves to the normal parity states, that is, states with the same parity as the valence space.  Since we are only considering even $p$-shell nuclei here, that implies we are only considering positive parity states.  Again, we perform a series of calculations at increasing values of $N_{\max}$ for a range of the HO parameters $\hbar\omega$ around the variational minimum, and apply the same extrapolation method as for the ground state energies. For narrow excited states, this extrapolation method seems to give results that are numerically reasonably stable, even for states that are above threshold.

\begin{table*}[t]
    \begin{ruledtabular}
    \begin{tabular*}{\textwidth}{l|c|c|cc|ccc}
                    & $^{6}$He$(2^+)$ & $^{6}$Li$(3^+)$ & $^{10}$Be$(2^+)_1$ & $^{10}$Be$(2^+)_2$ & $^{10}$B$(1^+)_1$ & $^{10}$B$(1^+)_2$ & $^{10}$B$(2^+)$ \\
\hline \\[-9pt]
 \multicolumn{8}{c}{$\Lambda = 450$ MeV} \\
\hline \\[-9pt]
 LO                 & $3.5(0.9)(*)$     & $5.3(0.8)(*)$     & $7.7(2.1)(*)$   & $6.1(1.7)(*)$   &$-6.7(1.6)(*)$   & $0.2(1.7)(*)$   &$-0.6(1.8)(*)$ \\
 NLO                & $1.10(0.31)(1.6)$ & $2.90(0.17)(2.0)$ & $3.5(0.8)(5.8)$ & $4.6(0.9)(5.8)$ &$-1.4(0.8)(6.9)$ & $1.8(0.8)(5.8)$ & $2.1(0.6)(5.8)$ \\
 N$^2$LO            & $2.10(0.15)(0.5)$ & $2.40(0.07)(0.6)$ & $3.3(0.5)(1.7)$ & $6.3(0.6)(1.7)$ & $1.7(1.0)(2.1)$ & $1.4(0.5)(1.7)$ & $3.4(0.5)(1.8)$ \\
 N$^3$LO            & $2.09(0.18)(0.5)$ & $2.41(0.09)(0.6)$ & $3.4(0.6)(1.7)$ & $5.6(0.7)(1.7)$ & $0.8(1.0)(2.3)$ & $1.4(0.6)(1.7)$ & $3.3(0.6)((1.8)$ \\
 N$^4$LO            & $2.07(0.17)(0.5)$ & $2.40(0.08)(0.6)$ & $3.4(0.6)(1.7)$ & $5.6(0.7)(1.7)$ & $0.8(1.0)(2.3)$ & $1.5(0.6)(1.7)$ & $3.3(0.6)((1.8)$ \\
 N$^4$LO$^+$           & $2.07(0.18)(0.5)$ & $2.42(0.09)(0.6)$ & $3.4(0.6)(1.7)$ & $5.6(0.7)(1.7)$ & $0.8(1.0)(2.3)$ & $1.4(0.6)(1.7)$ & $3.3(0.6)((1.8)$ \\
 \hline \\[-9pt]
\multicolumn{8}{c}{$\Lambda = 500$ MeV} \\
\hline \\[-9pt]
 LO                 & $3.6(1.0)(*)$     & $5.1(9)(*)$       & $8.1(2.5)(*)$   & $6.6(2.0)(*)$   &$-7.0(2.0)(*)$   & $0.2(2.0)(*)$   &$-0.8(2.1)(*)$ \\
 NLO                & $2.08(0.23)(1.7)$ & $3.93(0.14)(2.0)$ & $3.4(0.7)(6.8)$ & $4.2(0.7)(6.8)$ &$-1.6(0.6)(7.7)$ & $1.6(0.7)(6.6)$ & $1.6(0.5)(6.9)$ \\
 N$^2$LO            & $2.08(0.09)(0.5)$ & $2.41(0.07)(0.6)$ & $3.2(0.5)(2.2)$ & $6.2(0.6)(2.2)$ & $2.2(1.0)(2.5)$ & $1.9(0.5)(2.2)$ & $4.1(0.5)(2.3)$ \\
 N$^3$LO            & $2.10(0.14)(0.5)$ & $2.35(0.08)(0.6)$ & $3.4(0.6)(2.2)$ & $6.0(0.7)(2.2)$ & $1.3(1.0)(2.5)$ & $1.8(0.6)(2.2)$ & $3.8(0.6)(2.3)$ \\
 N$^4$LO            & $2.08(0.14)(0.5)$ & $2.34(0.07)(0.6)$ & $3.3(0.6)(2.2)$ & $5.9(0.7)(2.2)$ & $1.3(1.0)(2.5)$ & $1.7(0.6)(2.2)$ & $3.8(0.6)(2.3)$ \\
 N$^4$LO$^+$           & $2.09(0.13)(0.5)$ & $2.37(0.08)(0.6)$ & $3.3(0.6)(2.2)$ & $6.0(0.7)(2.2)$ & $1.3(1.0)(2.5)$ & $1.7(0.6)(2.2)$ & $3.7(0.6)(2.3)$ \\
\hline \\[-9pt]
Expt.               & $1.80$     & $2.19$     &  $3.37$   &  $5.96$  &  $0.72$     & $2.15$    & $3.59$ \\
     \end{tabular*}
\caption{\label{tab:ExA6A10} Excitation energies of low-lying normal parity states in $^6$He, $^6$Li, $^{10}$Be and $^{10}$B obtained in the NCSM for different chiral orders and cutoffs, SRG evolved to $\alpha = 0.08$~fm$^4$. Numbers in parenthesis are first the maximum of the estimated extrapolation uncertainties in the excited-state and  ground-state energies and then the correlated chiral truncation uncertainties at the 95\% confidence level, $(*)$ indicating no chiral truncation uncertainties at LO.}
\end{ruledtabular}   
\end{table*}

Our results for the obtained excitation energies (the difference of the extrapolated total energies) of the low-lying excited states of $A=6$ and $A=10$ are given in Table~\ref{tab:ExA6A10}, together with the corresponding experimental values.  Similar to the ground state energies in Tables~\ref{tab:EgsA4A6A8} and \ref{tab:EgsA10A12}, the first uncertainties are the numerical uncertainties associated with the extrapolation to the complete bases; for these uncertainties we use again the same procedure as in Refs.~\cite{Binder:2018pgl,Epelbaum:2018ogq,Maris:2020qne}, namely the maximum of the estimated extrapolation uncertainties of the total energies of the two states.  The second set of uncertainties correspond to the 95\% confidence interval of the chiral EFT truncation error, taking into account correlations between the ground state and the excited state, as discussed in subsection \ref{sec:correlated_model}, see Eq.~\eqref{eq:Gaussian_difference}.  

Again, as in the case of the ground state energies, the uncertainties of the obtained excitation energies are dominated by their chiral truncation uncertainties.  However, due to the strong correlations between the ground state and the excited state, the chiral truncation uncertainties of the excitation energies are significantly smaller than those of the ground state energies.  For almost all of the excited states shown in Table~\ref{tab:ExA6A10}, this uncertainty estimate is reduced by at least a factor of two compared to that of the corresponding ground state energies; the exception is the first $1^+$ state in $^{10}$B.  Furthermore, the excitation energies of most of these states are almost independent of the chiral order of the NN potential starting at N$^2$LO, again with the exception of the first $1^+$ state in $^{10}$B, and, at $\Lambda = 450$~MeV, the second $2^+$ state in $^{10}$Be.  Finally, note that for all of these excited states our results agree with experiment within the 95\% confidence interval, starting at NLO.

\begin{figure}[t]
  \includegraphics[width=0.53\columnwidth]{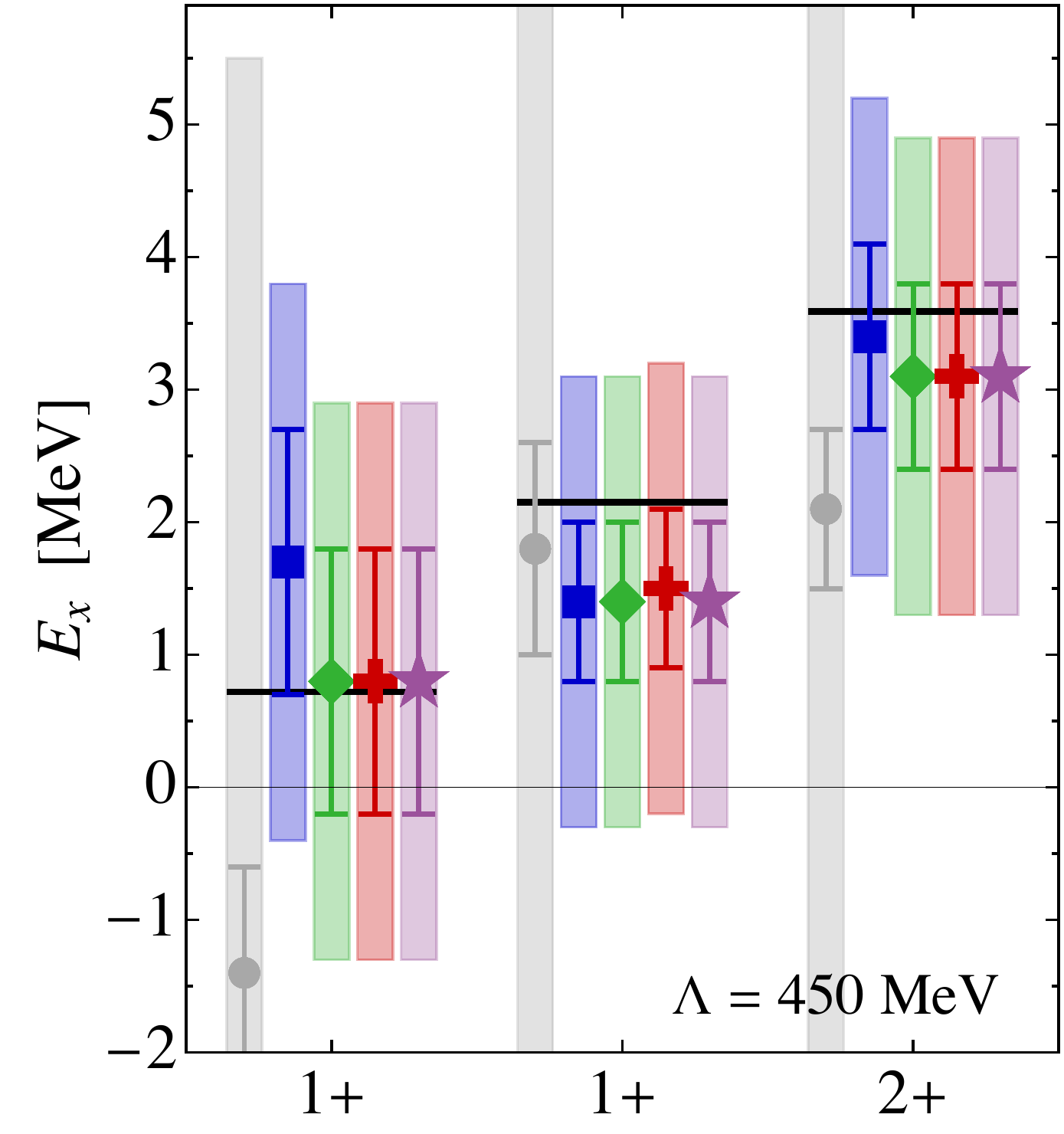}\hspace*{-2pt}
  \includegraphics[width=0.53\columnwidth]{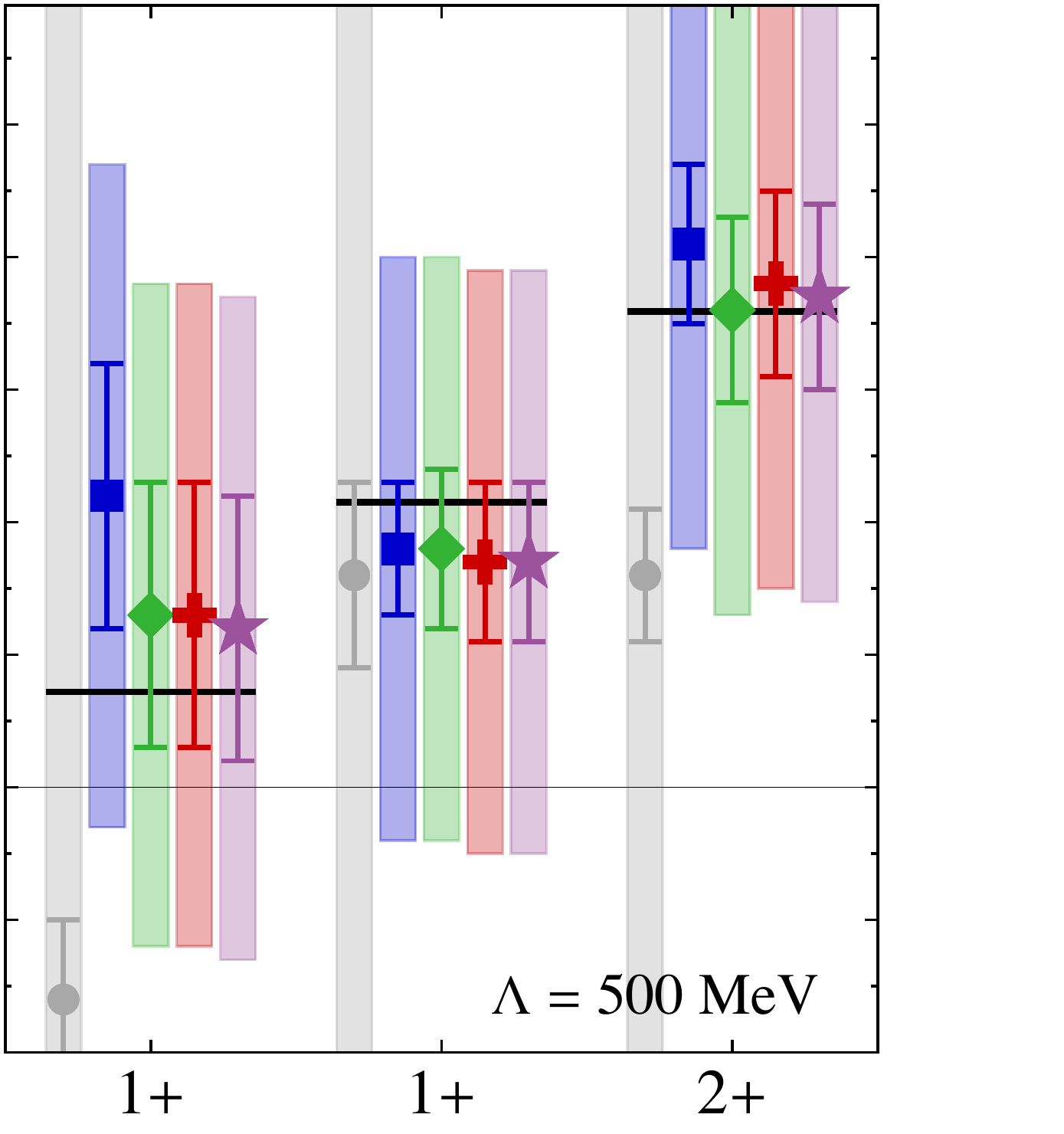}\\
  \caption{\label{fig:ex_10B} (Color online)
  Excitation energies of low-lying states in $^{10}$B with SMS interactions 
  from NLO (gray) to N$^4$LO$^+$ (purple) with $\Lambda=450\,\text{MeV}$ (left-hand panel) and
  $\Lambda=500\,\text{MeV}$ (right-hand panel), both using $\alpha=0.08\,\text{fm}^4$. 
  Error bars indicate the NCSM model-space uncertainties and shaded bands indicate the chiral truncation uncertainties at the 95\% confidence level.  
  Horizontal lines show the experimental excitation energies.}
\end{figure}

Both from Table~\ref{tab:ExA6A10} and in Fig.~\ref{fig:ex_10B}, we see that the energy level
of the first $1^+$ state of $^{10}$B is not actually strongly correlated to the ground state energy, but jumps around relative to the ground state: this state is the ground state at LO and NLO; at N$^2$LO including 3NFs it becomes the second excited $1^+$ state (without the 3NFs it remains the ground state at N$^2$LO~\cite{Maris:2020qne}); and it drops down again to become the first excited $1^+$ state if we use higher chiral orders for the NN potential. 

\begin{table*}[t]
    \begin{ruledtabular}
    \begin{tabular*}{\textwidth}{lrrrrr}
                    & \chead{$^{12}$B$(2^+)_1$} & \chead{$^{12}$B$(0^+)$} & \chead{$^{12}$B$(2^+)_2$} & \chead{$^{12}$B$(1^+)_2$} & \chead{$^{12}$B$(3^+)$} \\
\hline \\[-9pt]
\multicolumn{6}{c}{$\Lambda = 450$ MeV} \\
\hline \\[-9pt]
 LO                 & $4.4(1.3)(*)$   &$-1.3(1.3)(*)$   & $0.0(1.4)(*)$   & $2.1(1.6)(*)$   & $4.9(1.4)$\\
 NLO                & $1.2(0.8)(6.7)$ & $0.3(0.9)(6.8)$ & $1.8(0.9)(6.7)$ & $3.0(0.8)(6.7)$ & $3.8(0.9)(6.6)$  \\
 N$^2$LO            &$-0.9(0.4)(2.0)$ & $1.9(0.6)(2.0)$ & $3.4(0.6)(2.0)$ & $4.9(0.6)(2.0)$ & $5.3(0.7)(2.0)$  \\
 N$^3$LO            & $0.1(0.6)(2.0)$ & $1.5(0.7)(2.0)$ & $3.1(0.7)(2.0)$ & $4.4(0.7)(2.0)$ & $4.9(0.7)(2.0)$  \\
 N$^4$LO            &$-0.1(0.6)(2.0)$ & $1.6(0.7)(2.0)$ & $3.1(0.7)(2.0)$ & $4.5(0.7)(2.0)$ & $5.0(0.7)(2.0)$  \\
 N$^4$LO$^+$           & $0.0(0.7)(2.0)$ & $1.6(0.7)(2.0)$ & $3.1(0.7)(2.0)$ & $4.5(0.7)(2.0)$ & $5.0(0.7)(2.0)$  \\
 \hline \\[-9pt]
\multicolumn{6}{c}{$\Lambda = 500$ MeV} \\
\hline \\[-9pt]
 LO                 & $4.6(1.7)$(*)   &$-1.4(1.6)(*)$   & $0.0(1.7)(*)$   & $2.3(2.0)(*)$   & $5.2(1.8)(*)$\\
 NLO                & $1.4(0.7)(7.5)$ & $0.1(0.8)(7.8)$ & $1.5(0.8)(7.7)$ & $2.6(0.6)(7.7)$ & $3.5(0.8)(7.6)$  \\
 N$^2$LO            &$-1.1(0.4)(2.5)$ & $2.7(0.6)(2.6)$ & $4.1(0.6)(2.5)$ & $5.7(0.6)(2.5)$ & $6.1(0.7)(2.5)$  \\
 N$^3$LO            &$-0.2(0.6)(2.5)$ & $2.1(0.7)(2.6)$ & $3.6(0.7)(2.5)$ & $5.0(0.7)(2.5)$ & $5.4(0.7)(2.5)$  \\
 N$^4$LO            &$-0.3(0.6)(2.5)$ & $2.3(0.7)(2.6)$ & $3.8(0.7)(2.5)$ & $5.2(0.7)(2.5)$ & $5.6(0.7)(2.5)$  \\
 N$^4$LO$^+$           &$-0.2(0.7)(2.5)$ & $2.2(0.7)(2.6)$ & $3.7(0.7)(2.5)$ & $5.1(0.7)(2.5)$ & $5.6(0.7)(2.5)$  \\
\hline \\[-9pt]
Expt.               &  $0.95$   &  $2.72$   &  $3.76$ & $4.99$ & 5.61 \\
     \end{tabular*}
\caption{\label{tab:Ex12B} Excitation energies of low-lying normal parity states in $^{12}$B obtained in the NCSM for different chiral orders and cutoffs, SRG evolved to $\alpha = 0.08$~fm$^4$. 
Numbers in parenthesis are first the maximum of the estimated extrapolation uncertainties in the excited-state and  ground-state energies and then the correlated chiral truncation uncertainties at the 95\% confidence level, $(*)$ indicating no chiral truncation uncertainties at LO.}
\end{ruledtabular}   
\end{table*}
\begin{figure}[t]
  \includegraphics[width=0.53\columnwidth]{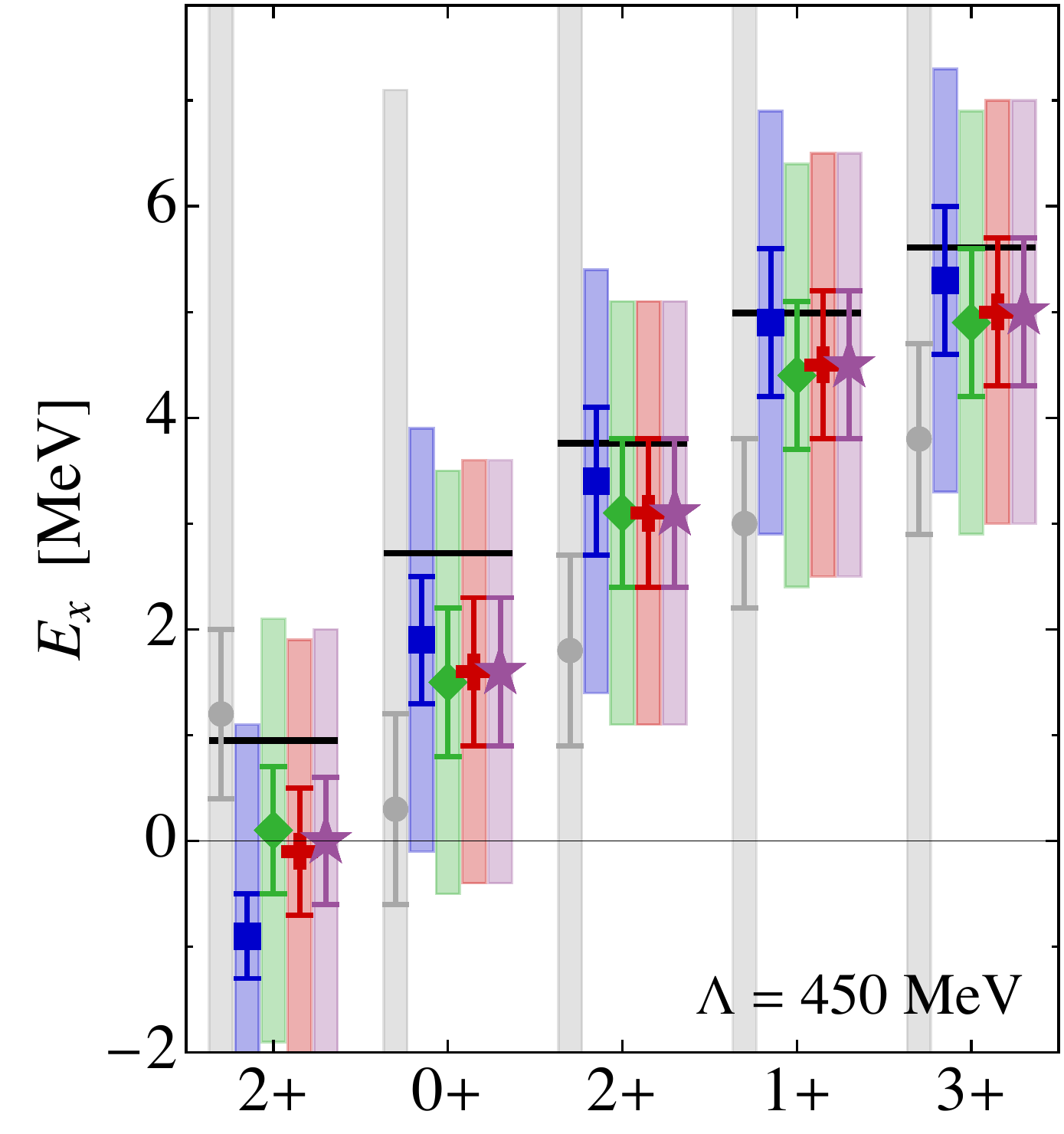}\hspace*{-2pt}
  \includegraphics[width=0.53\columnwidth]{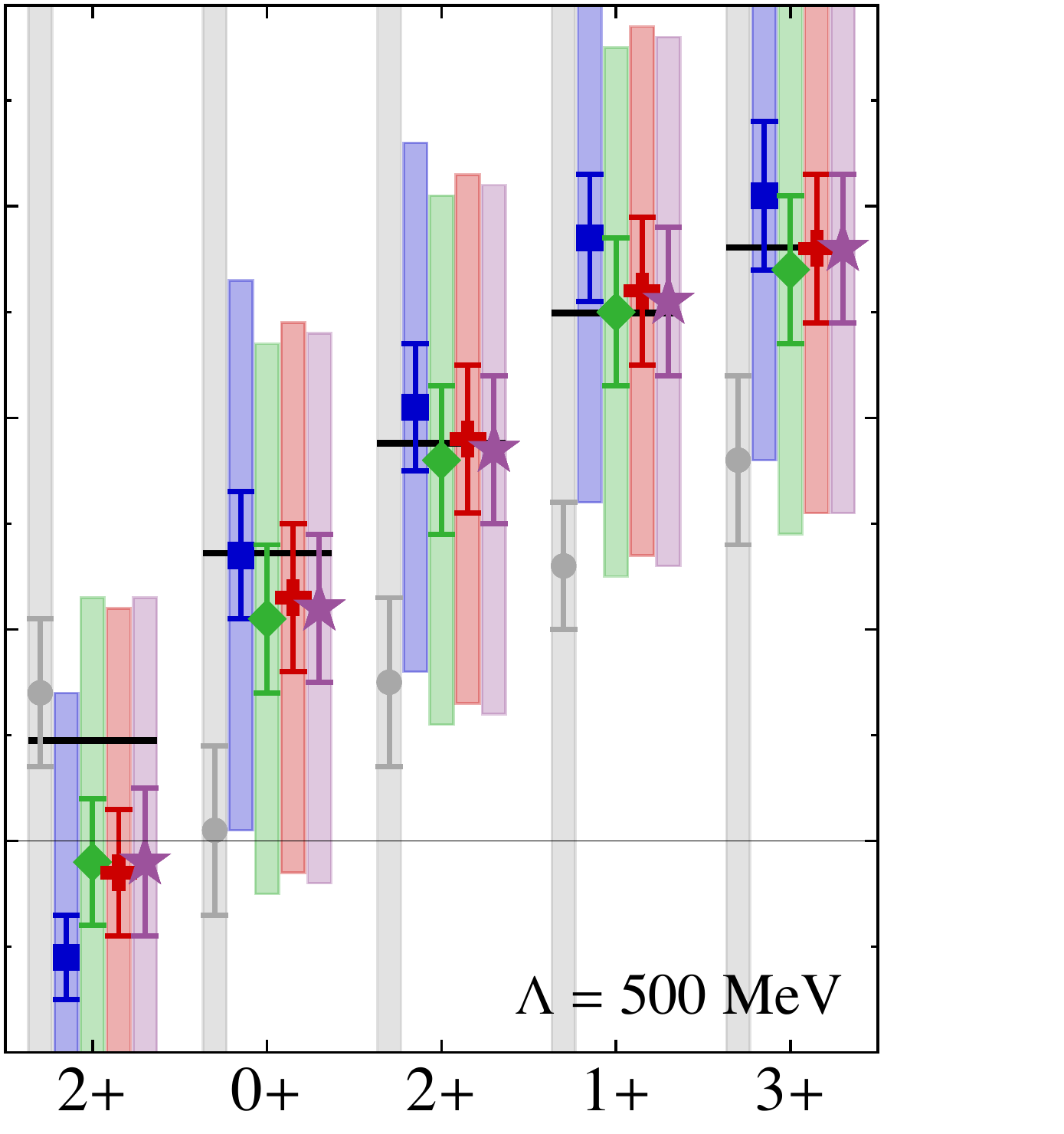}\\
  \caption{\label{fig:ex_12B} (Color online)
  Excitation energies of low-lying states in $^{12}$B with SMS interactions 
  from NLO (gray) to N$^4$LO$^+$ (purple) with $\Lambda=450\,\text{MeV}$ (left-hand panel) and
  $\Lambda=500\,\text{MeV}$ (right-hand panel), both using $\alpha=0.08\,\text{fm}^4$. 
  Error bars and bands are the same as in Fig.~\ref{fig:ex_10B}.
  Horizontal lines show the experimental excitation energies.}
\end{figure}

Also in $^{12}$B we see that the chiral truncation uncertainty estimate in the excitation energies is reduced by more than a factor of two compared to that of the corresponding ground state energy, due to strong correlations between the ground state and the excited states, see Table~\ref{tab:Ex12B}.  And, like in $^{10}$B, the lowest states jump around a bit at the lower chiral orders: at LO, the $0^+$ is the lowest state; at NLO lowest $1^+$ state is the lowest state, in agreement with experiment, but at N$^2$LO, the $2^+$ state becomes the lowest state.  Incorporating higher chiral orders for the NN potential, in combination with the N$^2$LO 3NFs, makes this $2^+$ state almost degenerate with the $1^+$ ground state.  On the other hand, the level ordering of the second $2^+$, the second $1^+$, and the lowest $3^+$ states is in agreement with experiment starting at NLO.  And again, for all these five excited states our theoretical excitation energies agree with experiment well within the 95\% confidence interval.

\begin{table}[t]
    \begin{ruledtabular}
    \begin{tabular*}{\textwidth}{lrrr}
                    & \chead{$^{12}$C$(2^+)$} & \chead{$^{12}$C$(1^+)$} & \chead{$^{12}$C$(4^+)$} \\
\hline \\[-9pt]
\multicolumn{4}{c}{$\Lambda = 450$ MeV} \\
\hline \\[-9pt]
 LO                 & $6.9(0.9)(*)$   & $31.3(1.2)(*)$   & $23.3(1.1)(*)$ \\
 NLO                & $3.4(0.5)(7.1)$ & $14.2(0.6)(8.6)$ & $12.2(0.7)(7.2)$ \\
 N$^2$LO            & $4.2(0.4)(2.1)$ &  $9.6(0.4)(2.6)$ & $13.7(0.4)(2.2)$ \\
 N$^3$LO            & $3.7(0.6)(2.1)$ & $10.9(0.6)(2.6)$ & $12.6(0.7)(2.2)$ \\
 N$^4$LO            & $3.7(0.6)(2.1)$ & $10.6(0.6)(2.6)$ & $12.6(0.7)(2.2)$ \\
 N$^4$LO$^+$           & $3.7(0.7)(2.1)$ & $10.5(0.7)(2.6)$ & $12.6(0.7)(2.2)$ \\
 \hline \\[-9pt]
\multicolumn{4}{c}{$\Lambda = 500$ MeV} \\
\hline \\[-9pt]
 LO                 & $7.5(1.3)(*)$   & $32.2(1.4)(*)$   & $24.6(2.0)(*)$ \\
 NLO                & $3.1(0.5)(8.2)$ & $13.6(0.6)(9.0)$ & $11.4(0.6)(8.4)$ \\
 N$^2$LO            & $4.5(0.4)(2.7)$ &  $9.9(0.4)(3.0)$ & $14.6(0.4)(2.8)$ \\
 N$^3$LO            & $3.6(0.6)(2.7)$ & $10.6(0.6)(3.0)$ & $12.9(0.7)(2.8)$ \\
 N$^4$LO            & $3.6(0.6)(2.7)$ & $10.1(0.6)(3.0)$ & $12.9(0.7)(2.8)$ \\
 N$^4$LO$^+$           & $3.6(0.6)(2.7)$ &  $9.9(0.6)(3.0)$ & $12.9(0.7)(2.8)$ \\
\hline \\[-9pt]
Expt.               &  $4.44$   &  $12.71$   &  $14.08$  \\
     \end{tabular*}
\caption{\label{tab:Ex12C} Excitation energies of low-lying normal parity states in $^{12}$C obtained in the NCSM for different chiral orders and cutoffs, SRG evolved to $\alpha = 0.08$~fm$^4$. 
Numbers in parenthesis are first the maximum of the estimated extrapolation uncertainties in the excited-state and  ground-state energies and then the correlated chiral truncation uncertainties at the 95\% confidence level, $(*)$ indicating no chiral truncation uncertainties at LO.}
\end{ruledtabular}   
\end{table}
\begin{figure}[t]
  \includegraphics[width=0.53\columnwidth]{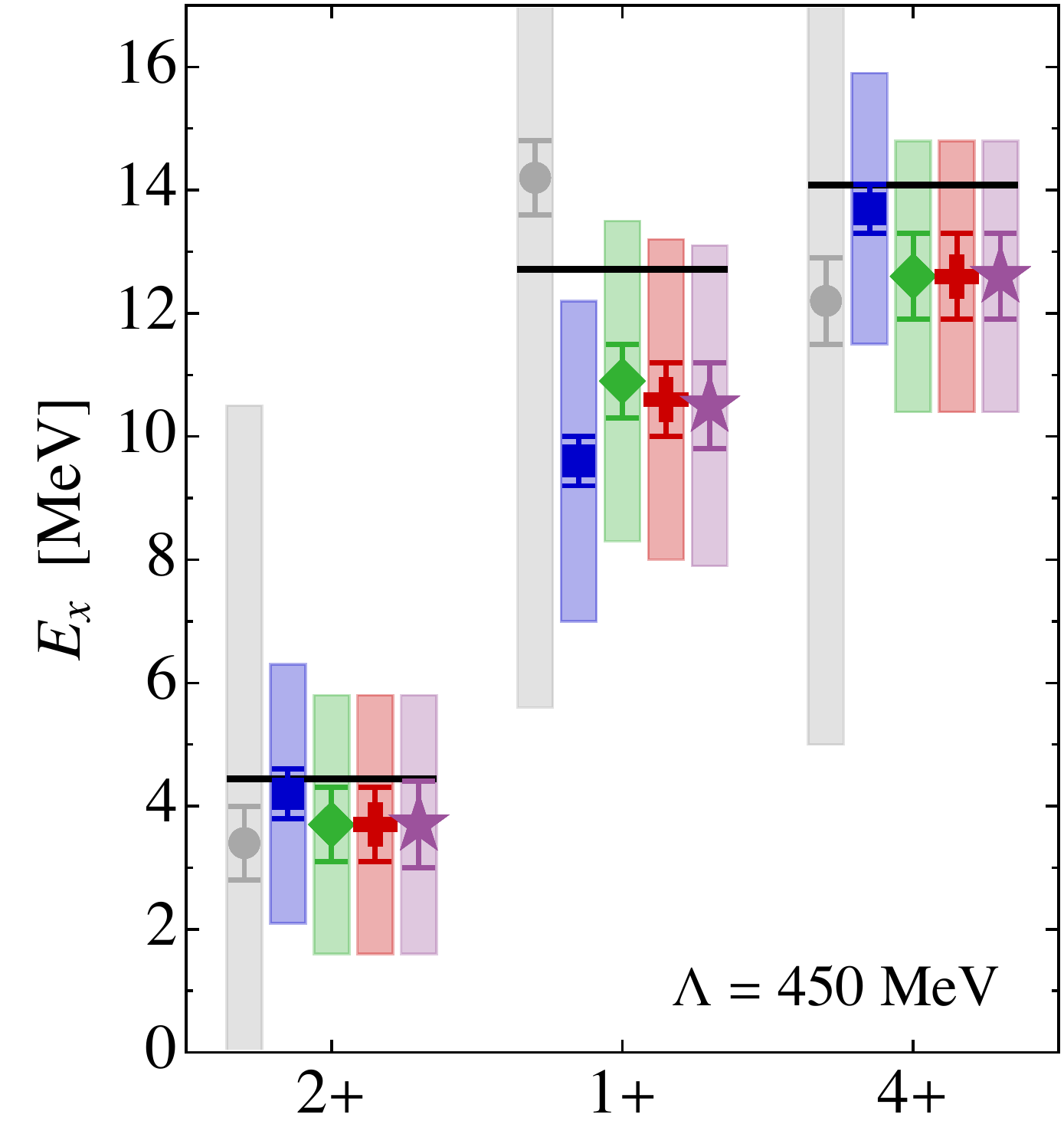}\hspace*{-2pt}
  \includegraphics[width=0.53\columnwidth]{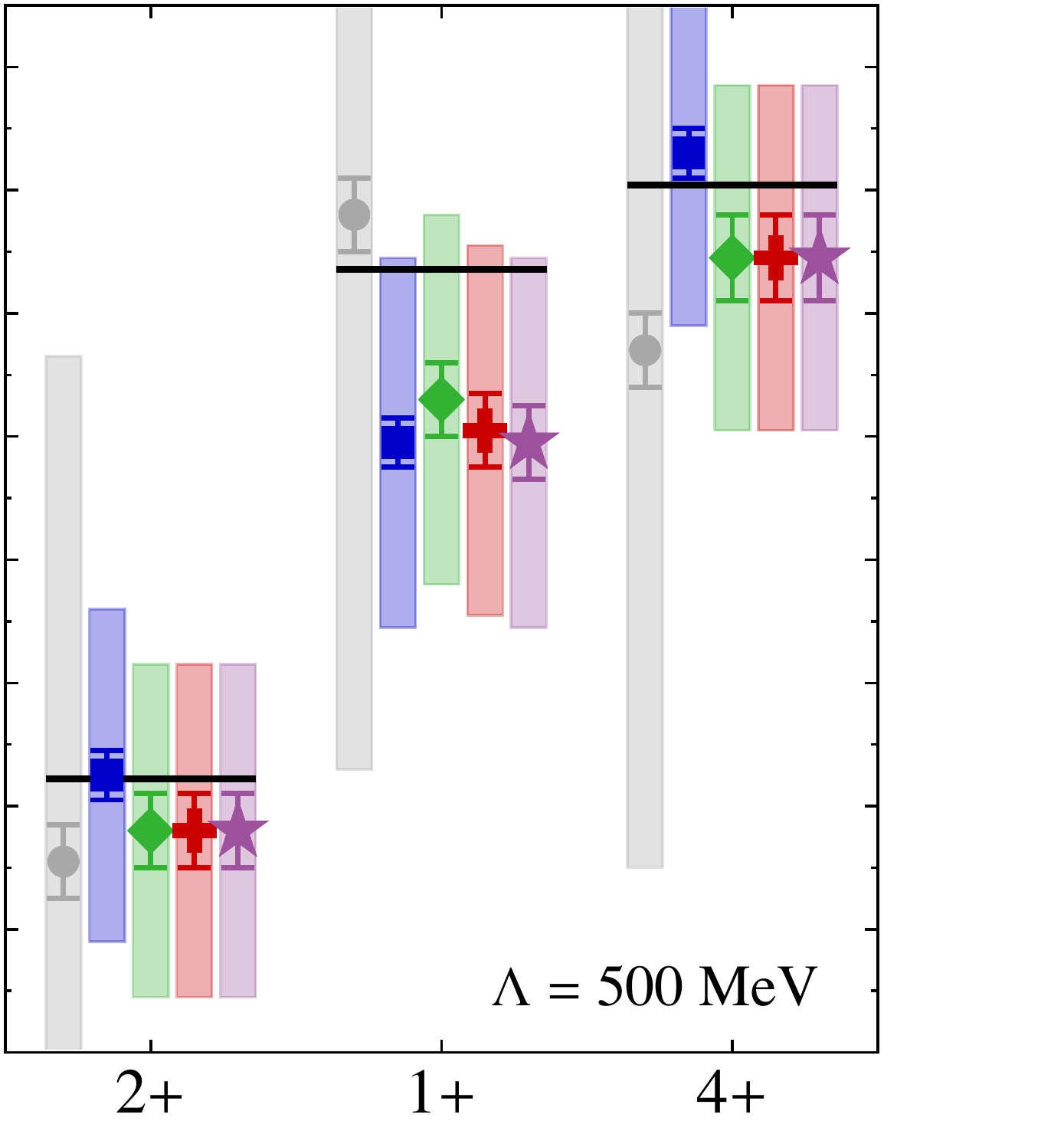}\\
  \caption{\label{fig:ex_12C} (Color online)
  Excitation energies of low-lying states in $^{12}$C with SMS interactions 
  from NLO (gray) to N$^4$LO$^+$ (purple) with $\Lambda=450\,\text{MeV}$ (left-hand panel) and
  $\Lambda=500\,\text{MeV}$ (right-hand panel), both using $\alpha=0.08\,\text{fm}^4$. 
  Error bars and bands are the same as in Fig.~\ref{fig:ex_10B}.
  Horizontal lines show the experimental excitation energies.}
\end{figure}

Finally, in Table~\ref{tab:Ex12C} and Fig.~\ref{fig:ex_12C}, we show our results for the $2^+$ and $4^+$ rotational excitations of the ground state of $^{12}$C, as well as the first excited $1^+$ state.  Not surprisingly, the rotational excitations of the ground state are strongly correlated, not only to the $0^+$ ground state, but also to each other.  More interesting is the $1^+$ excited state: of all the excited states considered here, it shows a noticeable dependence on the NN potential beyond N$^2$LO.  Furthermore, this is the only state for which  our theoretical calculations with the N$^2$LO and higher orders for the NN potential barely agree with the experimental value.  This will therefore be an important test once we incorporate consistent N$^3$LO 3NFs, which should significantly reduce the chiral uncertainties.


\begin{table}[p]
\renewcommand{\arraystretch}{0.9}
    \begin{ruledtabular}
    \begin{tabular*}{\textwidth}{@{\extracolsep{\fill}}lllll}
                    & \multicolumn{2}{c}{ $\Lambda=450$ MeV} & \multicolumn{2}{c}{ $\Lambda=500$ MeV}  \\
                    &  \chead{$E$ [MeV]}  & \chead{$R_{\text{p,rms}}$ [fm]}  &  \chead{$E$ [MeV]}  &  \chead{$R_{\text{p,rms}}$ [fm]}  \\
\hline \\[-6pt]
\multicolumn{5}{c}{$^{14}$O} \\
\hline
 LO                 &  $-152.3 $        &  $1.68$        &  $-149.4$         &  $1.74$        \\
 NLO                &  $-97.4(32.)$    &  $2.14(0.34)$  &  $-89.1(39.)$    &  $2.25(0.27)$  \\
 N$^2$LO            &  $-114.0(11.)$   &  $2.12(0.11)$  &  $-117.6(13.)$   &  $2.14(0.09)$  \\
 N$^3$LO            &  $-100.3(11.)$    &  $2.28(0.11)$  &  $-103.7(13.)$   &  $2.25(0.09)$  \\
 N$^4$LO            &  $-99.4(11.)$     &  $2.27(0.11)$  &  $-102.2(13.)$   &  $2.25(0.09)$  \\
 N$^4$LO$^+$           &  $-99.6(11.)$     &  $2.27(0.11)$  &  $-102.2(13.)$   &  $2.25(0.09)$  \\
 Expt.              &  $-98.7$          &  --            &  $-98.7 $         &  --            \\
\hline \\[-6pt]
\multicolumn{5}{c}{$^{16}$O} \\
\hline
 LO                 &  $-217.0 $        &  $1.51$        &  $-207.3$         &  $1.59$        \\
 NLO                &  $-134.1(42.)$   &  $2.08(0.38)$  &  $-122.6(50.)$   &  $2.21(0.31)$  \\
 N$^2$LO            &  $-148.3(14.)$   &  $2.12(0.13)$  &  $-152.5(17.)$   &  $2.15(0.10)$  \\
 N$^3$LO            &  $-130.8(14.)$   &  $2.28(0.13)$  &  $-134.3(17.)$   &  $2.27(0.10)$  \\
 N$^4$LO            &  $-129.1(14.)$   &  $2.28(0.13)$  &  $-131.7(17.)$   &  $2.27(0.10)$  \\
 N$^4$LO$^+$           &  $-129.2(14.)$   &  $2.28(0.13)$  &  $-131.6(17.)$   &  $2.28(0.10)$  \\
 Expt.              &  $-127.6$         &  $2.58$        &  $-127.6$         &  $2.58$        \\
\hline \\[-6pt]
\multicolumn{5}{c}{$^{18}$O} \\
\hline
 LO                 &  $-228.9$         &  $1.51$        &  $-217.1$         &  $1.61$        \\
 NLO                &  $-142.3(45.)$   &  $2.12(0.40)$  &  $-129.1(55.)$   &  $2.26(0.32)$  \\
 N$^2$LO            &  $-163.0(15.)$   &  $2.10(0.13)$  &  $-168.0(18.)$   &  $2.14(0.11)$  \\
 N$^3$LO            &  $-141.8(15.)$   &  $2.28(0.13)$  &  $-146.4(18.)$   &  $2.25(0.11)$  \\
 N$^4$LO            &  $-139.5(15.)$   &  $2.28(0.13)$  &  $-142.8(18.)$   &  $2.26(0.11)$  \\
 N$^4$LO$^+$           &  $-139.7(15.)$   &  $2.28(0.13)$  &  $-142.6(18.)$   &  $2.27(0.11)$  \\
 Expt.              &  $-139.8$         &  $2.66$        &  $-139.8$         &  $2.66$        \\
\hline \\[-6pt]
\multicolumn{5}{c}{$^{20}$O} \\
\hline
 LO                 &  $-240.0$         &  $1.53$        &  $-220.0$         &  $1.64$        \\
 NLO                &  $-150.6(49.)$   &  $2.11(0.34)$  &  $-135.7(60.)$   &  $2.27(0.27)$  \\
 N$^2$LO            &  $-180.4(16.)$   &  $2.07(0.11)$  &  $-187.2(20.)$   &  $2.10(0.09)$  \\
 N$^3$LO            &  $-153.8(16.)$   &  $2.26(0.11)$  &  $-159.9(20.)$   &  $2.23(0.09)$  \\
 N$^4$LO            &  $-151.0(16.)$   &  $2.27(0.11)$  &  $-155.3(20.)$   &  $2.24(0.09)$  \\
 N$^4$LO$^+$           &  $-151.2(16.)$   &  $2.27(0.11)$  &  $-154.9(20.)$   &  $2.25(0.09)$  \\
 Expt.              &  $-151.4$         &  --            &  $-151.4$         &  --            \\
\hline \\[-6pt]
\multicolumn{5}{c}{$^{22}$O} \\
\hline
 LO                 &  $-235.1$         &  $1.55$        &  $-219.6$         &  $1.69$        \\
 NLO                &  $-159.1(52.)$   &  $2.11(0.34)$  &  $-142.0(64.)$   &  $2.28(0.27)$  \\
 N$^2$LO            &  $-200.1(17.)$   &  $2.03(0.11)$  &  $-208.6(21.)$   &  $2.07(0.09)$  \\
 N$^3$LO            &  $-166.5(17.)$   &  $2.23(0.11)$  &  $-174.9(21.)$   &  $2.19(0.09)$  \\
 N$^4$LO            &  $-163.2(17.)$   &  $2.24(0.11)$  &  $-169.0(21.)$   &  $2.21(0.09)$  \\
 N$^4$LO$^+$           &  $-163.3(17.)$   &  $2.24(0.11)$  &  $-168.3(21.)$   &  $2.22(0.09)$  \\
 Expt.              &  $-162.0$         &  --            &  $-162.0$         &  --            \\
\hline \\[-6pt]
\multicolumn{5}{c}{$^{24}$O} \\
\hline
 LO                 &  $-235.1$         &  $1.57$        &  $-217.0$         &  $1.71$        \\
 NLO                &  $-166.9(55.)$   &  $2.11(0.35)$  &  $-147.8(66.)$   &  $2.29(0.27)$  \\
 N$^2$LO            &  $-214.5(18.)$   &  $2.04(0.12)$  &  $-222.4(22.)$   &  $2.08(0.09)$  \\
 N$^3$LO            &  $-174.7(18.)$   &  $2.25(0.12)$  &  $-183.9(22.)$   &  $2.21(0.09)$  \\
 N$^4$LO            &  $-171.2(18.)$   &  $2.26(0.12)$  &  $-177.2(22.)$   &  $2.23(0.09)$  \\
 N$^4$LO$^+$           &  $-171.5(18.)$   &  $2.27(0.12)$  &  $-176.6(22.)$   &  $2.24(0.09)$  \\
 Expt.              &  $-168.5$         &  --            &  $-168.5$         &  --            \\
\hline \\[-6pt]
\multicolumn{5}{c}{$^{26}$O} \\
\hline
 LO                 &  $-233.2$         &  $1.58$        &  $-212.4$         &  $1.75$        \\
 NLO                &  $-167.0(55.)$   &  $2.16(0.39)$  &  $-147.2(67.)$   &  $2.35(0.28)$  \\
 N$^2$LO            &  $-210.3(18.)$   &  $2.09(0.10)$  &  $-216.6(22.)$   &  $2.14(0.09)$\\
 N$^3$LO            &  $-170.8(18.)$   &  $2.32(0.12)$  &  $-178.6(22.)$   &  $2.28(0.09)$\\
 N$^4$LO            &  $-166.9(18.)$   &  $2.33(0.12)$  &  $-171.1(22.)$   &  $2.30(0.09$\\
 N$^4$LO$^+$           &  $-167.3(18.)$   &  $2.34(0.12)$  &  $-170.8(22.)$   &  $2.32(0.09)$  \\
 Expt.              &  $-168.4$         &  --            &  $-168.4$         &  --            \\
\end{tabular*}                
\vspace*{-2ex}
\caption{\label{tab:oxygen} Ground-state energies and point-proton rms-radii for the oxygen isotopes obtained in the IM-NCSM. Numbers in parenthesis indicate the chiral truncation uncertainties at the 95\% confidence level. 
}
\end{ruledtabular}   
\end{table}

\section{Beyond Light Nuclei}
\label{sec:mediummass}

\begin{figure*}[t]
  \includegraphics[width=0.8\columnwidth]{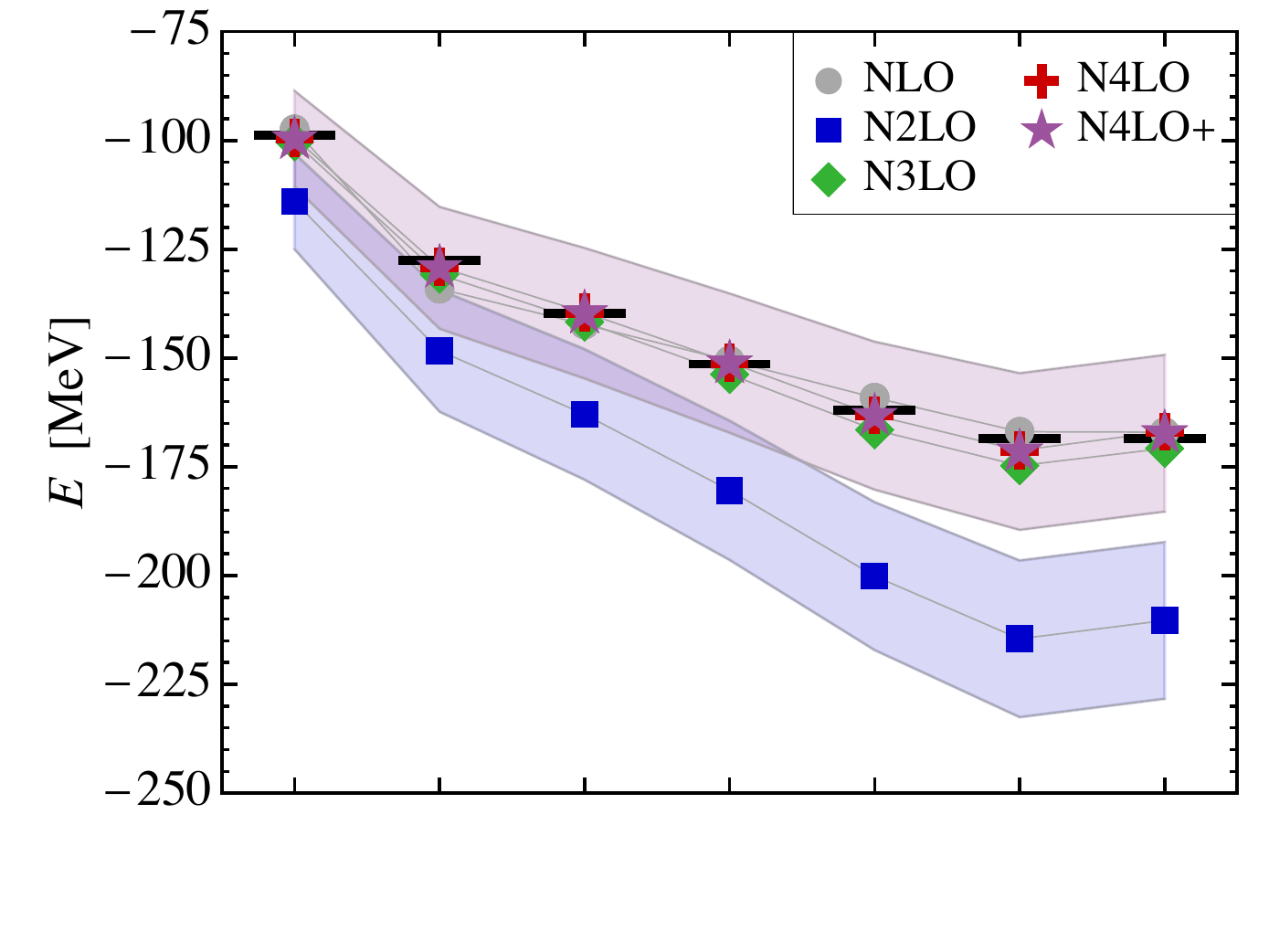}\hspace*{20pt}
  \includegraphics[width=0.8\columnwidth]{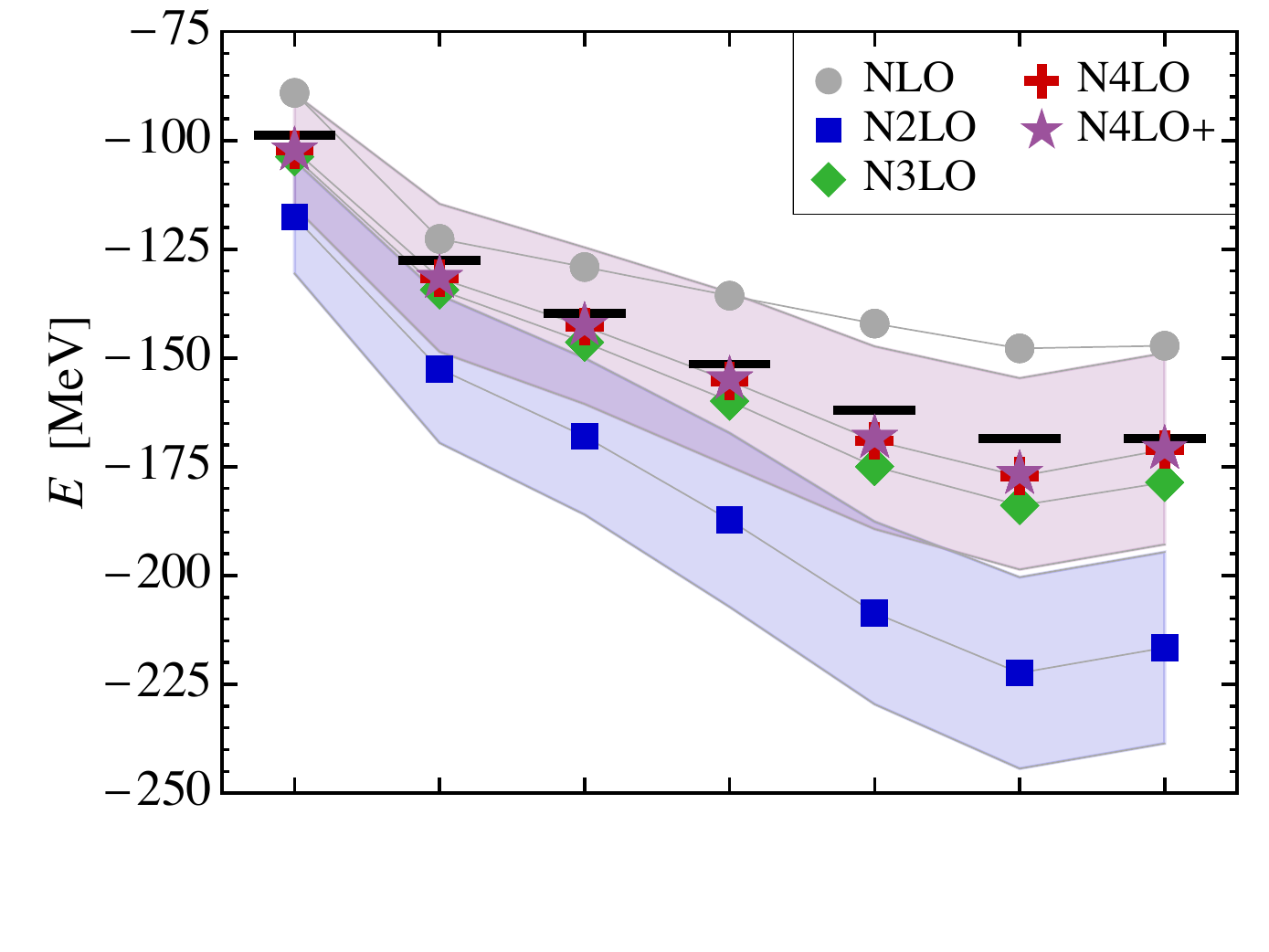} \\[-29pt]
  \includegraphics[width=0.8\columnwidth]{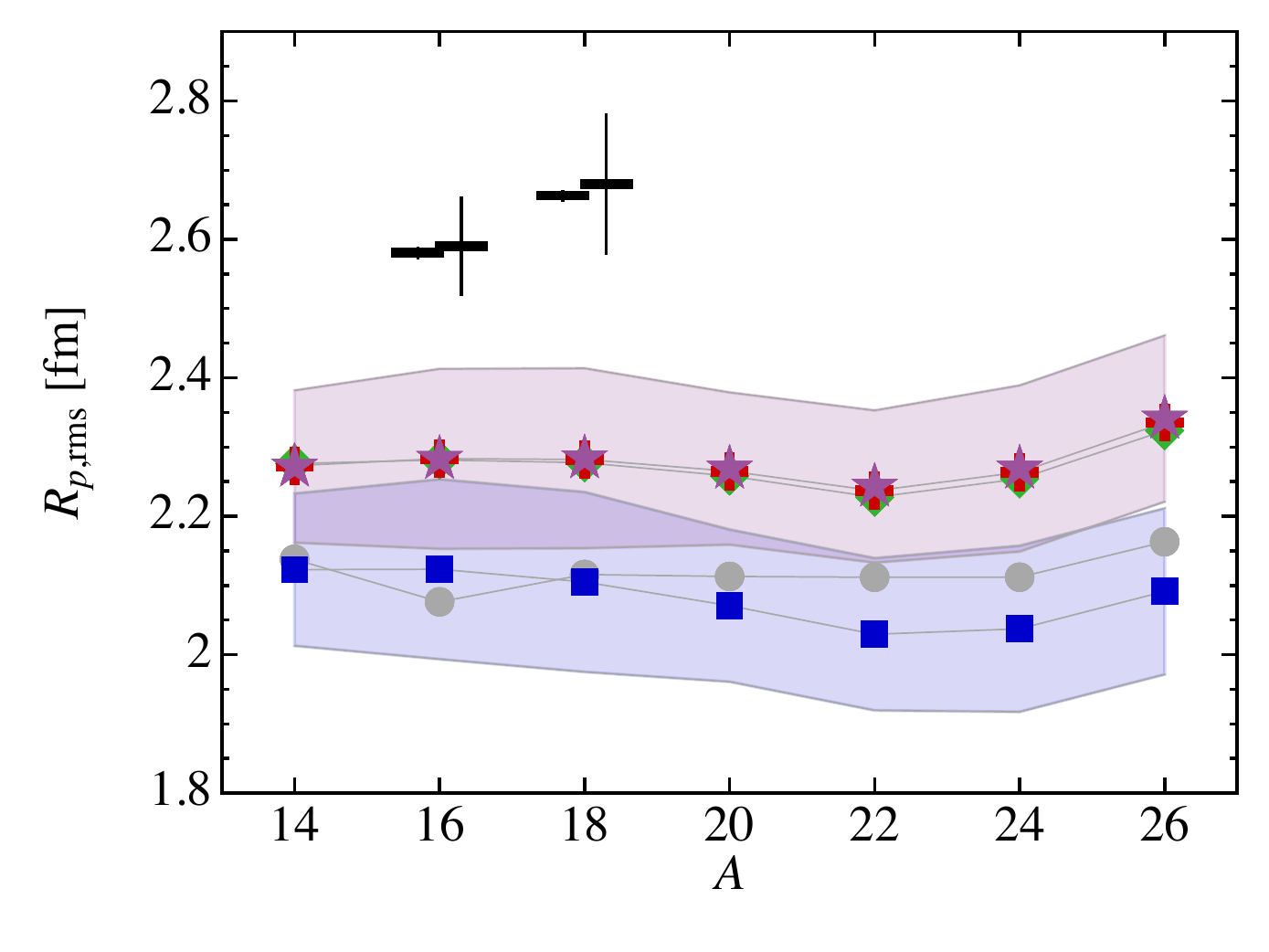}\hspace*{20pt}
  \includegraphics[width=0.8\columnwidth]{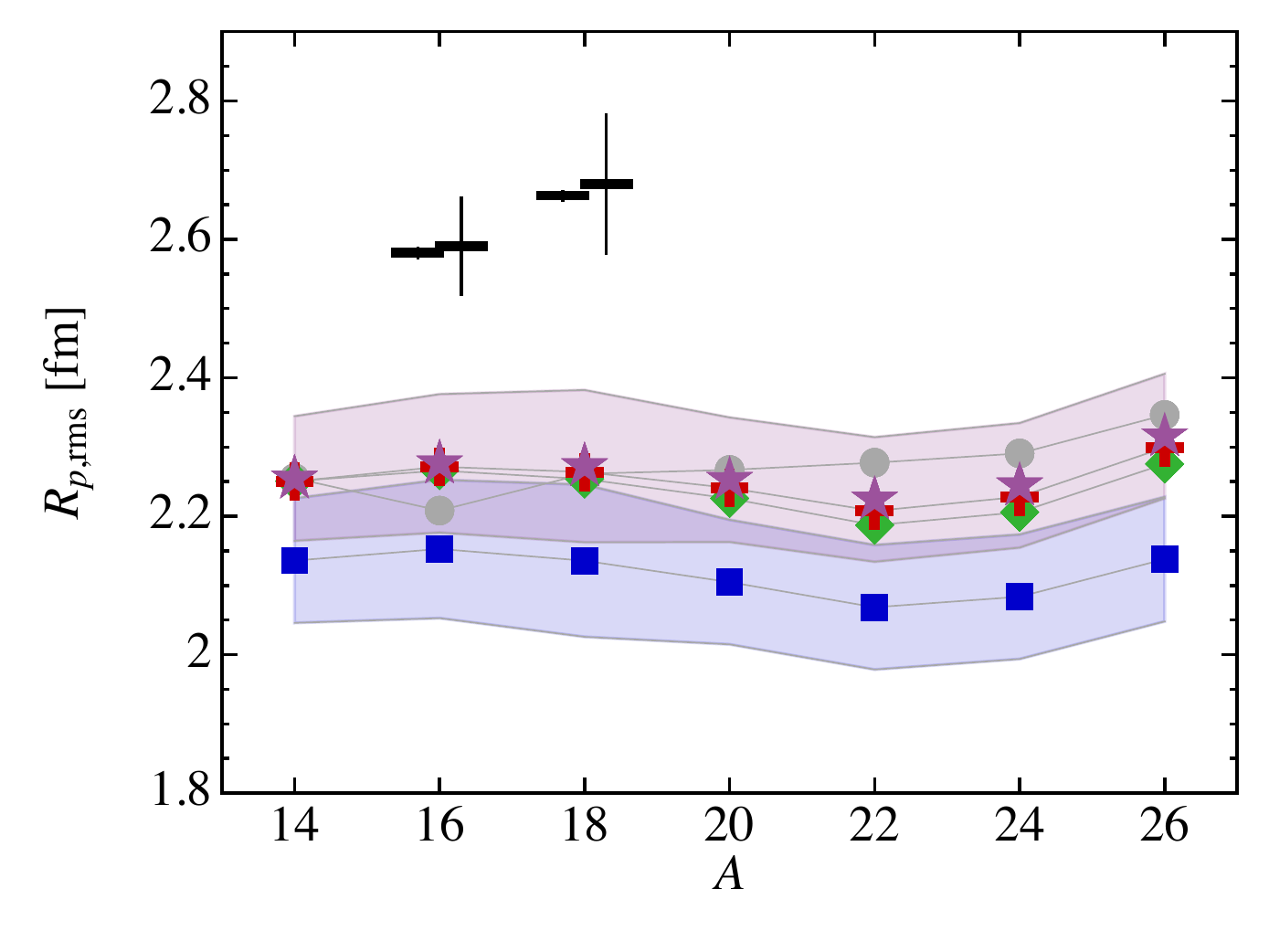} \\[-10pt]
  \caption{\label{fig:oxygen} (Color online)
   Ground-state energies and point-proton radii for even oxygen isotopes obtained in the IM-NCSM with SMS interactions from N$^2$LO to N$^4$LO$^+$ with $\Lambda=450\,\text{MeV}$ (left-hand panels) and $\Lambda=500\,\text{MeV}$ (right-hand panels) for flow-parameter $\alpha=0.08\,\text{fm}^4$. The error bands show the chiral truncation uncertainties at the 95\% confidence level obtained with the Bayesian model for N$^2$LO and N$^4$LO$^+$. 
}
\end{figure*}

Extending our analysis beyond the $p$-shell, we first consider the oxygen isotopic chain from $^{14}$O to $^{26}$O. We focus on the ground-state energies and point-proton rms-radii for the even oxygen isotopes and explore the systematics and uncertainties of these observables for different chiral orders and cutoffs along the lines of the previous section.  

For the ab initio solution of the many-body problem in this mass range, we employ the in-medium no-core shell model (IM-NCSM), which is a hybrid approach based on a multi-reference in-medium similarity renormalization group (IM-SRG) evolution of the Hamiltonian and the NCSM for the extraction of the many-body states and observables \cite{Gebrerufael:2016xih}. In a first step, a NCSM calculations in a small model space, the so called reference space characterized by the truncation parameter $N_{\max}^{\text{ref}}$, is used to extract a multi-determinantal reference state for the nucleus of choice. This reference state is used to set up a multi-reference normal ordering for all relevant operators and to formulate the multi-reference IM-SRG flow equations \cite{Gebrerufael:2015yig,Hergert:2014iaa,Hergert:2016etg}. The flow evolution is designed such that off-diagonal matrix elements of the Hamiltonian that couple the reference space to higher-lying basis states are suppressed, i.e., the flow evolution decouples the small $N_{\max}^{\text{ref}}$ reference space from the rest of the model space. This evolved Hamiltonian is then used in a final NCSM calculation with a truncation $N_{\max} \geq N_{\max}^{\text{ref}}$ to extract the energy eigenvalues and the eigenstates. The latter are used to compute additional observables, e.g., rms-radii, with matrix elements that are consistently evolved in the multi-reference IM-SRG. For reasons of efficiency, we use the Magnus formulation of the multi-reference IM-SRG truncated at the multi-reference normal-ordered two-body level. All calculations start from a natural-orbital single-particle basis obtained from a perturbatively corrected density matrix \cite{Tichai:2018qge} constructed in a large HO space including 13 oscillator shells. 

We use the same sequence of SMS interactions from LO to N$^4$LO$^+$ in the NN sector supplemented with a 3N interaction at N$^2$LO as in the previous sections for few-body systems and light nuclei for two different cutoffs $\Lambda=450$ MeV and $500$ MeV. As in the NCCI calculations, the Hamiltonian is subject to a free-space SRG evolution including three-body terms and we evolve the translationally-invariant radius operator consistently at the two-body level. 

The results of the IM-NCSM calculations for the oxygen isotopic chain are presented in Fig.~\ref{fig:oxygen} and in Tab.~\ref{tab:oxygen}. All calculations are done with a simple $N_{\max}^{\text{ref}}=0$ reference space and with and $N_{\max}=2$ model space for the final NCSM calculation. We have confirmed in all cases that the calculations are converged with respect to $N_{\max}$. In order to address the uncertainties of the many-body scheme, we probe the dependence of the observables on $N_{\max}$, $N_{\max}^{\text{ref}}$, and the IM-SRG flow parameter. As for most ground-state calculations a variation of the $N_{\max}^{\text{ref}}$ truncation parameter has the largest impact on the observables. Therefore, we use the difference between $N_{\max}^{\text{ref}}=0$ and $2$ to assess the many-body uncertainties, which are approximately $2$ MeV for ground-state energies and $0.05$ fm for radii. These uncertainty estimates are confirmed by the explicit comparison of the  $^{14}$O ground-state energies reported in Tab.~\ref{tab:oxygen} for the IM-NCSM with the conventional NCSM results presented in Tab.~\ref{tab:EgsA10A12}. For all orders and cutoffs we observe excellent agreement of the two many-body approaches well within their respective uncertainties.

To assess the uncertainties due to the truncation of the chiral expansion, we employ the correlated Bayesian statistical model described in Sec.~\ref{sec:correlated_model}.
The interaction uncertainties are significantly larger than the estimated many-body uncertainties, therefore, we only show the interaction uncertainties in Fig.~\ref{fig:oxygen} as colored bands for N$^2$LO and N$^4$LO$^+$. In Tab.~\ref{tab:oxygen} the interaction uncertainties for all orders starting with NLO are given in parentheses.

Let us first discuss the systematics of the ground-state energies for the two cutoffs shown in the upper panels of Fig.~\ref{fig:oxygen}. As expected, the LO interaction does not provide a realistic description with ground states overbound by 50 to 80 MeV (cf. Tab.~\ref{tab:oxygen}). But already the NN interaction at NLO provides energies in a reasonable range compared to experiment. The energies obtained at N$^2$LO again show a sizable overbinding and deviate from the general systematics. A similar effect was already observed for the mid-$p$-shell isotopes in Fig.~\ref{fig:EgsA10A12}. 
Starting from N$^3$LO the energies are very stable up to the highest order---within the estimated uncertainties they agree across the different chiral orders and the two cutoffs. And they are in excellent agreement with experiment for all isotopes. This is remarkable, since the underlying chiral interactions was determined strictly in the $A=2$ and $A=3$ sector, without any information on heavier nuclei.   
The lower panels in Fig.~\ref{fig:oxygen} show the corresponding result for point-proton rms radii. Again, the radii at LO are unrealistically small, but NLO already provides a significant improvement. In line with the overbinding observed when going to N$^2$LO, the radii decrease further. From N$^2$LO to N$^3$LO we observe a systematic increase of the radii, which exhausts or even exceeds the N$^2$LO uncertainty band. From N$^3$LO on, the radii are very stable and consistent within uncertainties across the different orders and the two cutoff values. While the pattern correlates with the pattern observed for the ground-state energies, the converged values of the radii are significantly smaller than the structure radii extracted from the experimental charge radii for $^{16}$O and $^{18}$O---despite the excellent agreement for the energies. 

\begin{figure}[t]
  \includegraphics[width=0.53\columnwidth]{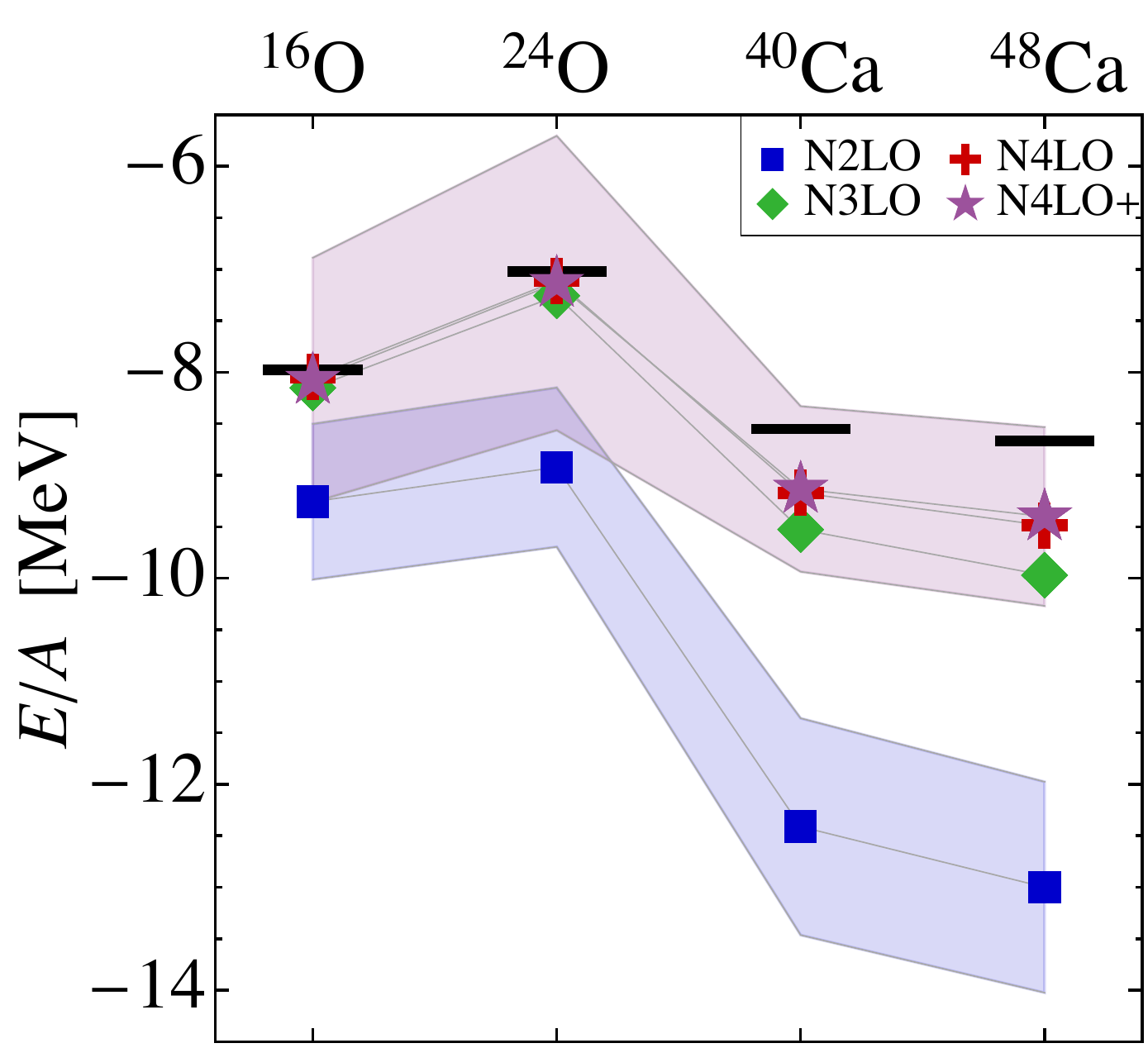}\hspace*{-3pt}
  \includegraphics[width=0.53\columnwidth]{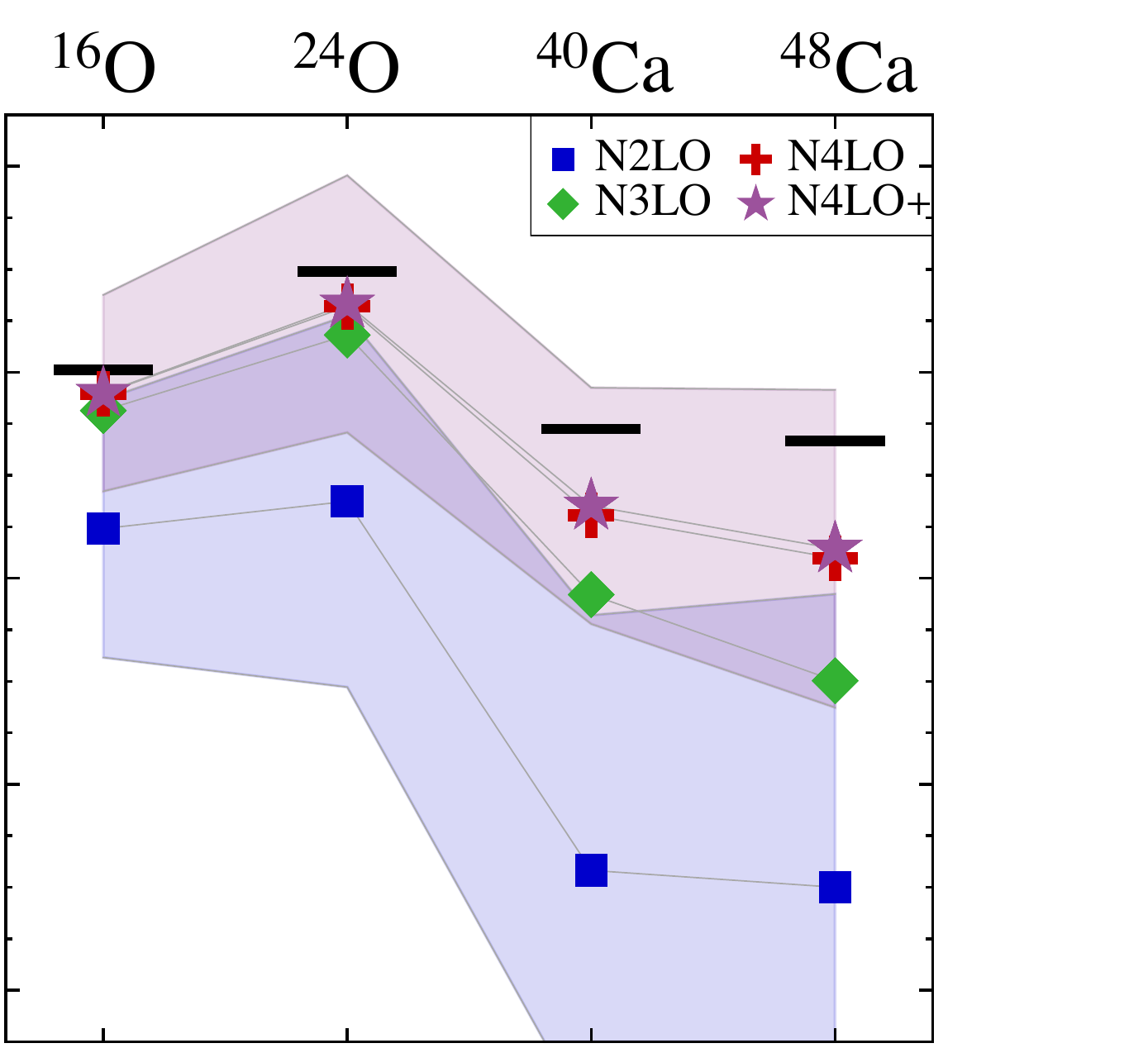}\\[-1pt]
  \includegraphics[width=0.53\columnwidth]{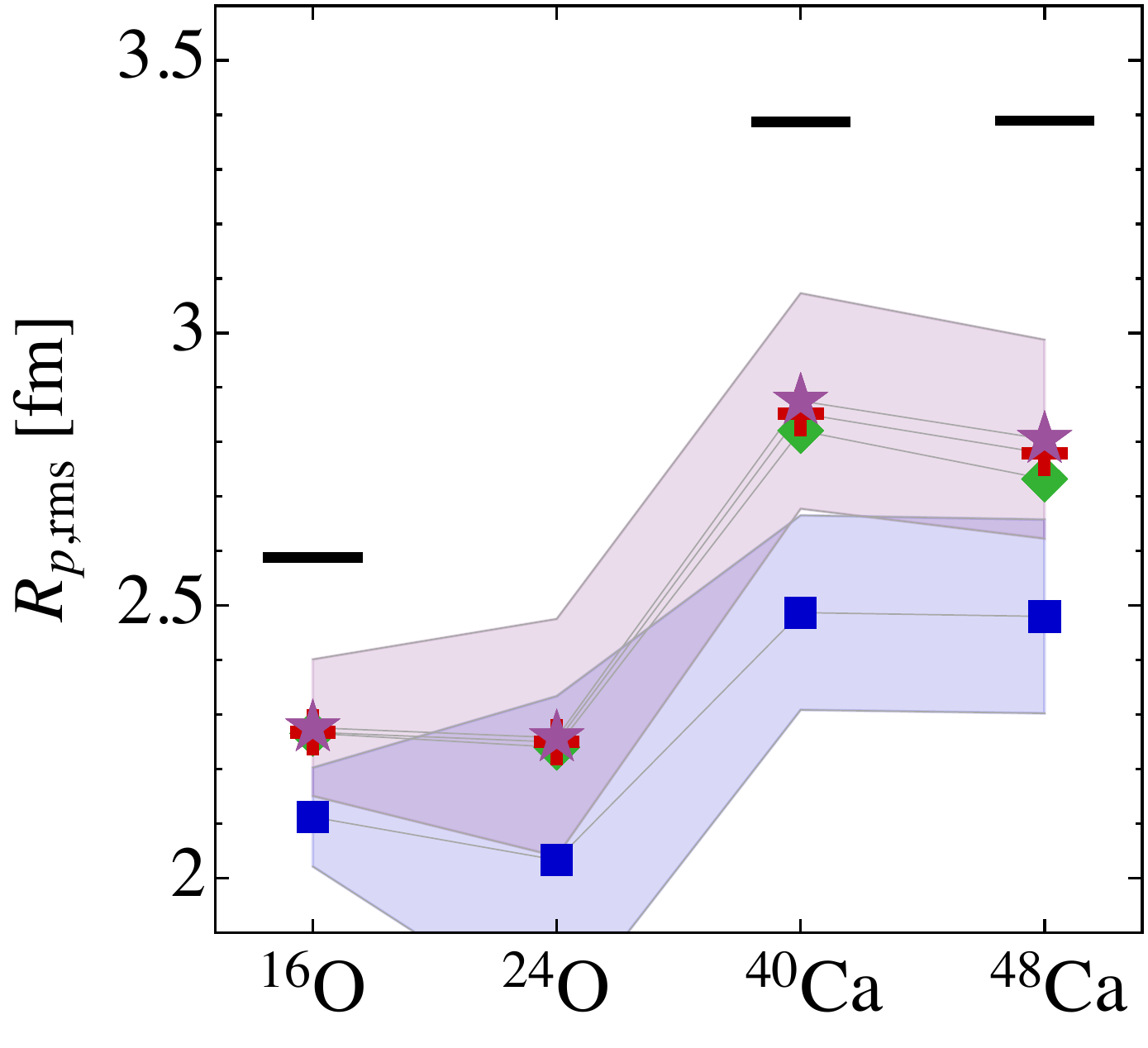}\hspace*{-3pt}
  \includegraphics[width=0.53\columnwidth]{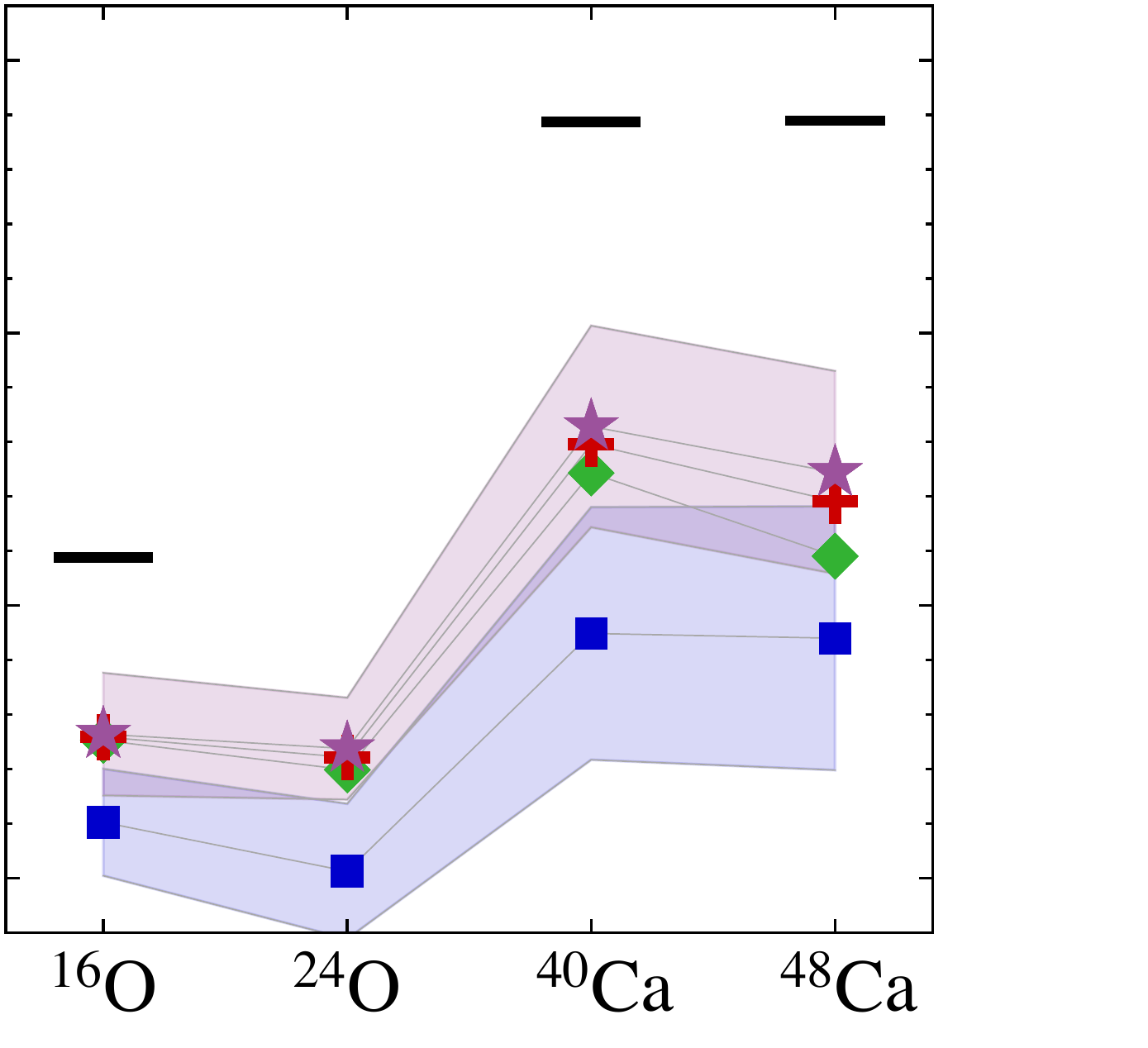}\\
  \caption{\label{fig:calcium} (Color online)
    Ground-state energies and point-proton radii for doubly-magic oxygen and calcium isotopes obtained in the IM-SRG with SMS interactions from NLO to N$^4$LO$^+$ for $\Lambda=450\,\text{MeV}$ (left-hand panels) and $\Lambda=500\,\text{MeV}$ (right-hand panels) with SRG flow parameter $\alpha=0.08\,\text{fm}^4$. The error bands show the chiral truncation uncertainties at the 95\% confidence level obtained with the pointwise Bayesian model for N$^2$LO and N$^4$LO$^+$. 
}
\end{figure}

These trends continue if we proceed to heavier nuclei. In Fig.~\ref{fig:calcium} we show the ground state energies and the rms radii of $^{16}$O and $^{24}$O as well as $^{40}$Ca and $^{48}$Ca obtained in single-reference IM-SRG calculations, which correspond to the $N_{\max}^{\text{ref}}=N_{\max}=0$ limit of the IM-NCSM for $^{16}$O and $^{40}$Ca. Also for the doubly-magic calcium isotopes, we observe a very nice convergence of the chiral expansion for both energies and radii. As before, N$^2$LO leads to  significant overbinding, but the higher orders stabilize quickly and agree within uncertainties. Though the ground-state energies are still in reasonable agreement with experiment, the underestimation of the radii is even more pronounced. For the calcium isotopes the radii at the highest chiral orders are by about $0.5$ fm too small compared to experiment, this corresponds to a reduction of the nuclear volume by almost 50\%. 

\begin{figure}[t]
  \includegraphics[width=0.8\columnwidth]{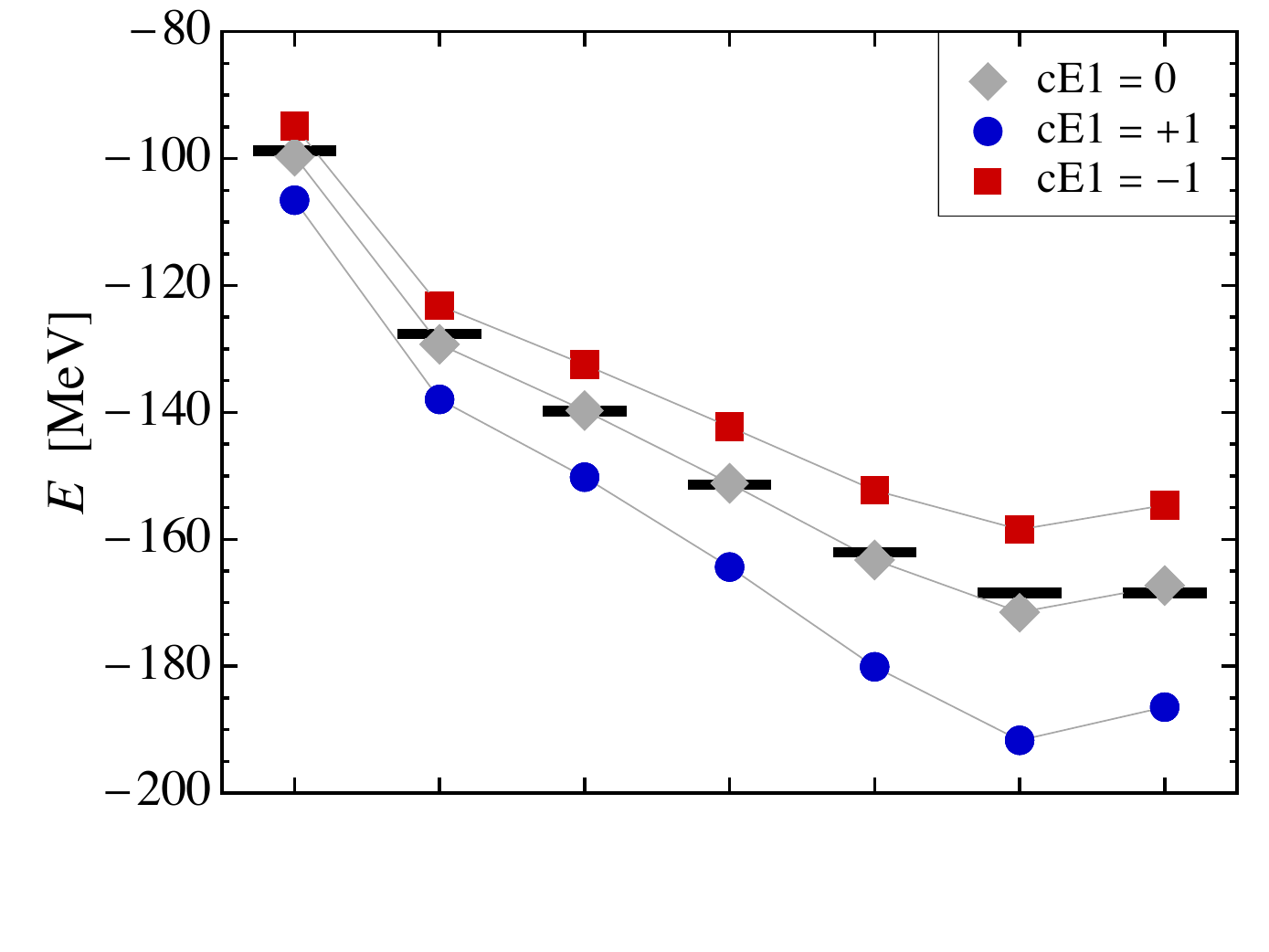}\\[-29pt]
  \includegraphics[width=0.8\columnwidth]{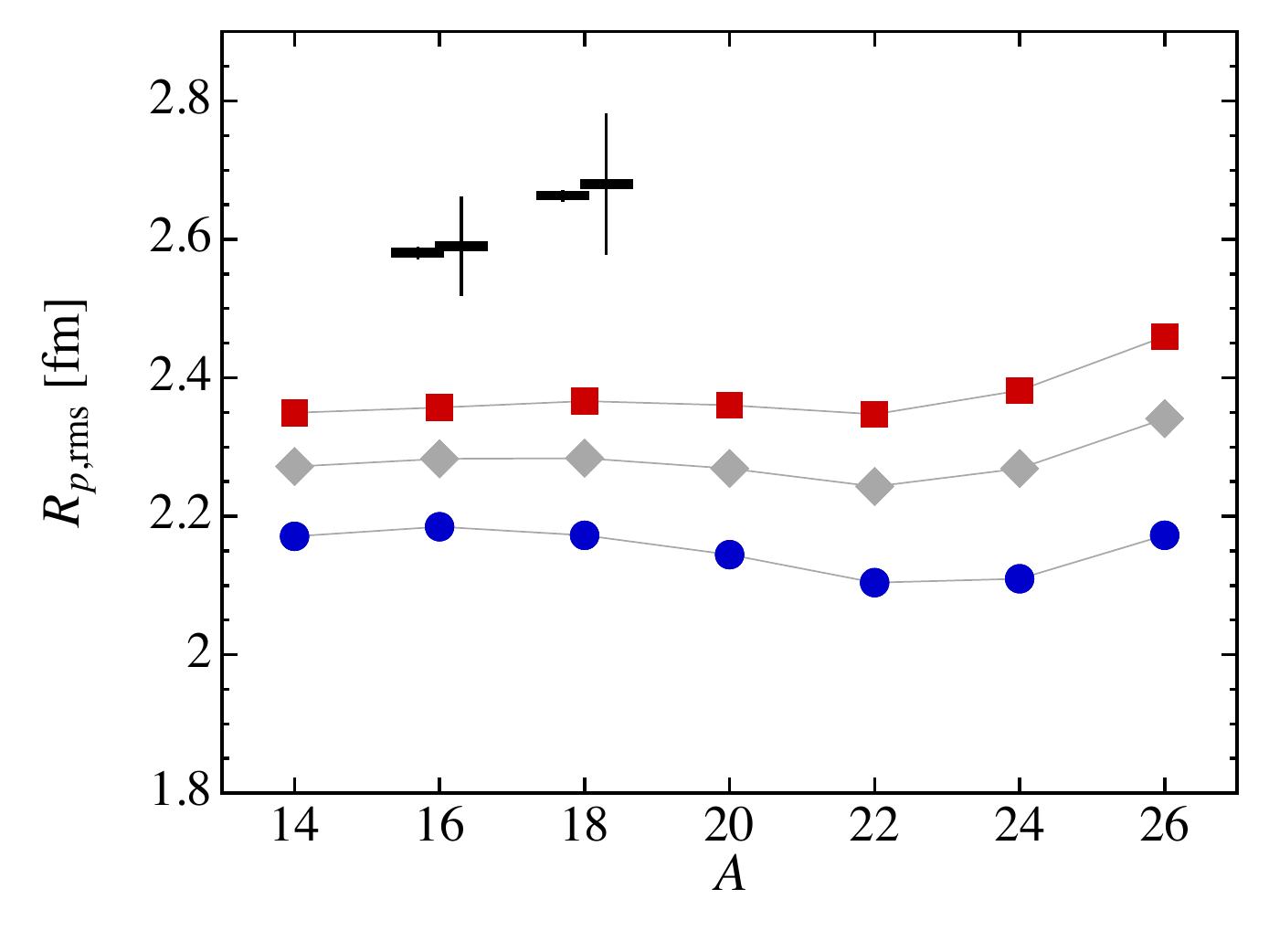}\\[-10pt]
  \caption{\label{fig:oxygen_cE1} (Color online)
  Ground-state energies and point-proton radii for even oxygen isotopes obtained in the IM-NCSM with the SMS interaction at N$^4$LO$^+$ for $\Lambda=450\,\text{MeV}$, supplemented by the $E1$ three-nucleon contact term at N$^4$LO with LEC values $c_{E1}=0,\pm1$.
}
\end{figure}

There are obvious limitations in the present calculations that might explain the systematic deviation for radii. Starting from N$^3$LO the 3N interaction is incomplete and while the additional 3N terms at N$^3$LO do not introduce additional LECs, the 3N terms at N$^4$LO come with a set of new 3N LECs. Work is in progress to derive all 3N contributions at N$^3$LO and N$^4$LO \cite{Bernard:2007sp,Bernard:2011zr,Girlanda:2011fh,Krebs:2012yv,Krebs:2013kha} and to compute the corresponding matrix elements in a partial-wave representation~\cite{Hebeler:2015wxa}. 
In order to probe the sensitivity of ground-state energies and radii to the sub-leading three-body contributions, particularly the terms with new LECs at N$^4$LO, we have selectively included the simplest, spin-isospin-independent contact term at N$^4$LO \cite{Girlanda:2011fh} with different values of the corresponding LEC $c_{E1}=-1,0,+1$ on top of the N$^4$LO$^+$ interaction. The resulting ground-state energies and radii for the oxygen isotopes obtained in the IM-NCSM are depicted in Fig.~\ref{fig:oxygen_cE1}. Clearly, these higher-order terms have the potential to significantly affect energies and radii. It remains to be seen whether the consistent inclusion of all terms will allow for a net change in the radii while keeping the good reproduction of the ground-state energies. 

Another limitation are the missing corrections to the charge density from exchange terms predicted in chiral EFT.
We are working on the consistent inclusion of these corrections to the charge densities and to the charge radius.

\section{Summary and conclusions}
\label{conclusion}

In this paper we have extended our earlier study \cite{Maris:2020qne} of few-nucleon
systems based on the SMS NN potentials along with the consistently regularized N$^2$LO 3NF by considering a broader range of
Nd scattering observables and heavier nuclei up to $^{48}$Ca. We have
also studied the role of higher-order corrections to the NN
interaction in connection with the systematic overbinding
trend for $A \gtrsim 10$ nuclei found in our earlier paper using the SMS N$^2$LO
NN potentials \cite{Maris:2020qne}. To quantify the contributions of
the NN interactions beyond N$^2$LO to various observables, we performed a
series of additional calculations using the SMS NN potentials at
N$^3$LO, N$^4$LO and N$^4$LO$^+$ orders of the EFT expansion in
combination with the N$^2$LO 3NF
and employing the same procedure to fix the LECs $c_D$ and $c_E$ from the
$^3$H binding energy and the differential cross section in Nd
scattering at $E = 70$~MeV. Clearly, from the point of view of
chiral EFT, the performed calculations can only be regarded complete to
N$^2$LO due to the missing contributions of many-body forces
beyond N$^2$LO. Yet, these results allowed us to get insights into
the convergence pattern of chiral EFT for light and medium-mass
nuclei and to extend and refine the Bayesian analysis of correlated truncation
errors. Moreover, they provide an important consistency check of our
previous calculations based on the SCS and SMS chiral EFT potentials. 
The main results of our paper can be summarized as
follows:
\begin{itemize}
\item
We have calculated selected Nd elastic scattering and breakup
observables. The obtained corrections to the neutron-deuteron total cross section at
$E =70$ and $135$~MeV stemming from the contributions to the NN force beyond
N$^2$LO agree well with expectations based on the power counting, as
revealed by  the estimated N$^2$LO truncation errors. We also found that
these corrections significantly reduce the residual cutoff dependence
of the calculated total cross sections. The predictions
based on the NN potentials at N$^3$LO and N$^4$LO$^+$ are consistent with the
experimental values within errors. This conclusion also holds for the
differential cross section and the vector and tensor
analyzing powers $A_Y({\rm d})$ and $A_{XX}- A_{YY}$ in elastic Nd
scattering at the considered energy of $E = 200$~MeV.
We have also calculated the differential cross section and the
analyzing powers $A_Y ({\rm N})$ and $A_{XX}$ for  selected breakup
configurations at $E = 135$ and $200$ MeV, finding again a
satisfactory agreement with the available experimental data.
\item
We have calculated the binding energies of the $A=3$ and $4$ nuclei
using  Faddeev-Yakubovsky equations as well as of selected $p$-shell nuclei with
$4 \leq A
\leq 14$ in the framework of the NCCI approach using  SRG
transformed two- and three-nucleon interactions. As already pointed
out in Ref.~\cite{Maris:2020qne},  the purely N$^2$LO calculations lead to a systematic
overbinding of nuclei with $A \gtrsim 10$ for both considered cutoff
values of $\Lambda = 450$~MeV and $500$~MeV, that increases with $A$
and reaches about $15\%$
for $^{14}$O. On the other hand, including the corrections to the NN
forces beyond N$^2$LO, the predicted ground-state energies of all
considered nuclei are found to be in very good agreement with the
experimental data.  We have also calculated  the
low-lying (narrow) excited state energies of the considered $p$-shell
nuclei, which are known to be strongly correlated \cite{Maris:2020qne}. To avoid an overestimation of
the truncation uncertainty, we performed a Bayesian analysis that
explicitly takes into account correlations between energy levels by using
Gaussian processes and learning the covariance structure from the
calculations. The larger set of results at different orders as compared to our earlier
paper 
Ref.~\cite{Maris:2020qne} allowed us to infer not only the characteristic variance
of the expansion coefficients $\bar c^2$ but also the value of the
dimensionless expansion parameter $Q$. For both considered cutoffs,  the posteriors are found
to be consistent with the prior value $Q \sim 0.3$ used in
Ref.~\cite{Maris:2020qne}.  All predicted excited state energies agree
with the data  within
errors, and we also observed little sensitivity in the spectra to the
higher-order corrections to the NN force. 
\item
We used IM-NCSM to study the ground state energies and point-proton
rms-radii of the even oxygen isotopes from $^{14}$O to $^{26}$O, while
the corresponding results for the calcium isotopes $^{40}$Ca and $^{48}$Ca
were obtained using the single-reference IM-SRG approach. For the
ground state energies, the resulting convergence pattern of the chiral
EFT expansion is similar to the one for lighter nuclei. In particular,
the strong overbinding observed when using both the NN interactions and the 3NF
at N$^2$LO is drastically reduced by taking into account the
contributions to the NN force beyond N$^2$LO. On the other hand, these
corrections still appear to be insufficient to reproduce the point-proton
radii for oxygen and calcium isotopes, which are significantly
underpredicted. The observed pattern is qualitatively similar to the
one reported in Ref.~\cite{Cipollone:2014hfa} using a different version of the chiral EFT
NN and 3N interactions in the framework of self-consistent Green's
function theory. To explore the impact of higher-order corrections to
the 3NF, we have included one particular short-range term  that
contributes to the 3NF at N$^4$LO with the corresponding
dimensionless LEC set to $c_{E1} = \pm 1$, see Ref.~\cite{Epelbaum:2019zqc} for a similar study in
Nd scattering.  We found sizable contributions to both the ground state
energies and radii, which also show a tendency to increase with $A$. These
results suggest that the employed Bayesian model for estimating
truncation uncertainties might become too optimistic for medium-mass
and heavier nuclei, see also Ref.~\cite{LENPIC:2018lzt} for similar conclusions and
a related discussion.
Finally, it is also worth pointing out that the short-range isoscalar NN
charge density operators were found to contribute
significantly to the deuteron charge and quadrupole form factors
\cite{Filin:2019eoe,Filin:2020tcs}, but their impact on the charge
radii of heavier nuclei and its scaling with $A$ have not
been investigated yet. 
\end{itemize}  
Clearly, to shed light on the remaining disagreement for the radii of
medium-mass nuclei it will be
necessary to perform complete calculations beyond N$^2$LO by taking
into account consistently regularized three- and four-nucleon forces
and the corresponding contributions to the charge density operator.
These studies would not only provide more accurate predictions for the
considered observables, but also increase the information about the
convergence pattern of the EFT expansion that can be used to refine
the Bayesian truncation model.  Work along these lines is in progress
by the LENPIC collaboration.

\section*{Acknowledgments}

This work was supported by BMBF (contract No.~05P21PCFP1 and 05P18RDFN1),
 by the DFG SFB 1245 (Projektnummer 279384907), 
 by the DFG and NSFC (DFG Project-ID 196253076 - TRR 110, NSFC Grant
 No. 11621131001), 
 by ERC NuclearTheory (grant No. 885150) and ERC EXOTIC (grant No. 101018170),
by the VolkswagenStiftung (Grant No. 93562),
by the EU Horizon 2020 research and innovation programme (STRONG-2020, grant agreement No. 824093),
by the US National Science Foundation under Grant NSF PHY--1913069, 
and by the US Department of Energy under Grants DE-FG02-87ER40371, 
DE-SC0018223 and DE-SC0018083.   
This research used resources of the National 
Energy Research Scientific Computing Center (NERSC) and
the Argonne Leadership Computing Facility (ALCF), which
are US Department of Energy Office of Science user facilities,
supported under Contracts No. DE-AC02-05CH11231 and
No. DE-AC02-06CH11357, and computing resources provided 
under the INCITE award `Nuclear Structure and Nuclear Reactions' from
the US Department of Energy, Office of Advanced Scientific Computing
Research. Further computing resources were provided on LICHTENBERG at 
the TU Darmstadt and on JURECA and the JURECA Booster 
of the J\"ulich Supercomputing Center, J\"ulich, Germany. 

\bibliographystyle{apsrev4-1}
\bibliography{lenpic_refs.bib}

\vfill

\end{document}